\documentclass[floatfix,twocolumn,showpacs,aps,tightenlines]{revtex4-1}
\usepackage{graphicx}
\usepackage{color}
\usepackage{psfrag}
\usepackage{dcolumn}
\usepackage{bm}

\usepackage{amssymb}
\usepackage{amsmath}
\usepackage{amsfonts}
\usepackage{longtable}
\usepackage{xspace}

\newcommand{\Spar}{\tilde S_\|}
\newcommand{\Dpar}{\tilde \Delta_\|}
\newcommand{\dmt}{\delta{m}}

\newcommand{\beq}{\begin{equation}}
\newcommand{\eeq}{\end{equation}}

\newcommand{\be}{\begin{equation}}
\newcommand{\ee}{\end{equation}}
\newcommand{\bea}{\begin{eqnarray}}
\newcommand{\eea}{\end{eqnarray}}
\newcommand{\bes}{\begin{subequations}}
\newcommand{\ees}{\end{subequations}}

\usepackage{bm}

\begin{document}

\title{Remnant of binary black-hole mergers:\\
New simulations and peak luminosity studies}

\author{James Healy}
\affiliation{Center for Computational Relativity and Gravitation,
School of Mathematical Sciences,
Rochester Institute of Technology, 85 Lomb Memorial Drive, Rochester,
New York 14623}

\author{Carlos O. Lousto}
\affiliation{Center for Computational Relativity and Gravitation,
School of Mathematical Sciences,
Rochester Institute of Technology, 85 Lomb Memorial Drive, Rochester,
New York 14623}

\date{\today}

\begin{abstract}
We present the results of 61 new simulations of
nonprecessing spinning black hole binaries with mass ratios 
$q=m_1/m_2$ in the range
$1/3\leq q\leq1$ and individual spins covering the parameter space
$-0.85\leq\alpha_{1,2}\leq0.85$.  We additionally perform 10
new simulations of nonspinning black hole binaries with mass ratios
covering the range $1/6\leq q<1$.
We follow the evolution for typically the last
ten orbits before merger down to the formation of the final remnant
black hole. This allows for assessment of the accuracy of our
previous empirical formulae for relating the binary parameters to 
the remnant final black hole mass, spin and recoil.
We use the new simulation to improve the fit to the above remnant
formulae and add a formula for the peak luminosity of gravitational
waves, produced around the merger of the two horizons into one.
We find excellent agreement (typical errors $\sim0.1-0.2\%$) for the 
mass and spin, and within $\sim5\%$ for the recoil and peak luminosity.
These formulae have direct application to parameter estimation
techniques applied to LIGO observations of gravitational waves from
binary black hole mergers.
\end{abstract}

 \maketitle

\section{Introduction}\label{sec:Intro}

The breakthroughs~\cite{Pretorius:2005gq,Campanelli:2005dd,Baker:2005vv}
in numerical relativity allow us to make detailed predictions for the
gravitational waves from the late inspiral, plunge, merger and
ringdown of black hole binary systems (BHB). The first generic, long-term
precessing binary black hole evolution without any symmetry was
performed in Ref.~\cite{Campanelli:2008nk}, including detailed comparison with 
post-Newtonian $\ell=2,3$ waveforms. 
Those predictions of numerical relativity simulations of BHB have
been recently confirmed 
by the first direct observation of gravitational
waves~\cite{Abbott:2016blz,Abbott:2016nmj,TheLIGOScientific:2016pea}
from such binary systems and by its comparison to targeted runs 
\cite{Abbott:2016apu,Lovelace:2016uwp}.
The observations are consistent with general relativity as the
correct theory for gravity as discussed in
\cite{TheLIGOScientific:2016src,TheLIGOScientific:2016pea}.

Numerical relativity techniques also allow us to explore the late
binary dynamics, beyond the post-Newtonian regime. 
For instance to study the {\it hangup}, i.e. the role individual black hole
spins play to delay or accelerate their merger \cite{Campanelli:2006uy},
the determination of the magnitude and direction 
of the {\it recoil} velocity of the final merged black hole
\cite{Campanelli:2007ew,Campanelli:2007cga,Lousto:2011kp},
and the {\it flip-flop} of individual spins during the orbital phase
\cite{Lousto:2014ida,Lousto:2015uwa,Lousto:2016nlp}.
The challenging regime of very high spins is now reachable
by numerical simulations~\cite{Lovelace:2014twa}.

In a hierarchical search or parameter estimation of gravitational
wave signals, aligned spin binaries, i.e. nonprecessing, play
an important role, since the main dynamics is given by the spin
components along the orbital angular momentum \cite{Lousto:2013vpa}
and the full 7 dimensional (assuming very small
eccentricities) parameter space of a binary is currently very
hard to cover with solely full numerical waveforms.

In Ref.~\cite{Healy:2014yta} we used 37 original runs and those available in 
the literature to determine fitting formulae that related aligned
spin binaries initial parameters, mass ratio and intrinsic individual
spins $(q=m_1/m_2,\alpha_1=S_1/m_1^2,\alpha_2=S_2/m_2^2)$ to
the final black hole mass, spin and recoil
$(m_f,\alpha_f,V_f)$. Here we revisit this scenario and perform
71 new runs that allow us to evaluate the errors of the previous
fitting formulae and then use the new results to improve the fittings.

The waveforms produced by these new simulations, covering mass
ratios in the range $1/3\leq q\leq1$ for the spinning binaries
and  $1/6\leq q<1$ for nonspinning ones will form the basis of
the new RIT's waveform catalog. The combination of RIT's and 
SXS \cite{Mroue:2013xna} and Gatech \cite{Jani:2016wkt}, 
catalogs can be used as a data bank for parameter
identification of gravitational wave signals \cite{Abbott:2016apu}.

The paper is organized as follows. Section \ref{sec:FN}
describes the methods and criteria for producing the new
simulations. Sec.~\ref{sec:Fits} describes the results
obtained for the remnant properties, i.e. final mass, spin, and
recoil, and the comparison with the predicted values from
the phenomenological formula of Ref. \cite{Healy:2014yta}.
We also introduce a fitting formula for the gravitational wave's
peak luminosity and use the new runs to better fit the 
mass, spin and recoil parameters. The new fitting parameters
are determined and provided in Tables~\ref{tab:fitparsms}
and \ref{tab:fitparsVL}.
In Sec.~\ref{sec:Discussion}, the use and potential extensions 
to this work to precessing binaries are discussed. In the appendices we give
tables containing the remnant information and fitting parameters,
as well as detailed studies of the numerical convergence and 
error estimates for those quantities.

\section{Full numerical evolutions}\label{sec:FN}


To compute the numerical initial data, we use the puncture
approach~\cite{Brandt97b} along with the {\sc
TwoPunctures}~\cite{Ansorg:2004ds} thorn.
We evolve the following BBH data sets using the {\sc
LazEv}~\cite{Zlochower:2005bj} implementation of the moving puncture
approach~\cite{Campanelli:2005dd,Baker:2005vv} with the conformal
function $W=\sqrt{\chi}=\exp(-2\phi)$ suggested by
Ref.~\cite{Marronetti:2007wz}.  For the run presented here, we use
centered, sixth-order finite differencing in
space~\cite{Lousto:2007rj} and a fourth-order Runge Kutta time
integrator. 
The use of a sixth-order spatial finite difference allow us to gain
a factor $\sim4/3$ with the respect to the eight-order implementation
due to the reduction of the ghost zones from 4 to 3.
We also allowed for a Courant factor $CFL=1/3$ instead of the
previous $CFL=0.25$ \cite{Zlochower:2012fk} 
gaining another speedup factor of 4/3. 
We verified that for this relaxing of the time integration step
we still conserve the horizon masses and spins of the individual
black holes during evolution and lock the phase of the gravitational
waveforms to acceptable levels (below $10^{-5}$). 
This, plus the use of the new
Xsede supercomputer {\it Comet} at 
SDSC~\footnote{https://portal.xsede.org/sdsc-comet}, lead to typical
evolution speeds of $250m/day$ on 16 nodes. Note that our
previous \cite{Lousto:2013oza,Lousto:2015uwa}
comparable simulations averages $\sim100m/day$.

Our code uses the {\sc EinsteinToolkit}~\cite{Loffler:2011ay,
einsteintoolkit} / {\sc Cactus}~\cite{cactus_web} /
{\sc Carpet}~\cite{Schnetter-etal-03b}
infrastructure.  The {\sc
Carpet} mesh refinement driver provides a
``moving boxes'' style of mesh refinement. In this approach, refined
grids of fixed size are arranged about the coordinate centers of both
holes.  The {\sc Carpet} code then moves these fine grids about the
computational domain by following the trajectories of the two BHs.

We use {\sc AHFinderDirect}~\cite{Thornburg2003:AH-finding} to locate
apparent horizons.  We measure the magnitude of the horizon spin using
the {\it isolated horizon} (IH) algorithm detailed in
Ref.~\cite{Dreyer02a} and as implemented in Ref.~\cite{Campanelli:2006fy}.
Note that once we have the horizon spin, we can calculate the horizon
mass via the Christodoulou formula 
${m_H} = \sqrt{m_{\rm irr}^2 + S_H^2/(4 m_{\rm irr}^2)}\,,$
where $m_{\rm irr} = \sqrt{A/(16 \pi)}$, $A$ is the surface area of
the horizon, and $S_H$ is the spin angular momentum of the BH (in
units of $m^2$, where $m=m_1+m_2$).  In the tables below, we use the variation in the
measured horizon irreducible mass and spin during the simulation as a
measure of the error in computing these quantities.  
We measure radiated energy,
linear momentum, and angular momentum, in terms of the radiative Weyl
Scalar $\psi_4$, using the formulas provided in
Refs.~\cite{Campanelli:1998jv,Lousto:2007mh}. However, rather than
using the full $\psi_4$, we decompose it into $\ell$ and $m$ modes and
solve for the radiated linear momentum, dropping terms with $\ell >
6$.  The formulas in Refs.~\cite{Campanelli:1998jv,Lousto:2007mh} are
valid at $r=\infty$.  We extract the radiated energy-momentum at
finite radius and extrapolate to $r=\infty$. We find that the new
perturbative extrapolation described in Ref.~\cite{Nakano:2015pta} provides the
most accurate waveforms. While the difference of fitting both linear and
quadratic extrapolations provides an independent measure of the error.

\subsection{New simulations}\label{subsec:newruns}

In order to supplement our previous work on aligned binaries
\cite{Healy:2014yta}
we consider a set of full numerical simulations with
initial configurations as described in Table~\ref{tab:ID}
and Table~\ref{tab:ID_nonspin}.
These simulations cover the parameter space of aligned 
(and counteraligned) spins for comparable mass ratios $q$ up
to 1:3. We have chosen comparable mass ratios 
and spin magnitude $\alpha_i\leq0.85$ as they are less 
demanding computationally than the more extreme cases.
In particular, they span over the median values of
mass ratio and spin parameters estimated for
the black hole binaries associated with the gravitational
wave signals GW150914 ($q=0.81^{+0.17}_{-0.20}$)
and GW151226 ($q=0.52^{+0.40}_{-0.29}$) and 
the transient LVT151012 ($q=0.57^{+0.38}_{-0.37}$) 
\cite{TheLIGOScientific:2016pea}.
We have also considered initial separations of the binary such 
that they produce at least 6 orbits prior to merger, with typical 
simulations producing 8-10 orbits and above. The study of higher
spins and longer waveforms (either through simulations or hybridization)
will be the subject of a forthcoming paper by the authors.

\begin{table*}
\caption{Initial data parameters for the quasi-circular
configurations with a smaller mass black hole (labeled 1),
and a larger mass spinning black hole (labeled 2). The punctures are located
at $\vec r_1 = (x_1,0,0)$ and $\vec r_2 = (x_2,0,0)$, with momenta
$P=\pm (P_r, P_t,0)$, spins $\vec S_i = (0, 0, S_i)$, mass parameters
$m^p/m$, horizon (Christodoulou) masses $m^H/m$, total ADM mass
$M_{\rm ADM}$, and dimensionless spins $a/m_H = S/m_H^2$.
The configuration are denoted by QX\_Y\_Z, where X gives the mass
ratio $m^H_1 / m^H_2$, Y gives the  spin of the smaller BH
($a_1/m_H^2$), and Z
gives the spin of the larger BH $(a_2/m_H^2)$.
}\label{tab:ID}
\begin{ruledtabular}
\begin{tabular}{lccccccccccccc}
Run   & $x_1/m$ & $x_2/m$  & $P_r/m$    & $P_t/m$ & $m^p_1/m$ & $m^p_2/m$ & $S_1/m^2$ & $S_2/m^2$ & $m^H_1/m$ & $m^H_2/m$ & $M_{\rm ADM}/m$ & $a_1/m_1^H$ & $a_2/m_2^H$\\
\hline

1 & -8.81 & 2.94 & -3.69e-04 & 0.06634 & 0.2405 & 0.6585 & 0 & -0.2812 & 0.25 & 0.75 & 0.9934 & 0 & -0.5 \\
2 & -7.69 & 2.56 & -4.73e-04 & 0.06929 & 0.2393 & 0.6575 & 0 & 0.2812 & 0.25 & 0.75 & 0.9924 & 0 & 0.5 \\
3 & -9.19 & 3.06 & 0 & 0.06508 & 0.2408 & 0.3912 & 0 & -0.4781 & 0.25 & 0.75 & 0.9937 & 0 & -0.85 \\
4 & -7.50 & 2.50 & -4.80e-04 & 0.0691 & 0.2391 & 0.3903 & 0 & 0.4781 & 0.25 & 0.75 & 0.9921 & 0 & 0.85 \\
5 & -9.00 & 3.00 & -3.34e-04 & 0.06503 & 0.2137 & 0.7226 & -0.03125 & -0.1406 & 0.25 & 0.75 & 0.9935 & -0.5 & -0.25 \\
6 & -8.25 & 2.75 & -4.03e-04 & 0.0674 & 0.2131 & 0.7219 & -0.03125 & 0.1406 & 0.25 & 0.75 & 0.9929 & -0.5 & 0.25 \\
7 & -8.25 & 2.75 & -4.30e-04 & 0.0682 & 0.213 & 0.7219 & 0.03125 & -0.1406 & 0.25 & 0.75 & 0.993 & 0.5 & -0.25 \\
8 & -7.69 & 2.56 & -4.89e-04 & 0.06974 & 0.2124 & 0.7213 & 0.03125 & 0.1406 & 0.25 & 0.75 & 0.9924 & 0.5 & 0.25 \\
9 & -9.38 & 3.12 & 0 & 0.06404 & 0.214 & 0.5871 & -0.03125 & -0.3656 & 0.25 & 0.75 & 0.9938 & -0.5 & -0.65 \\
10 & -7.69 & 2.56 & -4.69e-04 & 0.06921 & 0.2125 & 0.5857 & -0.03125 & 0.3656 & 0.25 & 0.75 & 0.9924 & -0.5 & 0.65 \\
11 & -8.62 & 2.88 & -4.02e-04 & 0.06741 & 0.2133 & 0.5866 & 0.03125 & -0.3656 & 0.25 & 0.75 & 0.9934 & 0.5 & -0.65 \\
12 & -7.50 & 2.50 & -4.90e-04 & 0.06938 & 0.2123 & 0.5856 & 0.03125 & 0.3656 & 0.25 & 0.75 & 0.9922 & 0.5 & 0.65 \\
13 & -8.81 & 2.94 & 0 & 0.06526 & 0.1495 & 0.7415 & -0.05 & 0 & 0.25 & 0.75 & 0.9934 & -0.8 & 0 \\
14 & -8.25 & 2.75 & 0 & 0.06687 & 0.1492 & 0.7409 & 0.05 & 0 & 0.25 & 0.75 & 0.9929 & 0.8 & 0 \\
15 & -9.38 & 3.12 & 0 & 0.06388 & 0.1498 & 0.659 & -0.05 & -0.2812 & 0.25 & 0.75 & 0.9938 & -0.8 & -0.5 \\
16 & -7.88 & 2.62 & 0 & 0.06869 & 0.1489 & 0.6576 & -0.05 & 0.2812 & 0.25 & 0.75 & 0.9926 & -0.8 & 0.5 \\
17 & -8.25 & 2.75 & 0 & 0.06841 & 0.1491 & 0.658 & 0.05 & -0.2812 & 0.25 & 0.75 & 0.9931 & 0.8 & -0.5 \\
18 & -7.50 & 2.50 & 0 & 0.0694 & 0.1486 & 0.6573 & 0.05 & 0.2812 & 0.25 & 0.75 & 0.9922 & 0.8 & 0.5 \\
19 & -10.12 & 3.38 & 0 & 0.06125 & 0.1502 & 0.462 & -0.05 & -0.45 & 0.25 & 0.75 & 0.9943 & -0.8 & -0.8 \\
20 & -7.12 & 2.38 & 0 & 0.07049 & 0.1483 & 0.4601 & 0.05 & 0.45 & 0.25 & 0.75 & 0.9918 & 0.8 & 0.8 \\
21 & -8.67 & 4.33 & -3.74e-04 & 0.0738 & 0.3233 & 0.4089 & 0 & -0.3556 & 0.3333 & 0.6667 & 0.9929 & 0 & -0.8 \\
22 & -7.33 & 3.67 & -4.99e-04 & 0.07755 & 0.3217 & 0.4079 & 0 & 0.3556 & 0.3333 & 0.6667 & 0.9914 & 0 & 0.8 \\
23 & -8.33 & 4.17 & -3.91e-04 & 0.07451 & 0.2868 & 0.6543 & -0.05556 & -0.04444 & 0.3333 & 0.6667 & 0.9925 & -0.5 & -0.1 \\
24 & -8.00 & 4.00 & -4.07e-04 & 0.07478 & 0.2865 & 0.654 & 0.05556 & 0.04444 & 0.3333 & 0.6667 & 0.9921 & 0.5 & 0.1 \\
25 & -6.67 & 3.33 & -7.42e-04 & 0.08443 & 0.2846 & 0.5813 & -0.05556 & 0.2222 & 0.3333 & 0.6667 & 0.9909 & -0.5 & 0.5 \\
26 & -7.33 & 3.67 & 0 & 0.08082 & 0.2855 & 0.5823 & 0.05556 & -0.2222 & 0.3333 & 0.6667 & 0.9916 & 0.5 & -0.5 \\
27 & -8.67 & 4.33 & -3.74e-04 & 0.07384 & 0.2871 & 0.5449 & -0.05556 & -0.2667 & 0.3333 & 0.6667 & 0.9929 & -0.5 & -0.6 \\
28 & -7.67 & 3.83 & -4.33e-04 & 0.07537 & 0.2861 & 0.544 & 0.05556 & 0.2667 & 0.3333 & 0.6667 & 0.9917 & 0.5 & 0.6 \\
29 & -5.71 & 4.29 & -9.10e-04 & 0.09328 & 0.4146 & 0.5435 & 0 & 0.08163 & 0.4286 & 0.5714 & 0.9899 & 0 & 0.25 \\
30 & -6.29 & 4.71 & 0 & 0.0895 & 0.4158 & 0.4965 & 0 & -0.1633 & 0.4286 & 0.5714 & 0.9908 & 0 & -0.5 \\
31 & -6.29 & 4.71 & 0 & 0.08653 & 0.4159 & 0.4966 & 0 & 0.1633 & 0.4286 & 0.5714 & 0.9905 & 0 & 0.5 \\
32 & -6.29 & 4.71 & 0 & 0.0904 & 0.4157 & 0.3477 & 0 & -0.2612 & 0.4286 & 0.5714 & 0.9909 & 0 & -0.8 \\
33 & -6.29 & 4.71 & 0 & 0.08564 & 0.416 & 0.3478 & 0 & 0.2612 & 0.4286 & 0.5714 & 0.9905 & 0 & 0.8 \\
34 & -5.71 & 4.29 & -9.31e-04 & 0.0939 & 0.404 & 0.5435 & -0.04592 & 0.08163 & 0.4286 & 0.5714 & 0.9899 & -0.25 & 0.25 \\
35 & -5.71 & 4.29 & -9.59e-04 & 0.09446 & 0.4039 & 0.5435 & 0.04592 & -0.08163 & 0.4286 & 0.5714 & 0.99 & 0.25 & -0.25 \\
36 & -6.29 & 4.71 & -7.22e-04 & 0.08936 & 0.3693 & 0.559 & -0.09184 & 0 & 0.4286 & 0.5714 & 0.9908 & -0.5 & 0 \\
37 & -6.29 & 4.71 & -6.54e-04 & 0.08722 & 0.3693 & 0.5591 & 0.09184 & 0 & 0.4286 & 0.5714 & 0.9906 & 0.5 & 0 \\
38 & -5.71 & 4.29 & 0 & 0.09422 & 0.3682 & 0.5435 & -0.09184 & 0.08163 & 0.4286 & 0.5714 & 0.9899 & -0.5 & 0.25 \\
39 & -5.71 & 4.29 & -9.34e-04 & 0.09383 & 0.3681 & 0.5436 & 0.09184 & -0.08163 & 0.4286 & 0.5714 & 0.99 & 0.5 & -0.25 \\
40 & -6.29 & 4.71 & 0 & 0.09061 & 0.3692 & 0.4964 & -0.09184 & -0.1633 & 0.4286 & 0.5714 & 0.9909 & -0.5 & -0.5 \\
41 & -6.29 & 4.71 & 0 & 0.08753 & 0.3694 & 0.4965 & -0.09184 & 0.1633 & 0.4286 & 0.5714 & 0.9906 & -0.5 & 0.5 \\
42 & -6.29 & 4.71 & 0 & 0.08839 & 0.3693 & 0.4965 & 0.09184 & -0.1633 & 0.4286 & 0.5714 & 0.9907 & 0.5 & -0.5 \\
43 & -6.29 & 4.71 & 0 & 0.08553 & 0.3694 & 0.4966 & 0.09184 & 0.1633 & 0.4286 & 0.5714 & 0.9905 & 0.5 & 0.5 \\
44 & -6.29 & 4.71 & 0 & 0.0866 & 0.3694 & 0.3478 & -0.09184 & 0.2612 & 0.4286 & 0.5714 & 0.9906 & -0.5 & 0.8 \\
45 & -6.29 & 4.71 & 0 & 0.0897 & 0.2586 & 0.559 & -0.1469 & 0 & 0.4286 & 0.5714 & 0.9909 & -0.8 & 0 \\
46 & -6.29 & 4.71 & 0 & 0.08633 & 0.2586 & 0.5592 & 0.1469 & 0 & 0.4286 & 0.5714 & 0.9906 & 0.8 & 0 \\
47 & -5.71 & 4.29 & 0 & 0.09286 & 0.2579 & 0.3469 & -0.1469 & 0.2612 & 0.4286 & 0.5714 & 0.99 & -0.8 & 0.8 \\
48 & -6.29 & 4.71 & 0 & 0.08779 & 0.2192 & 0.4479 & -0.1561 & 0.2082 & 0.4286 & 0.5714 & 0.9908 & -0.85 & 0.6375 \\
49 & -6.29 & 4.71 & 0 & 0.088 & 0.2192 & 0.448 & 0.1561 & -0.2082 & 0.4286 & 0.5714 & 0.9908 & 0.85 & -0.6375 \\
50 & -6.73 & 5.52 & -4.90e-04 & 0.08312 & 0.4021 & 0.5136 & -0.08932 & 0.09963 & 0.4505 & 0.5495 & 0.9914 & -0.44 & 0.33 \\
51 & -5.50 & 5.50 & 0 & 0.09112 & 0.4871 & 0.4327 & 0 & -0.125 & 0.5 & 0.5 & 0.9905 & 0 & -0.5 \\
52 & -5.50 & 5.50 & 0 & 0.0919 & 0.4871 & 0.303 & 0 & -0.2 & 0.5 & 0.5 & 0.9907 & 0 & -0.8 \\
53 & -6.00 & 6.00 & -5.50e-04 & 0.0856 & 0.4757 & 0.4882 & -0.0625 & 0 & 0.5 & 0.5 & 0.9912 & -0.25 & 0 \\
54 & -6.00 & 6.00 & -5.66e-04 & 0.08617 & 0.4757 & 0.4757 & -0.0625 & -0.0625 & 0.5 & 0.5 & 0.9912 & -0.25 & -0.25 \\
55 & -6.00 & 6.00 & -5.34e-04 & 0.08504 & 0.4758 & 0.4757 & -0.0625 & 0.0625 & 0.5 & 0.5 & 0.9911 & -0.25 & 0.25 \\
56 & -6.00 & 6.00 & -5.06e-04 & 0.08392 & 0.4758 & 0.4758 & 0.0625 & 0.0625 & 0.5 & 0.5 & 0.991 & 0.25 & 0.25 \\
57 & -5.50 & 5.50 & 0 & 0.09247 & 0.4326 & 0.4326 & -0.125 & -0.125 & 0.5 & 0.5 & 0.9907 & -0.5 & -0.5 \\
58 & -5.50 & 5.50 & -6.38e-04 & 0.08751 & 0.4328 & 0.4328 & 0.125 & 0.125 & 0.5 & 0.5 & 0.9903 & 0.5 & 0.5 \\
59 & -6.50 & 6.50 & 0 & 0.08277 & 0.3042 & 0.4556 & -0.2 & -0.1 & 0.5 & 0.5 & 0.992 & -0.8 & -0.4 \\
60 & -5.50 & 5.50 & 0 & 0.09411 & 0.3029 & 0.3029 & -0.2 & -0.2 & 0.5 & 0.5 & 0.991 & -0.8 & -0.8 \\
61 & -5.00 & 5.00 & -8.19e-04 & 0.09128 & 0.3023 & 0.3023 & 0.2 & 0.2 & 0.5 & 0.5 & 0.9895 & 0.8 & 0.8 \\
\end{tabular}
\end{ruledtabular}
\end{table*}

\begin{table*}
\caption{Initial data parameters for the quasi-circular
configurations as in Table \ref{tab:ID} but for the non-spinning
systems.}
\label{tab:ID_nonspin}
\begin{ruledtabular}
\begin{tabular}{lccccccccccccc}
Run   & $x_1/m$ & $x_2/m$  & $P_r/m$    & $P_t/m$ & $m^p_1/m$ & $m^p_2/m$ & $S_1/m^2$ & $S_2/m^2$ & $m^H_1/m$ & $m^H_2/m$ & $M_{\rm ADM}/m$ & $a_1/m_1^H$ & $a_2/m_2^H$\\
\hline

62 & -9.00 & 1.50 & -2.19e-04 & 0.0459 & 0.1358 & 0.8511 & 0 & 0 & 0.1429 & 0.8571 & 0.9952 & 0 & 0 \\
63 & -8.96 & 1.79 & -2.55e-04 & 0.05116 & 0.1589 & 0.8266 & 0 & 0 & 0.1667 & 0.8333 & 0.9947 & 0 & 0 \\
64 & -8.80 & 2.20 & -3.08e-04 & 0.05794 & 0.1913 & 0.7923 & 0 & 0 & 0.2 & 0.8 & 0.994 & 0 & 0 \\
65 & -8.44 & 2.81 & -3.83e-04 & 0.06677 & 0.2401 & 0.7411 & 0 & 0 & 0.25 & 0.75 & 0.993 & 0 & 0 \\
66 & -8.04 & 3.21 & -4.50e-04 & 0.07262 & 0.2751 & 0.7045 & 0 & 0 & 0.2857 & 0.7143 & 0.9924 & 0 & 0 \\
67 & -7.33 & 3.67 & -5.72e-04 & 0.0802 & 0.3216 & 0.6557 & 0 & 0 & 0.3333 & 0.6667 & 0.9916 & 0 & 0 \\
68 & -7.19 & 4.31 & -5.46e-04 & 0.08206 & 0.3632 & 0.6138 & 0 & 0 & 0.375 & 0.625 & 0.9914 & 0 & 0 \\
69 & -7.05 & 4.70 & -5.29e-04 & 0.08281 & 0.3883 & 0.5887 & 0 & 0 & 0.4 & 0.6 & 0.9913 & 0 & 0 \\
70 & -6.29 & 4.71 & -6.86e-04 & 0.08828 & 0.4159 & 0.5591 & 0 & 0 & 0.4286 & 0.5714 & 0.9907 & 0 & 0 \\
71 & -6.49 & 5.51 & -5.29e-04 & 0.08448 & 0.4477 & 0.529 & 0 & 0 & 0.4595 & 0.5405 & 0.9912 & 0 & 0 \\

\end{tabular}
\end{ruledtabular}
\end{table*}

\begin{figure}
\includegraphics[angle=270,width=0.49\columnwidth]{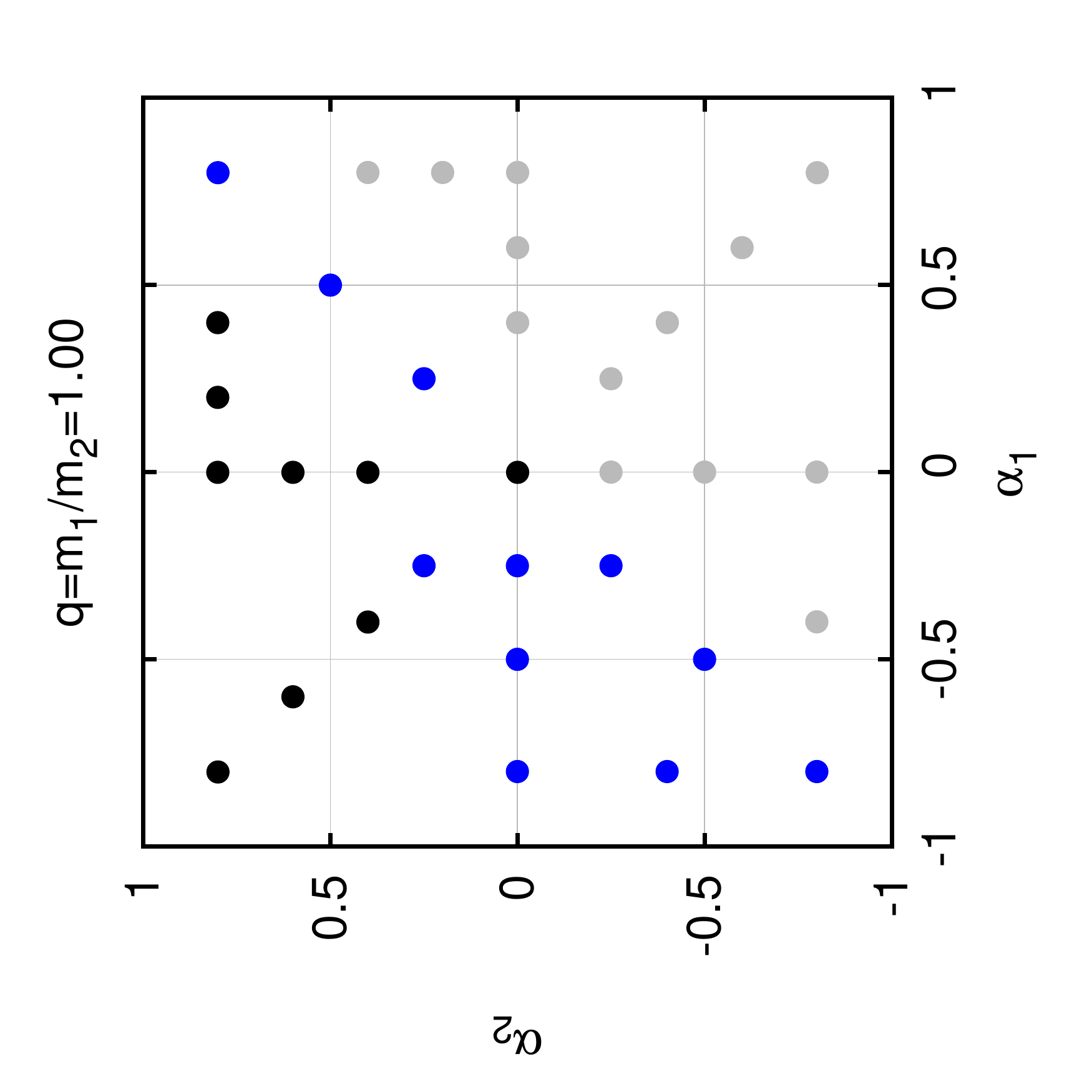}
\includegraphics[angle=270,width=0.49\columnwidth]{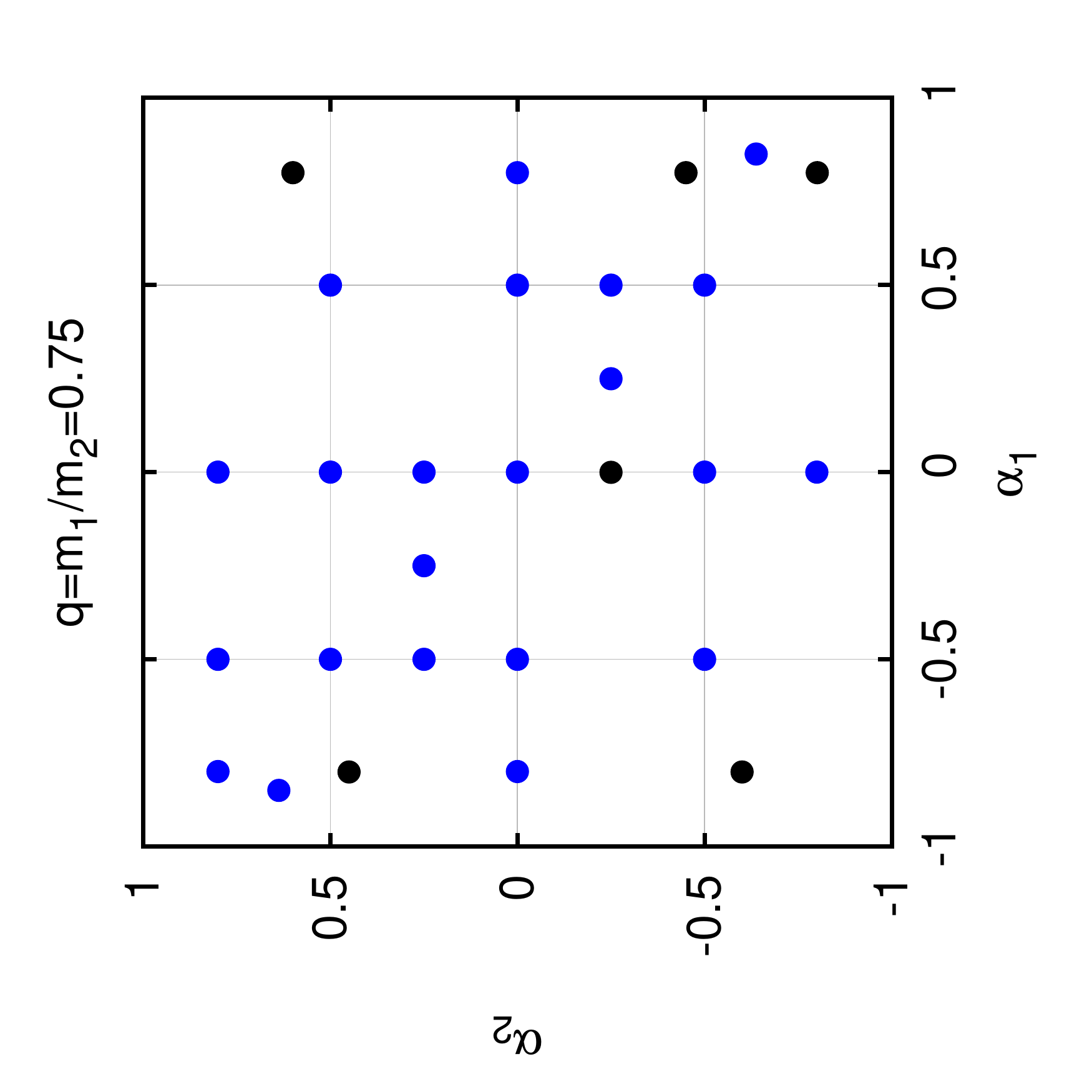}\\
\includegraphics[angle=270,width=0.49\columnwidth]{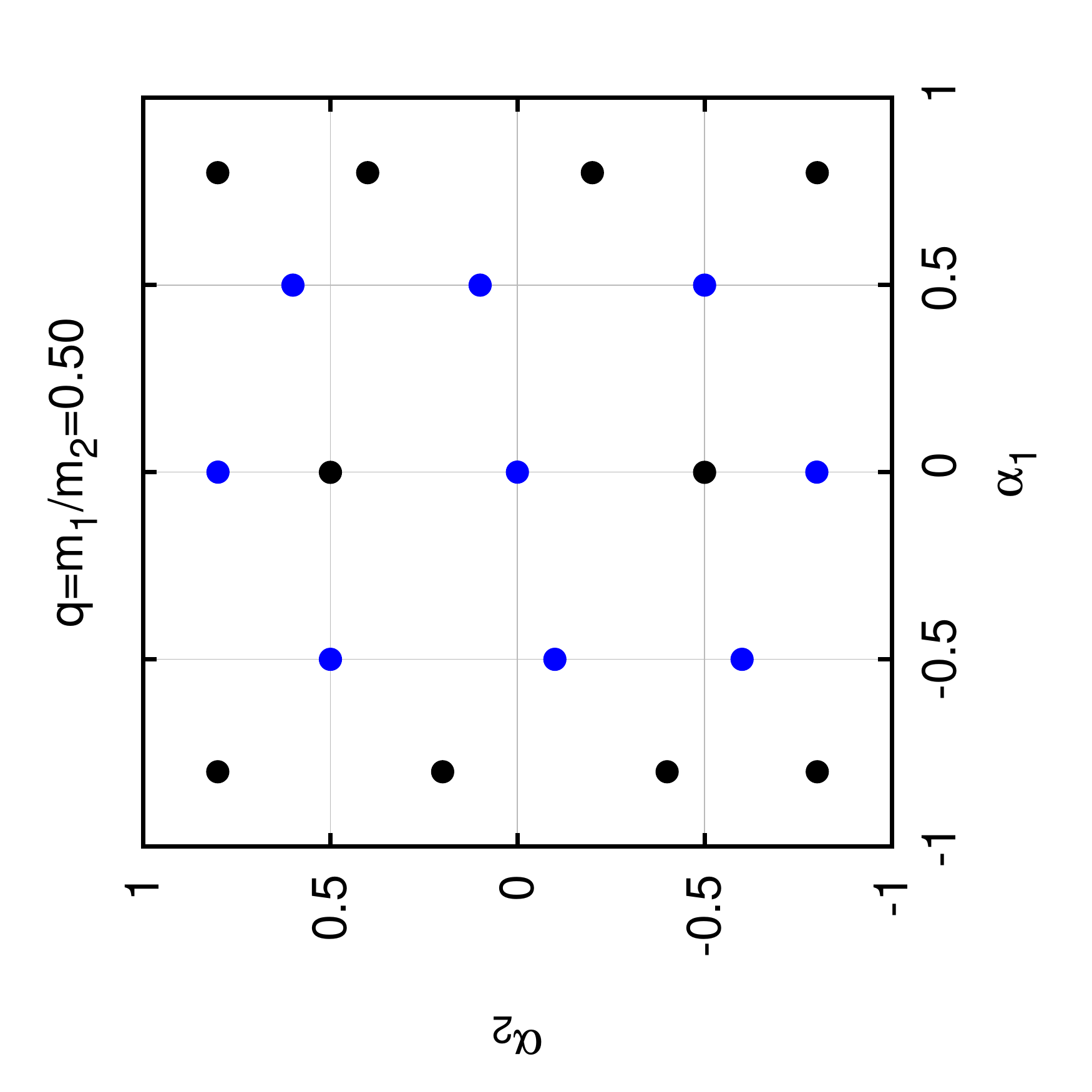}
\includegraphics[angle=270,width=0.49\columnwidth]{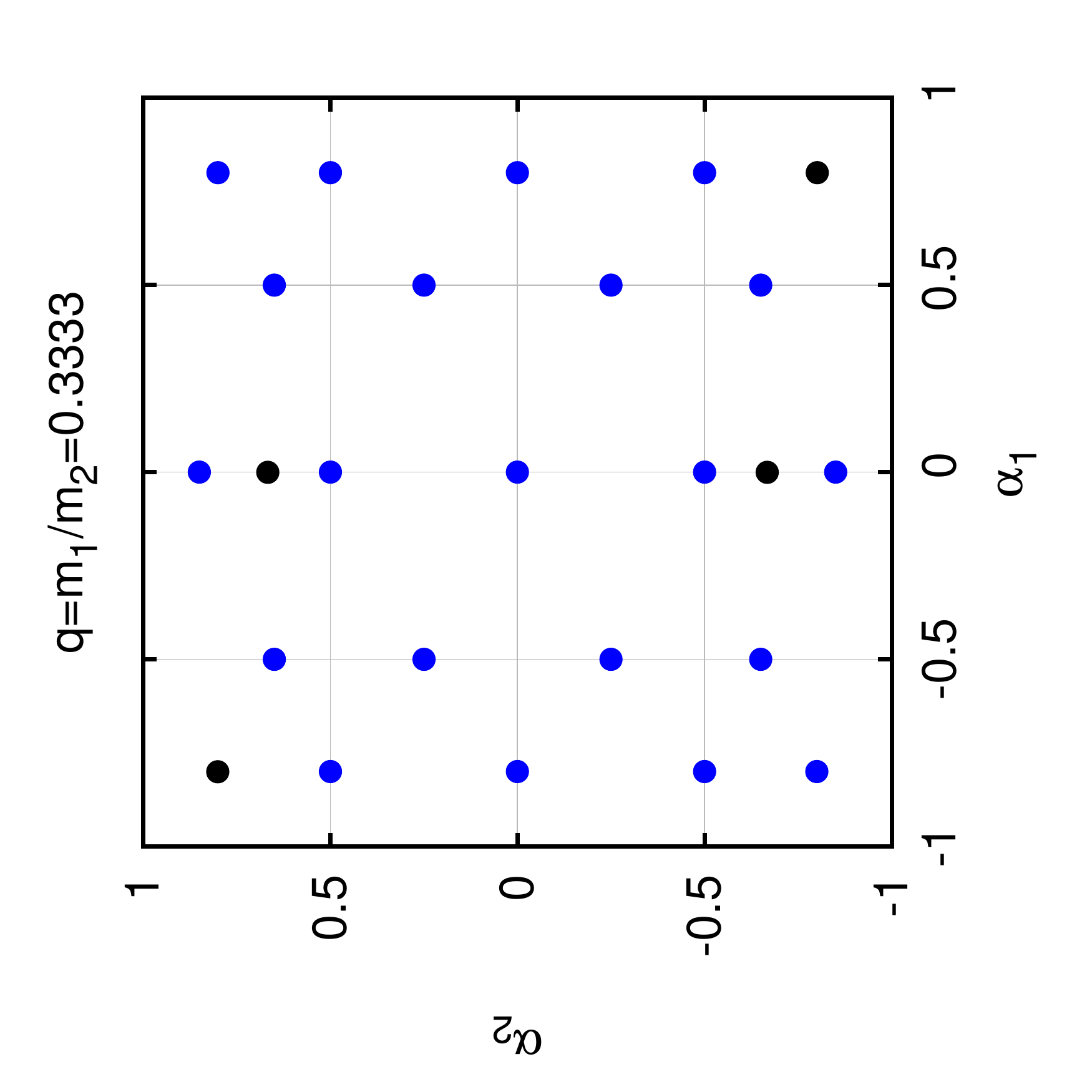}
\caption{Initial configuration of the simulated binaries. The 
black dots represent the previous simulations reported in 
\cite{Healy:2014yta} and those in blue are the new simulations of this
paper. Those points in light gray are obtained by the symmetry 
$1\leftrightarrow2$ of the case $q=1$.
\label{fig:configs}}
\end{figure}

Note that our new runs supplement the previous study as 
depicted in Fig.~\ref{fig:configs}. Those runs have been in part
selected to cover the parameter space of aligned binaries on surfaces
of equal $\tilde{S}_{0}=(\alpha_2+q\,\alpha_1)/(1+q)$ and in part motivated
to better model the gravitational wave events mentioned above.
They also supplement other independent studies, thus providing a tighter grid
of simulations that have been used to directly estimate the parameters
of the black hole binary that produced GW150914\cite{Abbott:2016apu}.
One of the runs provided here (with $q=0.82$) was compared with a 
totally independent simulation by the SXS collaboration finding 
excellent agreement for the waveform as well as the final remnant parameters
\cite{Lovelace:2016uwp}.

\section{Remnant and Luminosity fitting formulae}\label{sec:Fits}

Besides the interest in producing waveforms for direct comparison
with observation, the simulation of orbiting black hole binaries produce
information about the final remnant of the merger of the two holes.
This was already the subject of studies using the Lazarus approach
\cite{Baker:2003ds} previous to the breakthrough that allowed 
longer accurate computations
and lead to numerous empirical formulae relating the initial parameters
(individual masses and spins) of the binary to those of the final
remnant, such as its mass, spin, and recoil velocity
\cite{Barausse:2012qz,Rezzolla:2007rz,Hofmann:2016yih,Jimenez-Forteza:2016oae,Lousto:2009mf,Lousto:2013wta,Healy:2014yta,Zlochower:2015wga}, or algebraic properties of the final metric
\cite{Campanelli:2008dv,Owen:2010vw}.
The computation of the peak luminosity has also recently been the
subject of interest in relation to the observation of gravitational waves
\cite{TheLIGOScientific:2016wfe,TheLIGOScientific:2016pea,Keitel:2016krm}.

In general, the mapping of the 7-dimensional space (assuming quasicircular
orbits) of a binary into the 4-dimensional space of the final black hole 
mass and spin is
a complicated process because of the task of accurately accounting 
for the precession of the individual spins particularly 
during the latest stages.
In the case studied here, aligned binaries, the symmetry of the problem
prevents orbital plane precession and we have obtained accurate
representations of the empirical remnant formulae by a fourth order
polynomial expansion in the appropriate binary 
variables~\cite{Lousto:2012gt,Lousto:2013wta}
and imposing the particle limit analytically~\cite{Healy:2014yta}.

In order to verify the predictions of these formulae we cover the 
parameter space of comparable mass binaries with a new set of 71 simulations
that nearly doubles the original 37 used in \cite{Healy:2014yta}
and we extend the fittings to also model the peak luminosity.
We summarize the results of the 
evolution of the new 61 spinning and 10 nonspinning binaries 
in tables \ref{tab:ecc}-\ref{tab:kicks_paper1} given in the appendix.

Table \ref{tab:ecc} for the 61 spinning cases and 
table \ref{tab:ecc_nonspin} for the 10 nonspinning ones
display the orbital initial frequency of the binary as well
as its initial measured eccentricity (small but not zero as
we used PN quasicircular orbits \cite{Baker:2001sf,Husa:2007rh},
but did not seek to iterate the parameters towards zero as
in \cite{Pfeiffer:2007yz}). We also provide the eccentricity at merger
(as the binary circularizes by emitting gravitational waves) and
number of orbits before the merger actually takes place in
the simulation (as measured by the formation of a common (apparent) horizon). 
Finally, for cross reference with other papers
using these runs 
we provide the technical name used to identify the run.

Because the puncture initial data we use \cite{Brandt97b}
assumes a conformally flat 3-metric it has a non-physical 
(albeit generally small) wave content that gets radiated away
or absorbed by the holes. In order to provide a more physical
set of initial parameters to be used into the remnant formulae,
we extract the parameters of the black holes after the
values of the horizon masses and spin settles. We observe
that evaluating those at a time $t=150m$ after the beginning of
the simulation is very 
accurate and we provide those values in Table \ref{tab:IDr}
and Table \ref{tab:IDr_nonspin} for the 61 spinning and 10 nonspinning
binaries respectively.

The remnant properties are displayed in Tables \ref{tab:spinerad} 
and \ref{tab:spinerad_nonspin}
for the final mass and spin of the product of the merger
of the 61 spinning and 10 nonspinning binaries. We compare in
those tables the determination of the final mass and spin from
the direct measure of the isolated horizon quantities using
\cite{Dreyer02a} or from the conservation of the ADM mass and momentum
of the system, i.e. subtracting from them the radiated energy 
and momentum carried by the waveforms to obtain the remnant mass and spin of
the final black hole. The level of agreement of those quantities
provides an estimate of the errors in determining the final
mass and spin of the hole. In our experience using the isolated
horizon quantities provides an excellent approximation with
a much smaller error than the above estimate.

Finally we also provide Tables \ref{tab:kicks} -   \ref{tab:kicks_nonspin}
 (for the new runs)
\ref{tab:kicks_paper1} (for the previous \cite{Healy:2014yta} runs)
with the recoil
velocity and peak luminosity as directly computed from the 
waveforms (adding up to $\ell=6$ modes). Also included are
the weights used in the least-square fittings.  
We study in 
appendix~\ref{app:errors} 
the error estimates of in those quantities produced by the
finite difference resolutions we used in the simulations and
incorporate this information into the fitting formulae below.

\subsection{Fitting final mass and spin}\label{subsec:finalms}

We compare the results of these new simulations with the 
predictions of the previous fit to
evaluate the errors of the previous fitting given in 
Ref.~\cite{Healy:2014yta}.

The fitting formula for $M_{\rm rem}$ is given by,
\begin{eqnarray}\label{eq:4mass}
\frac{M_{\rm rem}}{m} = (4\eta)^2\,\Big\{M_0 + K_1 \Spar+ K_{2a}\,\Dpar\dmt +
                     K_{2b}\,\Spar^2+ \nonumber\\
                     K_{2c}\,\Dpar^2+
                     K_{2d}\,\dmt^2 +
                     K_{3a}\,\Dpar\Spar\dmt+ \nonumber\\
                     K_{3b}\,\Spar\Dpar^2+
                     K_{3c}\,\Spar^3+ \nonumber\\
                     K_{3d}\,\Spar\dmt^2+
                     K_{4a}\,\Dpar\Spar^2\dmt + \nonumber\\
                     K_{4b}\,\Dpar^3\dmt +
                     K_{4c}\,\Dpar^4+
                     K_{4d}\,\Spar^4+\nonumber\\
                     K_{4e}\,\Dpar^2 \Spar^2+
                     K_{4f}\,\dmt^4+
                     K_{4g}\,\Dpar\dmt^3+\nonumber\\
                     K_{4h}\,\Dpar^2\dmt^2 +
                     K_{4i}\,\Spar^2\dmt^2\Big\}+\nonumber\\
                     \left[1+\eta(\tilde{E}_{\rm ISCO}+11)\right]\dmt^6,\quad\,
\end{eqnarray}

and the fitting formula for the final spin has the form,

\begin{eqnarray}\label{eq:4spin}
\alpha_{\rm rem} = \frac{S_{\rm rem}}{M^2_{\rm rem}} =
                     (4\eta)^2\Big\{L_0 + L_{1}\,\Spar+\nonumber\\ 
                     L_{2a}\,\Dpar\dmt+
                     L_{2b}\,\Spar^2+
                     L_{2c}\,\Dpar^2+
                     L_{2d}\,\dmt^2+\nonumber\\
                     L_{3a}\,\Dpar\Spar\dmt+
                     L_{3b}\,\Spar\Dpar^2+
                     L_{3c}\,\Spar^3+\nonumber\\
                     L_{3d}\,\Spar\dmt^2+
                     L_{4a}\,\Dpar\Spar^2\dmt+
                     L_{4b}\,\Dpar^3\dmt+\nonumber\\
                     L_{4c}\,\Dpar^4+
                     L_{4d}\,\Spar^4+
                     L_{4e}\,\Dpar^2\Spar^2+\nonumber\\
                     L_{4f}\,\dmt^4+
                     L_{4g}\,\Dpar\dmt^3+\nonumber\\
                     L_{4h}\,\Dpar^2\dmt^2+
                     L_{4i}\,\Spar^2\dmt^2\Big\}+\nonumber\\
                     \Spar(1+8\eta)\dmt^4+\eta\tilde{J}_{\rm ISCO}\dmt^6.
\end{eqnarray}
Note that the two formulae above impose the particle limit by including
the ISCO dependencies (See Ref. \cite{Healy:2014yta,Ori:2000zn} for the explicit expressions).

Here, as in  Ref.~\cite{Healy:2014yta}, we use the notation
\begin{eqnarray*}
  m = m_1 + m_2,\\
\delta m = \frac{m_1 - m_2}{m},\\
 \tilde{S} = (\vec S_1 + \vec S_2)/m^2,\\
  \tilde{\Delta}  = (\vec S_2/m_2 - \vec S_1/m_1)/m,
\end{eqnarray*}
where $m_i$ is the mass of BH $i=1,2$ and $\vec S_i$ is the spin of BH
$i$.
We also use the auxiliary variables
\begin{eqnarray*}
 \eta = \frac{m_1 m_2}{m^2},\\
 q=\frac{m_1}{m_2},\\
 \vec \alpha_i = \vec S_i/m_i^2,
\end{eqnarray*}
where $|\vec \alpha_i| \leq 1$ is the dimensionless spin of BH $i$,
and we use the convention that $m_1 \leq m_2$ and hence $q\leq 1$.
Here the  index $\perp$ and $\|$
refer to components perpendicular to and parallel to
 the orbital angular momentum.

\begin{figure}
\includegraphics[width=0.95\columnwidth]{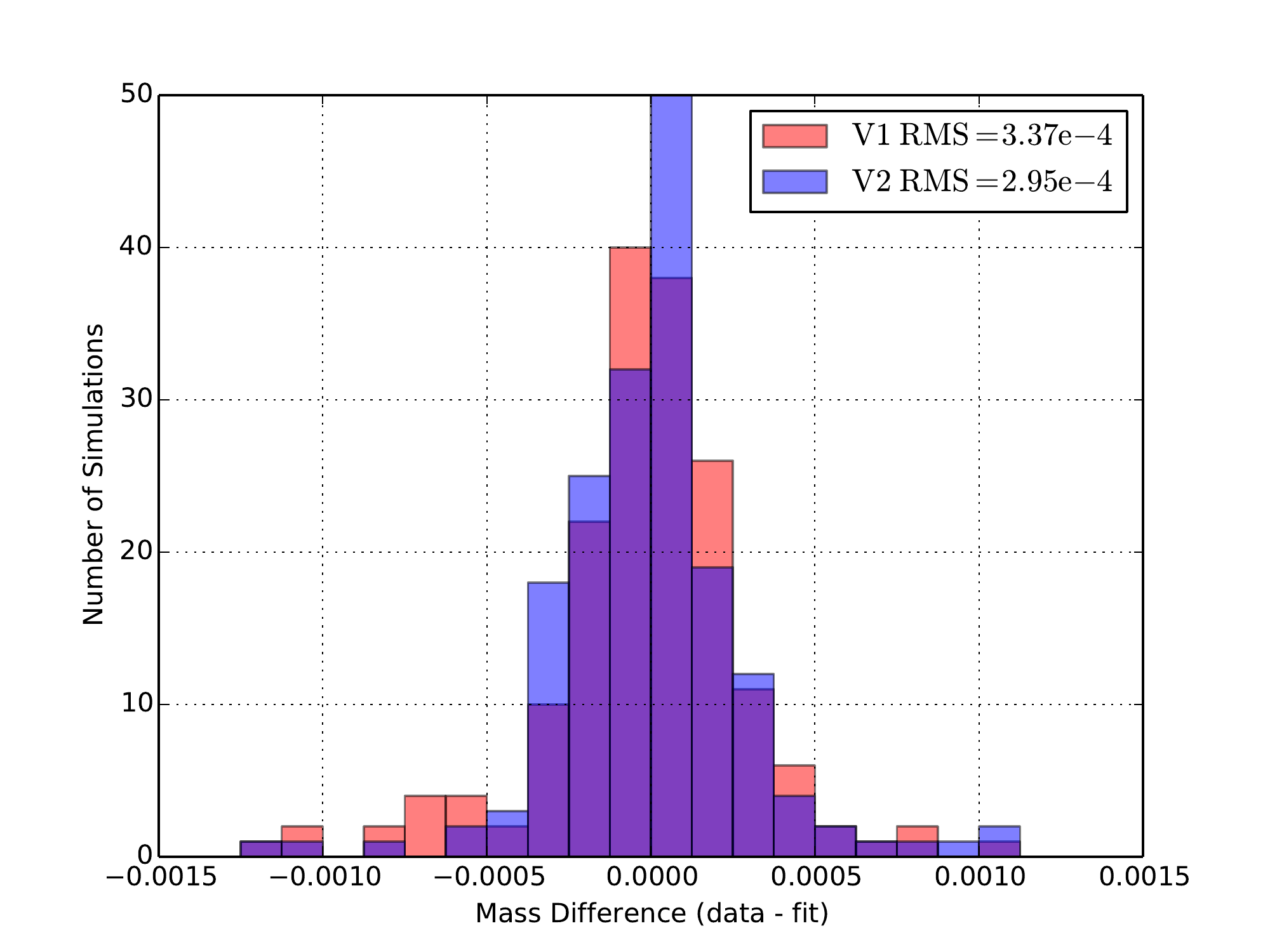}\\
\includegraphics[width=0.95\columnwidth]{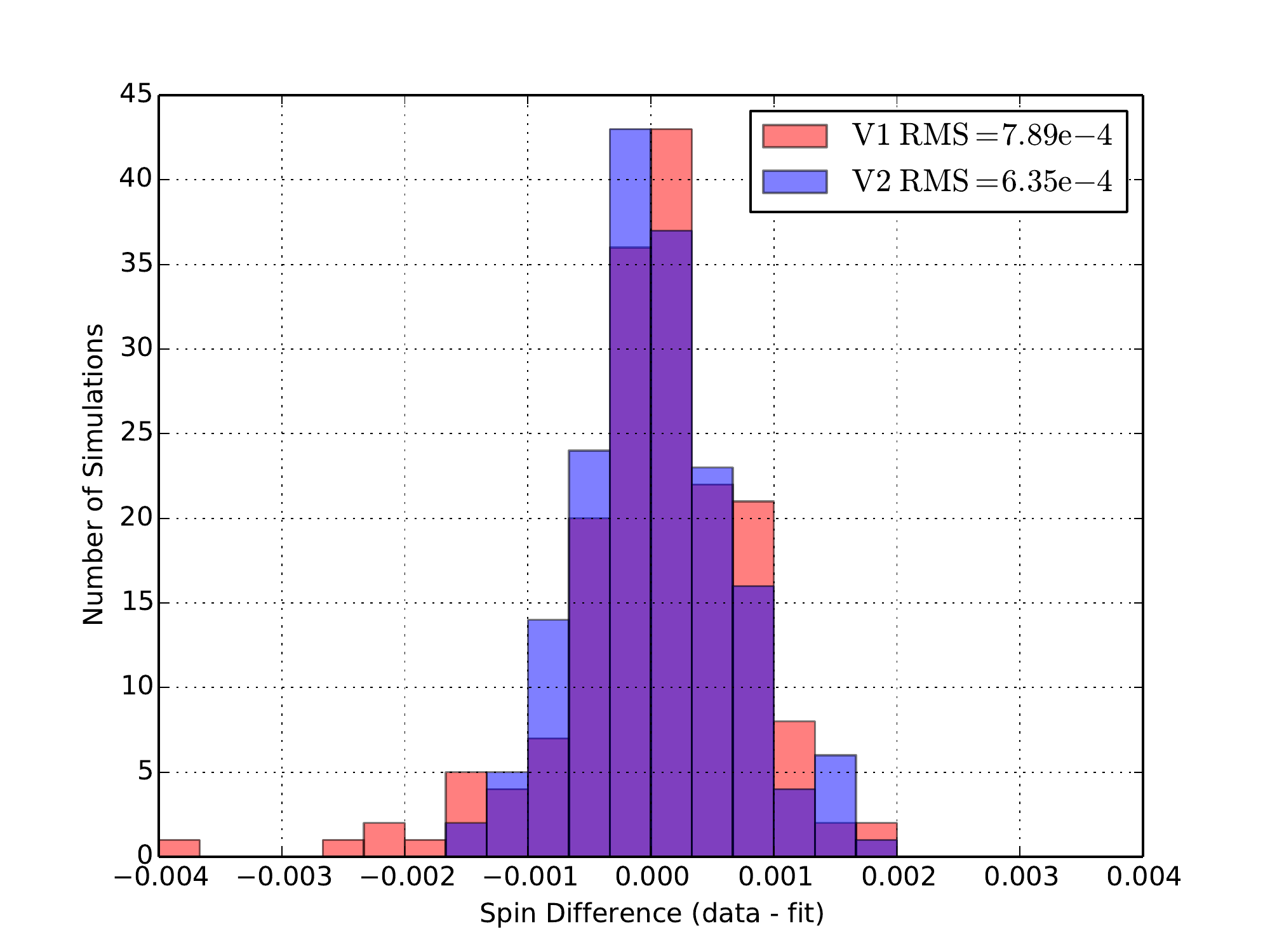}
\caption{Counts of the number of runs whose residual (data - fit) 
falls within a particular
bin. Both cases have 25 bins over the visible range with the final mass
shown in the top panel, and final spin in the bottom panel. 
V1 labels the fitting from Ref.~\cite{Healy:2014yta} 
and V2 labels the fitting from this paper.
\label{fig:res1}}
\end{figure}

The results of the
comparison of all available data in Ref.~\cite{Healy:2014yta} plus the
new data and SXS data for the mass and spins (175 runs in total)
with the previous formulae are displayed as the red histograms 
in Fig.~\ref{fig:res1} and are labeled as $V1$.
We also compare these same data points with the new fitted parameters (See
below) and the results are summarized by the blue
histograms labeled as $V2$.

From Fig.~\ref{fig:res1}, 
we observe that the predictions using the previous formulae 
for the final mass and spin are very accurate even for the new
runs. Note that in general the new runs started 
at longer separations and used a different numerical evolution system,
as mentioned in the previous section. 
Given this success we will continue to use the form of
the fitting formulae as in Ref.~\cite{Healy:2014yta} but will include
now all runs available to improve the statistical errors in
the fitting of the 19 parameters in each formula.

We thus provide a full updated set of fitting parameters using all
the 71 new runs, an additional set of 68 SXS spin-aligned runs
from their BBH catalog \cite{SXS:catalog},
 and the 36 runs used in Ref.~\cite{Healy:2014yta}
to produce a more accurate set of remnant formulae. The results
are summarized in Table~\ref{tab:fitparsms} for the final mass and spin.


\begin{table*}
\caption{
Table of fitting parameters for the mass, and spin formulas.}\label{tab:fitparsms}
\begin{ruledtabular}
\begin{tabular}{lr|lr}
M0 &  $0.951659 \pm 0.000022$ & 	L0 & $0.686732 \pm 0.000023$ \\ 
K1 &  $-0.051130 \pm 0.000131$ & 	L1 & $0.613285 \pm 0.000114$ \\ 
K2a &  $-0.005699 \pm 0.000318$ & 	L2a & $-0.148530 \pm 0.000311$ \\ 
K2b &  $-0.058064 \pm 0.000459$ & 	L2b & $-0.113826 \pm 0.000458$ \\ 
K2c &  $-0.001867 \pm 0.000160$ & 	L2c & $-0.003240 \pm 0.000178$ \\ 
K2d &  $1.995705 \pm 0.000292$ & 	L2d & $0.798011 \pm 0.000297$ \\ 
K3a &  $0.004991 \pm 0.001479$ & 	L3a & $-0.068782 \pm 0.001820$ \\ 
K3b &  $-0.009238 \pm 0.000809$ & 	L3b & $0.001291 \pm 0.000442$ \\ 
K3c &  $-0.120577 \pm 0.000769$ & 	L3c & $-0.078014 \pm 0.000674$ \\ 
K3d &  $0.016417 \pm 0.001121$ & 	L3d & $1.557286 \pm 0.001018$ \\ 
K4a &  $-0.060721 \pm 0.002814$ & 	L4a & $-0.005710 \pm 0.001826$ \\ 
K4b &  $-0.001798 \pm 0.000714$ & 	L4b & $0.005920 \pm 0.000612$ \\ 
K4c &  $0.000654 \pm 0.000225$ & 	L4c & $-0.001706 \pm 0.000255$ \\ 
K4d &  $-0.156626 \pm 0.002114$ & 	L4d & $-0.058882 \pm 0.002072$ \\ 
K4e &  $0.010303 \pm 0.001919$ & 	L4e & $-0.010187 \pm 0.001971$ \\ 
K4f &  $2.978729 \pm 0.000691$ & 	L4f & $0.964445 \pm 0.000706$ \\ 
K4g &  $0.007904 \pm 0.001385$ & 	L4g & $-0.110885 \pm 0.001400$ \\ 
K4h &  $0.000631 \pm 0.000409$ & 	L4h & $-0.006821 \pm 0.001316$ \\ 
K4i &  $0.084478 \pm 0.002097$ & 	L4i & $-0.081648 \pm 0.002222$ \\ 
\end{tabular}
\end{ruledtabular}
\end{table*}

\begin{figure}
\includegraphics[angle=270,width=0.95\columnwidth]{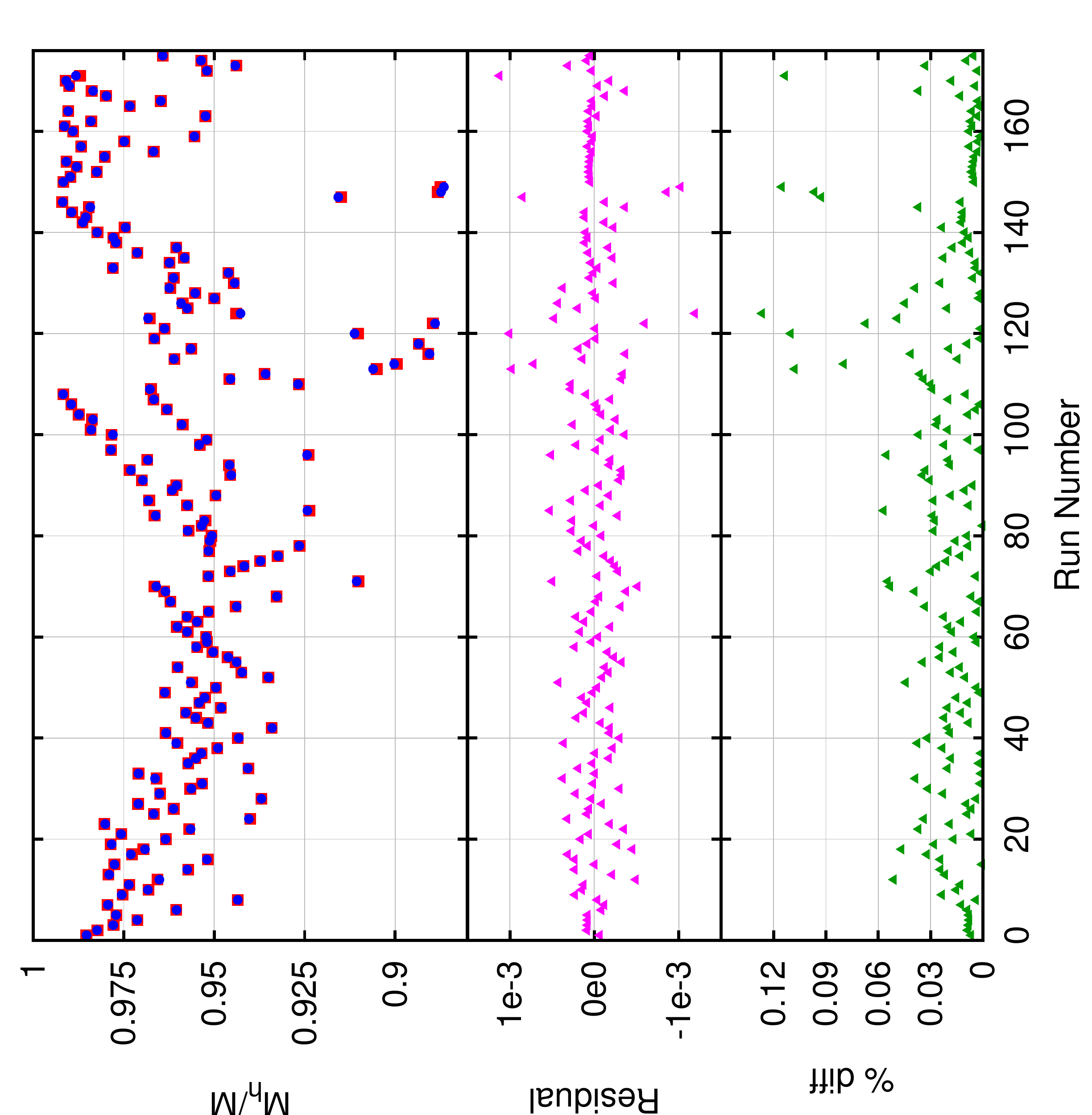}
\caption{Predicted values and residuals of the runs compared to the new fitting
for the final mass.
\label{fig:fitM}}
\end{figure}

The results of the new fits can be summarized in Fig.~\ref{fig:fitM}
where we observe the values of the fit compared to those computed
by the simulations for the final mass and the residuals of the fit
in both the differences and percentage differences. Notably
the residuals are limited to differences in the masses of 
$\pm1.2\times10^{-3}$, corresponding to percentages of less than $0.13\%$.

The minimum mass (and hence the maximum radiated energy) as a function
of the mass ratio and spins ($M_{rem}(q,\alpha_1,\alpha_2)$) occurs for
equal mass binaries bearing maximum spins along the orbital angular
momentum. This corresponds also to the maximum hangup effect \cite{Campanelli:2006uy}.
Evaluation of the above fitted formula gives us $M_{rem}(1,1,1)= 0.88672$
which corresponds to radiating above $11\%$ of the initial ADM mass of
the binary system. The minimum energy radiated occurs in the particle
limit, but we can evaluate the minimum energy radiated for the equal-mass
case.  In this case, minimum radiation occurs due to the anti-hangup, 
when both spins are anti-aligned 
with the orbital angular momentum and maximally spinning. In that
case only about $3\%$ of the mass is radiated into gravitational waves
with $M_{rem}( 1, -1, -1 ) = 0.96799$.

\begin{figure}
\includegraphics[angle=270,width=0.95\columnwidth]{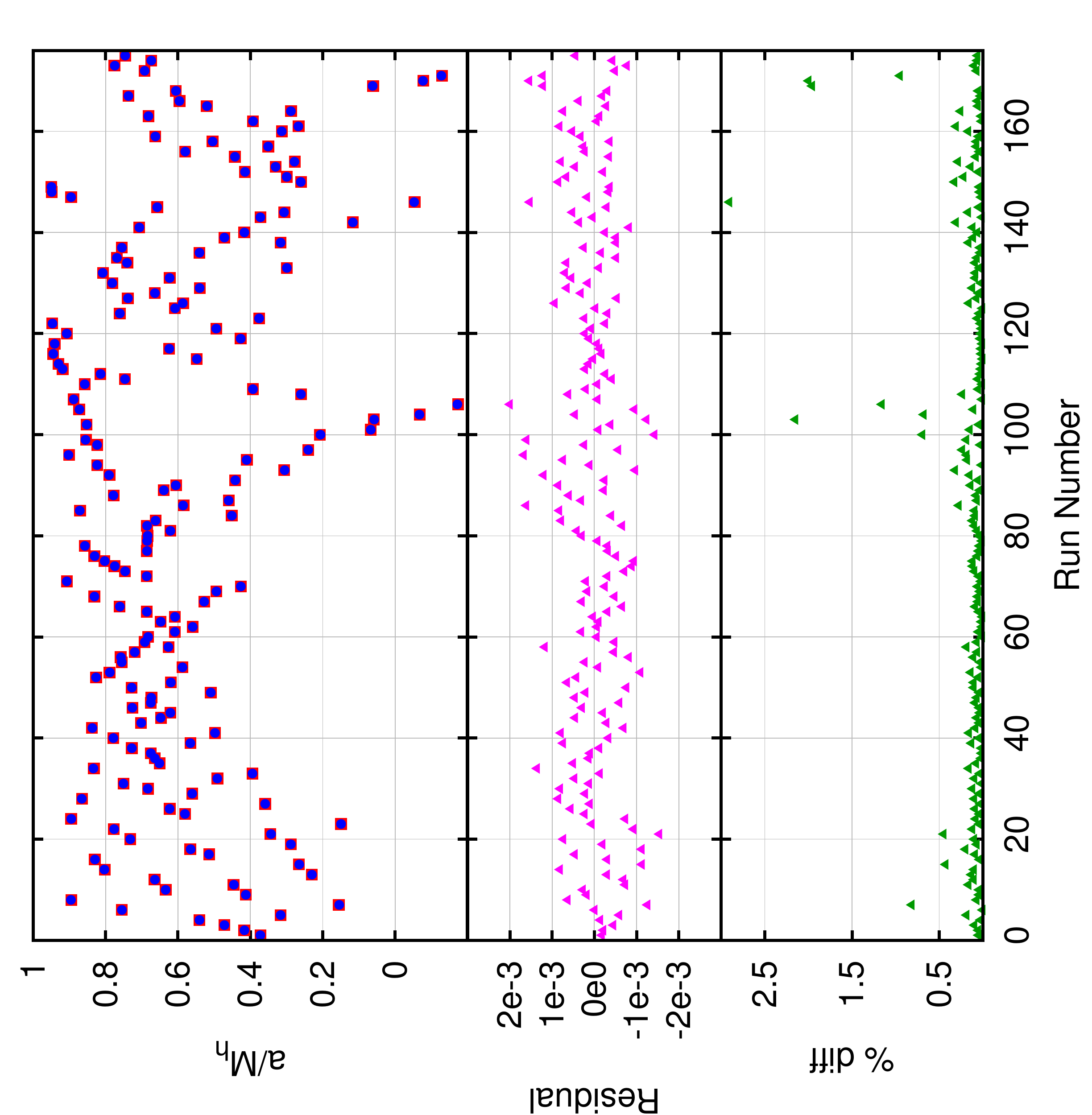}
\caption{Predicted values and residuals of the runs compared to the new fitting
for the final spin.
\label{fig:fitS}}
\end{figure}

A similar analysis for the fits to the final remnant spin is shown
in Fig.~\ref{fig:fitS} where the residuals are bounded by 
$\pm2\times10^{-3}$ and typical percentages differences of less than
 $0.5\%$ (except for those cases where the final spin is close to zero).

The maximum (minimum) remnant spin, $\alpha_{rem}(q,\alpha_1,\alpha_2)$,
are achieved in the particle limit 
for maximal Kerr aligned (anti-aligned) spinning black holes,
but if we consider equal mass binaries then the maximum and minimum 
final spin of the remnant are $\alpha_{rem}( 1, 1, 1 ) = 0.95149$
and $\alpha_{rem}(1,-1,-1)= 0.35771$.

A final remnant black hole with vanishing spin is also achievable
according to these remnant formulae for $q\leq0.3$ as shown in
Fig. 16 of Ref.~\cite{Healy:2014yta}.  The current iteration of the
fit does not change this substantially.

The comparative analysis displayed in Fig.~\ref{fig:res1} of the 
old and new formulae shows a slight improvement of the new formulae
from the statistical point of view. This improvement is due to a larger
(and slightly more accurate) sample of data.
The improvement is not
large in the final mass since the old formula was already providing an excellent approximation.
The improvements are more notable for final spin and the recoil 
velocities. We have also 
introduced a peak luminosity fit as it follows in the next subsection.

\subsection{Modeling final recoils and peak luminosities}\label{subsec:finalVL}

Given the success of the above formulae we will propose a new fitting
formula for the peak luminosity of gravitational waves with a similar
form to the mass fitting. The motivation for the above formulae
comes from a combination of a Taylor expansion in terms of the seven
binary parameters, limited by the symmetry properties of the binary
under parity and exchange of the black hole labels. We have also found
that the use of the variables $\tilde{\vec{S}}$, $\tilde{\vec{\Delta}}$,
and $\dmt$ has a direct connection with PN variables and provides an excellent
fit~\cite{Lousto:2012gt,Lousto:2013wta,Healy:2014yta,Zlochower:2015wga}. 
This can be extended to the peak luminosity that corresponds to the
peak power of gravitational radiation and will share the symmetry properties
used for the total final mass of the remnant. 

For both the recoil velocity and the peak luminosity fittings we use the
original 36 simulations from Ref~\cite{Healy:2014yta} and the 71 new
simulations for a total of 107 simulations.  

\subsubsection{Peak luminosity fits}

We propose

\begin{eqnarray}\label{eq:4plum}
L_{\rm peak} = (4\eta)^2\,\Big\{N_0 + N_1 \Spar+ N_{2a}\,\Dpar\dmt +
                     N_{2b}\,\Spar^2+ \nonumber\\
                     N_{2c}\,\Dpar^2+
                     N_{2d}\,\dmt^2 +
                     N_{3a}\,\Dpar\Spar\dmt+ \nonumber\\
                     N_{3b}\,\Spar\Dpar^2+
                     N_{3c}\,\Spar^3+ \nonumber\\
                     N_{3d}\,\Spar\dmt^2+
                     N_{4a}\,\Dpar\Spar^2\dmt + \nonumber\\
                     N_{4b}\,\Dpar^3\dmt +
                     N_{4c}\,\Dpar^4+
                     N_{4d}\,\Spar^4+\nonumber\\
                     N_{4e}\,\Dpar^2 \Spar^2+
                     N_{4f}\,\dmt^4+
                     N_{4g}\,\Dpar\dmt^3+\nonumber\\
                     N_{4h}\,\Dpar^2\dmt^2 +
                     N_{4i}\,\Spar^2\dmt^2\Big\}.
\end{eqnarray}

Note that the radiated power in the particle limit scales as $\eta^2$
(See Ref.~\cite{Fujita:2014eta}, Eq. (16) and (20); evaluated at the ISCO
for its peak value).
Thus we do not include this explicitly in the fitting above, with the 
same true in the recoil velocity case below, since radiative terms would
need to be fitted anyway.

\begin{figure}
\includegraphics[width=0.95\columnwidth]{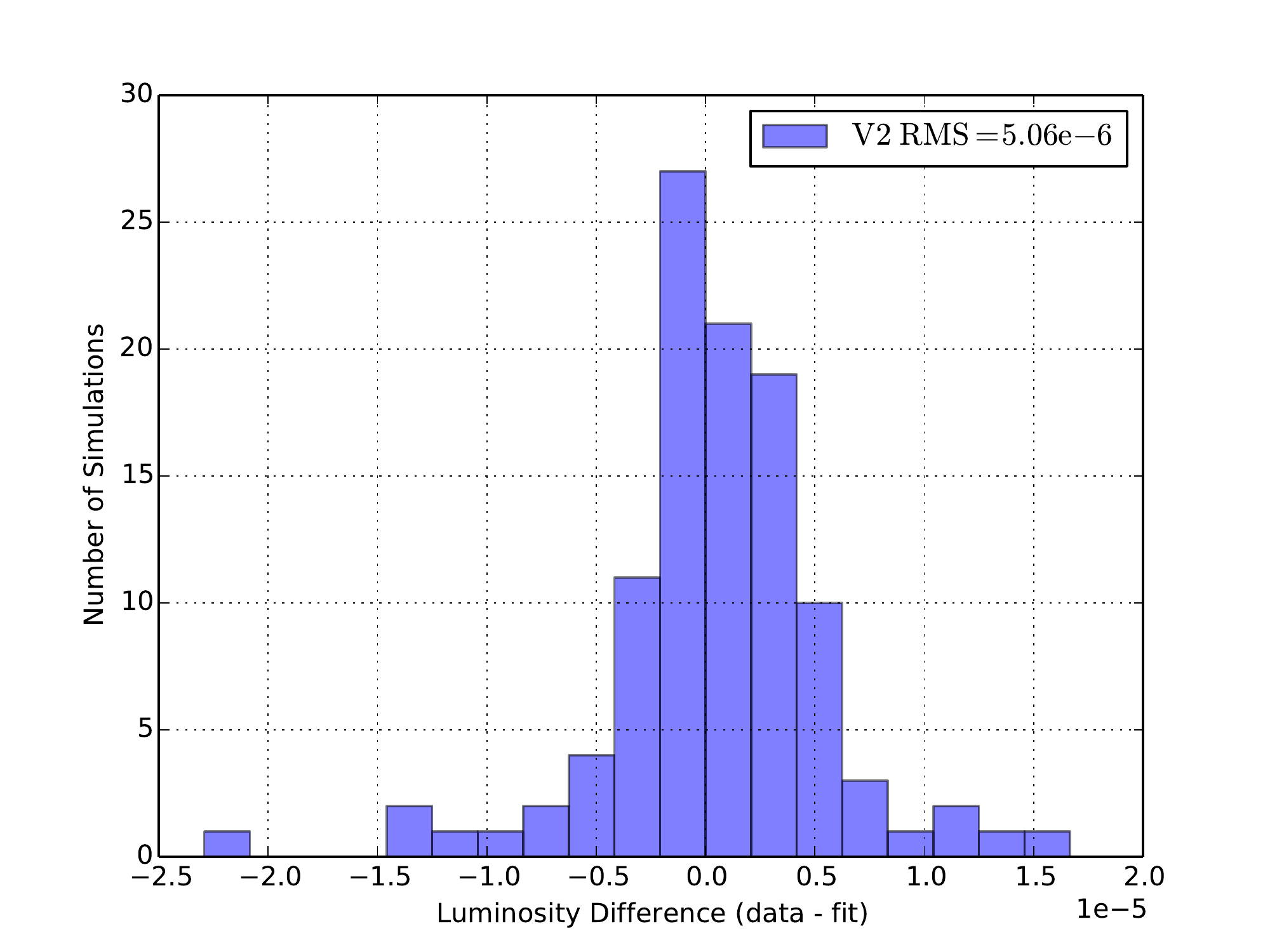}\\
\includegraphics[width=0.95\columnwidth]{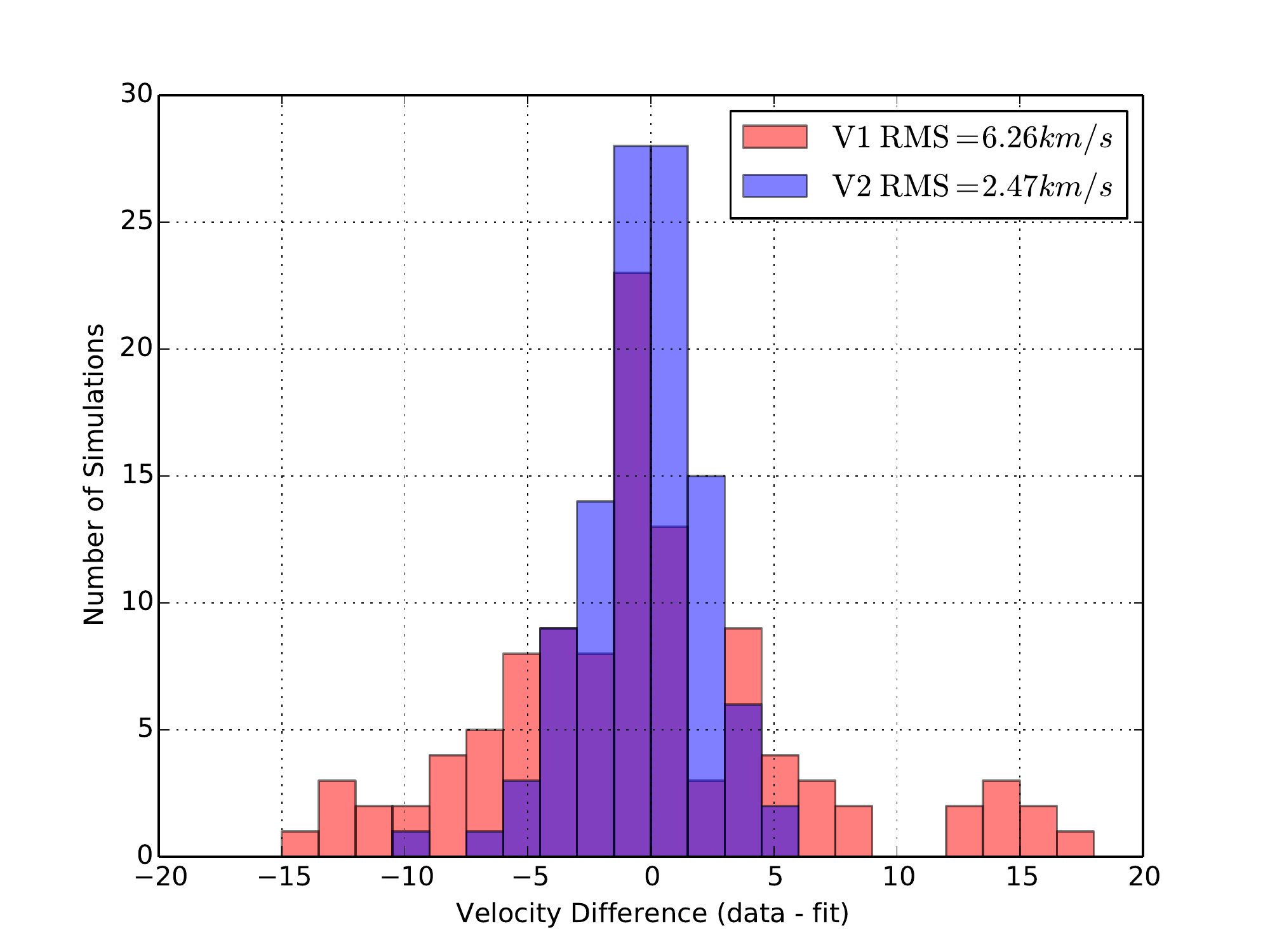}
\caption{Counts of the number of runs whose residual (data - fit)
falls within a particular
bin. Both cases have 25 bins over the visible range with the recoil velocity
shown in the top panel, and recoil velocity in the bottom panel.
V1 labels the fitting from Ref.~\cite{Healy:2014yta}
and V2 labels the fitting from this paper.
\label{fig:res2}}
\end{figure}

\begin{figure}
\includegraphics[angle=270,width=0.95\columnwidth]{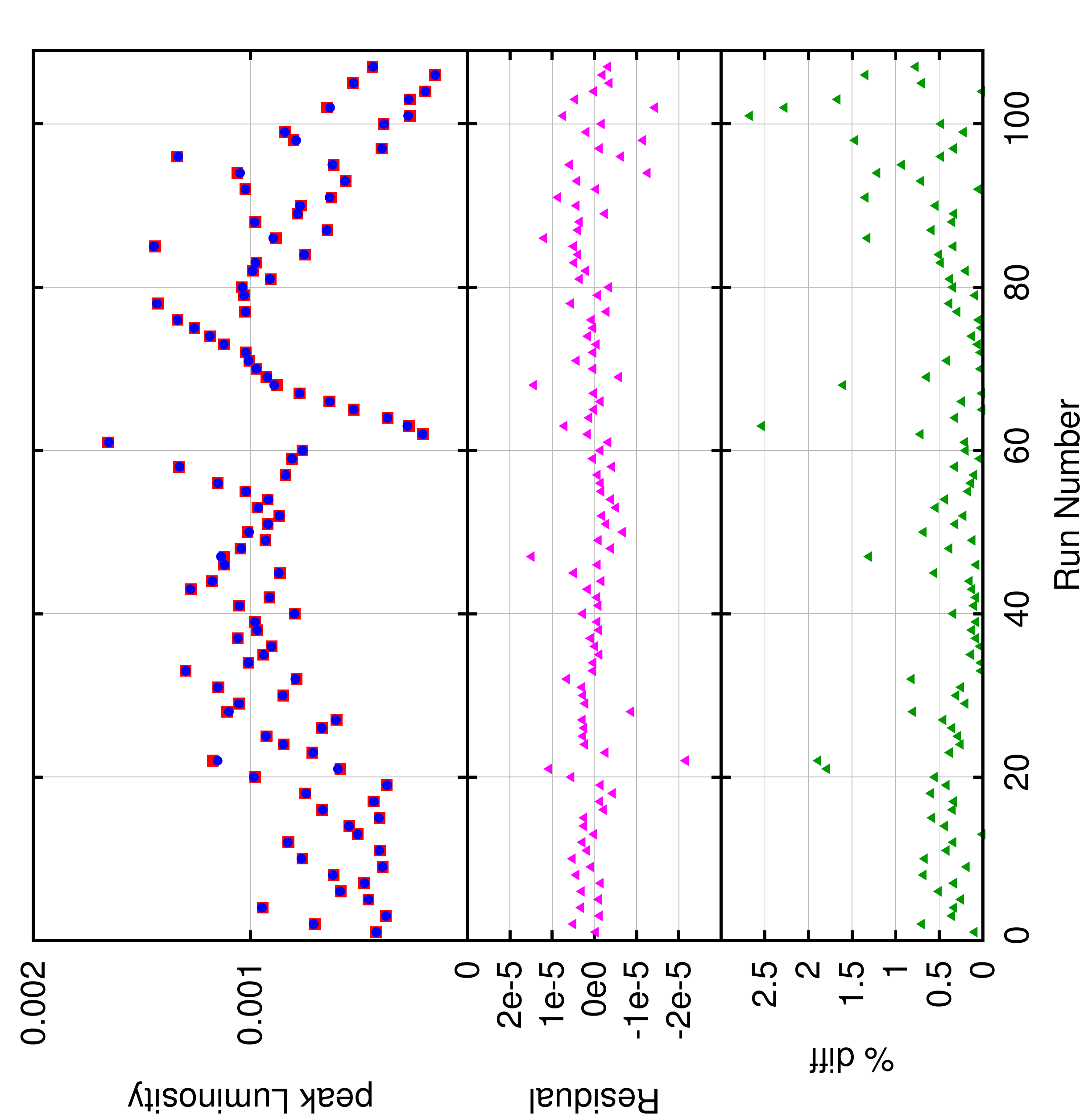}
\caption{Predicted values and residuals of the runs compared to the new fitting
for the peak luminosity.
\label{fig:fitL}}
\end{figure}

The best fitted parameters are displayed in Table~\ref{tab:fitparsVL} below,
including the fitting statistical uncertainty (note that some of the
coefficient are statistically compatible with taking a zero value).
The residuals are shown in the top panel of Fig.~\ref{fig:res2} and 
the comparison of the fits to data are displayed in Fig.~\ref{fig:fitL}.
Fig.~\ref{fig:fitL} also displays residuals in the middle panel with values 
of less than $2\times10^{-5}$ in dimensionless
units which corresponds to typical errors of less than $3\%$.

Note that the percentile errors of the luminosity (and this will also be
true for the recoil velocity) are an order of magnitude larger than
for the mass and spin of the remnant. This is due to the fact that
while the final mass and spin can be obtained very accurately from
the isolated horizon formulae applied to the final black hole,
the peak luminosity is computed directly from the waveforms at infinity
and have errors associated with the finite difference used during
the evolution (as well as finite extraction radii and modes used) as 
detailed in Appendix \ref{app:errors}. We have used a method of
weighted least squares to take into account the different available
resolutions for the pool of simulations used in this fit. 
A maximum of $\sim-5\%$ difference in the peak luminosity is observed
with respect to
\footnote{F. Jim\'enez Forteza et al., LIGO Document T1600018,
https://dcc.ligo.org/LIGO-T1600018/public}. Such a large difference only occurs for the largest
peak luminosities occurring for near equal-mass systems with large aligned
spins. (The two fits agree to better than 0.5\% about the peak
luminosity for an equal-mass nonspinning binary, for instance.) 

By direct evaluation of the fitting formula for the luminosity we see that
the maximum value occurs for equal mass binaries and maximum spinning black
holes oriented along the orbital angular momentum. In this case we obtain
for the peak luminosity $L_{\mathrm{peak}}(1,1,1)=0.001967$ in dimensionless units, 
equivalent to $7.1368\times10^{56} ergs/sec$.


\begin{table*}
\caption{
Table of fitting parameters (left) for the recoil (in Km/s)
and (right) peak luminosity formulas.}\label{tab:fitparsVL}
\begin{ruledtabular}
\begin{tabular}{lr|lr}
H & $7528.531080 \pm 60.544382$ & 	N0 &  1.021017e-03 $\pm$ 8.905891e-07 \\ 
H2a & $-1.795874 \pm 0.114360$ & 	N1 &  8.974289e-04 $\pm$ 1.892220e-05 \\ 
H2b & $-0.615667 \pm 0.061626$ & 	N2a & -9.774672e-05 $\pm$ 4.050598e-05 \\ 
H3a & $-0.447651 \pm 0.132894$ & 	N2b & 9.208838e-04 $\pm$ 4.258906e-05 \\ 
H3b & $-0.771102 \pm 0.382234$ & 	N2c & 1.869827e-05 $\pm$ 1.740217e-05 \\ 
H3c & $-1.700807 \pm 0.177238$ & 	N2d & -3.913170e-04 $\pm$ 1.368557e-05 \\ 
H3d & $-0.021333 \pm 0.017638$ & 	N3a & -1.202141e-04 $\pm$ 1.496747e-04 \\ 
H3e & $-0.753230 \pm 0.216550$ & 	N3b & 1.481022e-04 $\pm$ 9.213429e-05 \\ 
H4a & $-0.585791 \pm 0.413974$ & 	N3c & 1.379015e-03 $\pm$ 1.492745e-04 \\ 
H4b & $-1.524603 \pm 0.929478$ & 	N3d & -4.937306e-04 $\pm$ 9.776929e-05 \\ 
H4c & $0.969809 \pm 0.650891$ & 	N4a & 8.847927e-04 $\pm$ 4.470162e-04 \\ 
H4d & $0.788852 \pm 0.664358$ & 	N4b & 3.292542e-07 $\pm$ 9.428275e-07 \\ 
H4e & $-1.701279 \pm 0.835741$ & 	N4c & 1.701707e-05 $\pm$ 3.024644e-05 \\ 
H4f & $-0.017526 \pm 0.224595$ & 	N4d & 1.514840e-03 $\pm$ 2.847964e-04 \\ 
a & $2.463283 \pm 0.027020$ & 		N4e & -1.484568e-04 $\pm$ 1.737503e-04 \\ 
b & $1.466051 \pm 0.152762$ & 		N4f & 1.366596e-04 $\pm$ 6.530259e-05 \\ 
c & $0.554279 \pm 0.189169$ & 		N4g & 1.603431e-04 $\pm$ 1.673143e-04 \\ 
   &                         &		N4h & -6.185306e-05 $\pm$ 1.113796e-04 \\ 
   &                         &		N4i & -1.036025e-03 $\pm$ 2.504408e-04 \\ 
\end{tabular}
\end{ruledtabular}
\end{table*}


\subsubsection{Recoil velocities}

We model the total recoil as~\cite{Healy:2014yta}
\begin{equation}\label{eq:empirical}
\vec{V}_{\rm recoil}(q,\vec\alpha_i)=v_m\,\hat{e}_1+
v_\perp(\cos(\xi)\,\hat{e}_1+\sin(\xi)\,\hat{e}_2),
\end{equation}
$\hat{e}_1,\hat{e}_2$ are orthogonal unit vectors in the
orbital plane, and $\xi$ measures the angle between the ``unequal mass''
and ``spin'' contributions to the recoil velocity in the orbital plane, and
with,
\begin{eqnarray}\label{eq:4recoil}
v_\perp = H\eta^2\left(\Dpar+ H_{2a} \Spar\dmt    
                     + H_{2b} \Dpar \Spar
                     + H_{3a} \Dpar^2\dmt\right.\nonumber\\
                     \left.+ H_{3b} \Spar^2\dmt
                     + H_{3c} \Dpar\Spar^2
                     + H_{3d} \Dpar^3
                     + H_{3e} \Dpar \dmt^2\right.\nonumber\\
                     + H_{4a} \Spar\Dpar^2\dmt
                     \left.+ H_{4b} \Spar^3 \dmt
                     + H_{4c} \Spar \dmt^3\right.\nonumber\\
                     \left.+ H_{4d} \Dpar \Spar \dmt^2
                     + H_{4e} \Dpar \Spar^3
                     + H_{4f} \Spar \Dpar^3\right),\\
\xi=a+b\, \Spar +c\, \dmt\Dpar.\nonumber
\end{eqnarray}
Where 
\begin{equation}\label{eq:vm}
v_m=\eta^2 \delta m\left(A+B\,\delta m^2\right).
\end{equation}
and according to Ref.~\cite{Gonzalez:2007hi} we have $A=9210\,km/s$,
and $B=2790\,km/s$.

In the bottom panel of Fig.~\ref{fig:res2}, the residuals between the first
fitting formula and the new fitting formula on the full set of 107 simulations
are shown.  While the first fitting formula performed reasonably well, with the
largest deviations from the data on the order of $15\,km/s$, the new
fitting formula is a marked improvement benefiting from a sample size of our
simulations almost tripled from the first version.  

\begin{figure}
\includegraphics[angle=270,width=0.95\columnwidth]{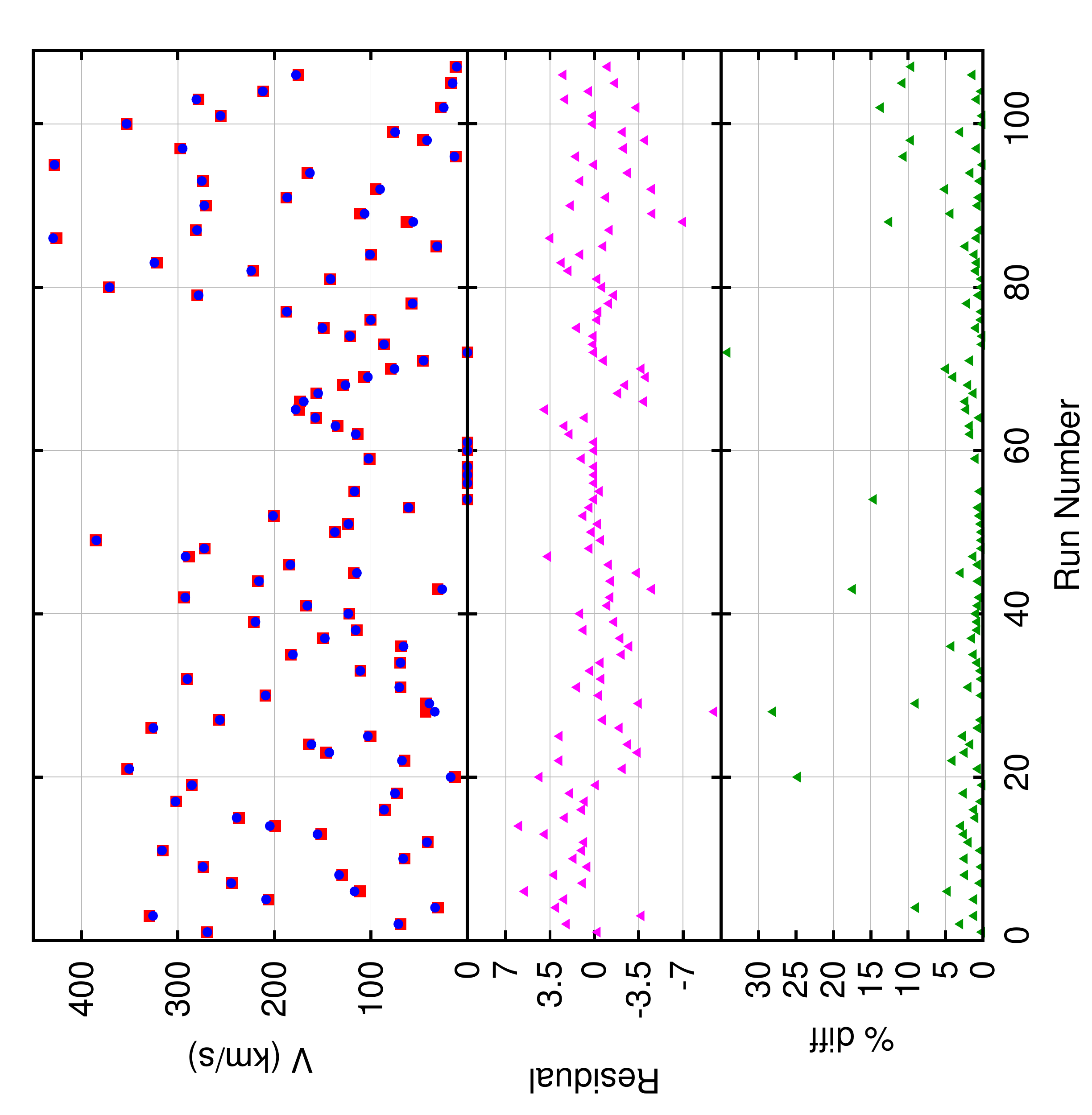}
\caption{Predicted values and residuals (in km/s) of the runs compared to the new fitting
for the recoil velocity.
\label{fig:fitP}}
\end{figure}

The best fitted parameters are displayed in Table~\ref{tab:fitparsVL} below.
The comparison of the fits to data are displayed in Fig.~\ref{fig:fitP},
that essentially displays residuals of less than $7\,km/s$, corresponding 
to typical errors of $5-10\%$ for moderate to large recoils.  

As for the luminosity,
the percentile errors of the recoil velocity are an order of magnitude larger than
for the mass and spin of the remnant. We note again that this is due to the fact that
while the final mass and spin can be obtained very accurately from
the isolated horizon formulae applied to the final black hole,
the recoil velocity is computed directly from (different modes of) 
the waveforms at infinity (See though Ref.~\cite{Krishnan:2007pu} for a 
horizon evaluation of the linear momentum)
and have errors associated with the finite difference used during
the evolution (as well as finite extraction radii and modes) as detailed
in Appendix \ref{app:errors}. We have used a method of
weighted least squares to take into account the different available
resolutions for the pool of simulations used in this fit.

The current analysis with a more extensive set of runs (including the new 71 runs
and the former 36 run) allow us to make a statistically more robust determination
of the 17 fitting parameters. 
We can also determine that the maximum recoil velocity does not occur for equal mass binaries but for $q\approx2/3$ in an up/down configuration. We find
$V_{max}=516.58 \pm 26.33 km/s$ for $q_{max}=0.6628 \pm 0.05661$ 
with $\alpha_1^{max}=1.0$ and  $\alpha_2^{max}=-1.0$

In addition to the particle limit and the equal mass, equal spins cases 
there are other configurations that lead to zero recoil velocity as 
predicted by the above fitting formulae. We find that the results 
displayed in Fig.~14 of Ref.~\cite{Healy:2014yta} is not changed 
substantially with the updated formulae.



\section{Conclusions and Discussion}\label{sec:Discussion}

In this paper we have revisited the scenario of nonprecessing binary
black hole mergers extending our previous work by tripling the number
of our simulations. We have thus verified that our formulae for the
final remnant mass and spin were accurate well within a fraction of 
$1\%$ level, and used the new runs to improve those fittings.
We observe in the comparative analysis of Fig.~\ref{fig:res1} that
the new fit particularly improves the residuals of the final spin. The
wings of the residuals now
halved and are more symmetric around zero
than for the old fit, as expected for random errors.
The improvement is not as evident for the final mass, since the original
fit was already extremely good.
The comparative analysis of the fittings for 
the recoil velocity displayed in Fig.~\ref{fig:res2}
shows that the typical residuals are halved by the newly fitted formula. 
Thus reducing the error of the remnant recoil to within $5\%$. 
Finally, we see in the bottom panel of Fig.~\ref{fig:fitL}
that the typical residuals of the luminosity fit lie within $2\%$ values.
Note that we also improved the assessment of systematic errors in those 
results by studying the observer location and finite
difference dependences of the results, particularly for the recoil
velocity and peak luminosity, since the final mass and spin can be
obtained more accurately from the horizon evaluated measurements.

These formulae are particularly relevant for LIGO gravitational
wave signal analysis. In fact, the formulae of our previous paper 
\cite{Healy:2014yta}
has been used to analyze the aLIGO O1 events
\cite{Abbott:2016blz,TheLIGOScientific:2016wfe,TheLIGOScientific:2016src,Abbott:2016izl,Abbott:2016nmj,TheLIGOScientific:2016pea}.
The updated formulae in this paper (including that of the peak luminosity) 
are well suited for implementation and use in an improved analysis of the 
forthcoming observational runs of advanced LIGO.

While the above formulae are strictly valid for nonprecessing systems,
they represent an important basis to model precessing systems as they
provide several of the expansion terms to be included in such
formulae \cite{Lousto:2012gt,Lousto:2013wta,Zlochower:2015wga}
and they represent a reasonable approximation for most applications.
For instance, Ref.~
\footnote{N.K. Johnson-McDaniel et al., LIGO Document T1600168, 
  https://dcc.ligo.org/LIGO-T1600168/public}
uses a simple augmentation of
the final spin in \cite{Healy:2014yta}
to improve its accuracy for precessing systems and shows that the final
mass formula of \cite{Healy:2014yta} has good accuracy even for precessing
systems.

\acknowledgments 
The authors thank M. Campanelli, N.K.Johnson-McDaniel, H. Nakano,
R. O'Shaughnessy and Y. Zlochower for discussions on this work.
The authors gratefully acknowledge the NSF for financial support from Grants
PHY-1607520, PHY-1305730, PHY-1212426, PHY-1229173,
AST-1028087, PHY-0969855, OCI-0832606, and
DRL-1136221. Computational resources were provided by XSEDE allocation
TG-PHY060027N, and by NewHorizons and BlueSky Clusters 
at Rochester Institute of Technology, which were supported
by NSF grant No. PHY-0722703, DMS-0820923, AST-1028087, and PHY-1229173.

\appendix
\section{Summary of properties of the new simulations}\label{app:tables}

In this appendix, we supplement the initial data parameters given in 
Tables \ref{tab:ID}-\ref{tab:ID_nonspin} with more information
about the simulations.  

In Tables \ref{tab:ecc} and \ref{tab:ecc_nonspin}
we give the initial orbital frequency, number of orbits
to merger (as measured by the formation of a common apparent horizon and
monitored by the time of peak luminosity), and initial and final 
eccentricities estimated as
\cite{Campanelli:2008nk}
$e = r^2 \ddot r/m$.
\begin{table*}
\caption{Table of the initial orbital frequency $m\omega_i$,
number of orbits to merger, $N$, and the initial and final eccentricities,
$e_i$ and $e_f$ for the spinning cases.}\label{tab:ecc}
\begin{ruledtabular}
\begin{tabular}{llcccc}
Run & Config.   & $m\omega_i$ & $N$ & $e_i$ & $e_f$ \\
\hline
1 & Q0.3333\_0.0000\_-0.5000 &	$0.0236$ & $8.7$ & $0.008$ & $0.002$\\
2 & Q0.3333\_0.0000\_0.5000 &	$0.0269$ & $10.5$ & $0.006$ & $0.002$\\
3 & Q0.3333\_0.0000\_-0.8500 &	$0.0241$ & $7.6$ & $0.016$ & $0.006$\\
4 & Q0.3333\_0.0000\_0.8500 &	$0.0274$ & $12.1$ & $0.007$ & $0.002$\\
5 & Q0.3333\_-0.5000\_-0.2500 &	$0.0226$ & $10.0$ & $0.008$ & $0.001$\\
6 & Q0.3333\_-0.5000\_0.2500 &	$0.0249$ & $10.3$ & $0.006$ & $0.001$\\
7 & Q0.3333\_0.5000\_-0.2500 &	$0.0252$ & $8.9$ & $0.006$ & $0.002$\\
8 & Q0.3333\_0.5000\_0.2500 &	$0.0271$ & $9.7$ & $0.006$ & $0.002$\\
9 & Q0.3333\_-0.5000\_-0.6500 &	$0.0229$ & $8.6$ & $0.015$ & $0.005$\\
10 & Q0.3333\_-0.5000\_0.6500 &	$0.0270$ & $10.8$ & $0.005$ & $0.002$\\
11 & Q0.3333\_0.5000\_-0.6500 &	$0.0244$ & $8.0$ & $0.008$ & $0.003$\\
12 & Q0.3333\_0.5000\_0.6500 &	$0.0275$ & $11.3$ & $0.005$ & $0.003$\\
13 & Q0.3333\_-0.8000\_0.0000 &	$0.0240$ & $9.9$ & $0.012$ & $0.004$\\
14 & Q0.3333\_0.8000\_0.0000 &	$0.0257$ & $10.2$ & $0.009$ & $0.003$\\
15 & Q0.3333\_-0.8000\_-0.5000 &	$0.0228$ & $8.9$ & $0.016$ & $0.005$\\
16 & Q0.3333\_-0.8000\_0.5000 &	$0.0277$ & $9.8$ & $0.010$ & $0.004$\\
17 & Q0.3333\_0.8000\_-0.5000 &	$0.0273$ & $7.5$ & $0.013$ & $0.005$\\
18 & Q0.3333\_0.8000\_0.5000 &	$0.0287$ & $10.6$ & $0.005$ & $0.002$\\
19 & Q0.3333\_-0.8000\_-0.8000 &	$0.0205$ & $9.7$ & $0.017$ & $0.005$\\
20 & Q0.3333\_0.8000\_0.8000 &	$0.0306$ & $11.1$ & $0.006$ & $0.001$\\
21 & Q0.5000\_0.0000\_-0.8000 &	$0.0204$ & $9.6$ & $0.010$ & $0.003$\\
22 & Q0.5000\_0.0000\_0.8000 &	$0.0244$ & $11.9$ & $0.006$ & $0.002$\\
23 & Q0.5000\_-0.5000\_-0.1000 &	$0.0211$ & $10.6$ & $0.009$ & $0.001$\\
24 & Q0.5000\_0.5000\_0.1000 &	$0.0218$ & $12.0$ & $0.006$ & $0.001$\\
25 & Q0.5000\_-0.5000\_0.5000 &	$0.0290$ & $7.6$ & $0.006$ & $0.001$\\
26 & Q0.5000\_0.5000\_-0.5000 &	$0.0273$ & $7.0$ & $0.014$ & $0.004$\\
27 & Q0.5000\_-0.5000\_-0.6000 &	$0.0204$ & $9.6$ & $0.011$ & $0.003$\\
28 & Q0.5000\_0.5000\_0.6000 &	$0.0229$ & $13.1$ & $0.006$ & $0.001$\\
29 & Q0.7500\_0.0000\_0.2500 &	$0.0295$ & $6.8$ & $0.007$ & $0.003$\\
30 & Q0.7500\_0.0000\_-0.5000 &	$0.0278$ & $6.2$ & $0.015$ & $0.005$\\
31 & Q0.7500\_0.0000\_0.5000 &	$0.0258$ & $9.2$ & $0.011$ & $0.003$\\
32 & Q0.7500\_0.0000\_-0.8000 &	$0.0289$ & $5.4$ & $0.017$ & $0.005$\\
33 & Q0.7500\_0.0000\_0.8000 &	$0.0255$ & $10.2$ & $0.010$ & $0.002$\\
34 & Q0.7500\_-0.2500\_0.2500 &	$0.0299$ & $6.3$ & $0.007$ & $0.003$\\
35 & Q0.7500\_0.2500\_-0.2500 &	$0.0302$ & $5.9$ & $0.007$ & $0.003$\\
36 & Q0.7500\_-0.5000\_0.0000 &	$0.0263$ & $6.9$ & $0.010$ & $0.003$\\
37 & Q0.7500\_0.5000\_0.0000 &	$0.0252$ & $8.9$ & $0.007$ & $0.001$\\
38 & Q0.7500\_-0.5000\_0.2500 &	$0.0318$ & $5.7$ & $0.013$ & $0.003$\\
39 & Q0.7500\_0.5000\_-0.2500 &	$0.0297$ & $6.3$ & $0.008$ & $0.002$\\
40 & Q0.7500\_-0.5000\_-0.5000 &	$0.0289$ & $5.3$ & $0.016$ & $0.006$\\
41 & Q0.7500\_-0.5000\_0.5000 &	$0.0264$ & $8.1$ & $0.013$ & $0.004$\\
42 & Q0.7500\_0.5000\_-0.5000 &	$0.0270$ & $7.2$ & $0.014$ & $0.003$\\
43 & Q0.7500\_0.5000\_0.5000 &	$0.0254$ & $10.3$ & $0.010$ & $0.002$\\
44 & Q0.7500\_-0.5000\_0.8000 &	$0.0261$ & $9.1$ & $0.012$ & $0.003$\\
45 & Q0.7500\_-0.8000\_0.0000 &	$0.0281$ & $6.1$ & $0.016$ & $0.005$\\
46 & Q0.7500\_0.8000\_0.0000 &	$0.0258$ & $9.3$ & $0.011$ & $0.003$\\
47 & Q0.7500\_-0.8000\_0.8000 &	$0.0313$ & $6.5$ & $0.013$ & $0.005$\\
48 & Q0.7500\_-0.8500\_0.6375 &	$0.0270$ & $7.8$ & $0.014$ & $0.004$\\
49 & Q0.7500\_0.8500\_-0.6375 &	$0.0271$ & $7.4$ & $0.014$ & $0.002$\\
50 & Q0.8200\_-0.4400\_0.3300 &	$0.0215$ & $10.4$ & $0.009$ & $0.001$\\
51 & Q1.0000\_0.0000\_-0.5000 &	$0.0276$ & $6.4$ & $0.015$ & $0.005$\\
52 & Q1.0000\_0.0000\_-0.8000 &	$0.0285$ & $5.7$ & $0.016$ & $0.002$\\
53 & Q1.0000\_-0.2500\_0.0000 &	$0.0225$ & $9.1$ & $0.009$ & $0.002$\\
54 & Q1.0000\_-0.2500\_-0.2500 &	$0.0228$ & $8.5$ & $0.010$ & $0.002$\\
55 & Q1.0000\_-0.2500\_0.2500 &	$0.0223$ & $9.8$ & $0.007$ & $0.001$\\
56 & Q1.0000\_0.2500\_0.2500 &	$0.0219$ & $11.1$ & $0.007$ & $0.001$\\
57 & Q1.0000\_-0.5000\_-0.5000 &	$0.0289$ & $5.3$ & $0.017$ & $0.007$\\
58 & Q1.0000\_0.5000\_0.5000 &	$0.0248$ & $10.3$ & $0.006$ & $0.001$\\
59 & Q1.0000\_-0.8000\_-0.4000 &	$0.0212$ & $8.4$ & $0.017$ & $0.004$\\
60 & Q1.0000\_-0.8000\_-0.8000 &	$0.0314$ & $4.0$ & $0.018$ & $0.003$\\
61 & Q1.0000\_0.8000\_0.8000 &	$0.0281$ & $9.9$ & $0.006$ & $0.001$\\
\end{tabular}
\end{ruledtabular}
\end{table*}

\begin{table*}
\caption{Table of the initial orbital frequency $m\omega_i$,
number of orbits to merger, $N$, and the initial and final eccentricities,
$e_i$ and $e_f$ for nonspinning systems.}\label{tab:ecc_nonspin}
\begin{ruledtabular}
\begin{tabular}{llcccc}
Run & Config.   & $m\omega_i$ & $N$ & $e_i$ & $e_f$ \\
\hline
62 & Q0.1667\_0.0000\_0.0000 &	$0.0260$ & $12.6$ & $0.006$ & $0.002$\\
63 & Q0.2000\_0.0000\_0.0000 &	$0.0254$ & $11.4$ & $0.005$ & $0.001$\\
64 & Q0.2500\_0.0000\_0.0000 &	$0.0248$ & $10.8$ & $0.005$ & $0.002$\\
65 & Q0.3333\_0.0000\_0.0000 &	$0.0242$ & $10.1$ & $0.006$ & $0.001$\\
66 & Q0.4000\_0.0000\_0.0000 &	$0.0244$ & $9.4$ & $0.006$ & $0.002$\\
67 & Q0.5000\_0.0000\_0.0000 &	$0.0254$ & $8.4$ & $0.006$ & $0.001$\\
68 & Q0.6000\_0.0000\_0.0000 &	$0.0238$ & $9.1$ & $0.007$ & $0.001$\\
69 & Q0.6667\_0.0000\_0.0000 &	$0.0228$ & $10.0$ & $0.007$ & $0.002$\\
70 & Q0.7500\_0.0000\_0.0000 &	$0.0257$ & $7.9$ & $0.007$ & $0.002$\\
71 & Q0.8500\_0.0000\_0.0000 &	$0.0223$ & $9.9$ & $0.007$ & $0.001$\\
\end{tabular}
\end{ruledtabular}
\end{table*}

The values of the individual masses and spins, as well as the 
recombinations of mass ratio and the two spins used in the fitting
functions are given in Tables \ref{tab:IDr} and \ref{tab:IDr_nonspin}.
These relaxed values are measured at evolution time $t=150m$, suitably
long enough after the start of the merger for the initial ansatz
content to settle.
\begin{table*}
\caption{The mass and spin of the BHBs in Table~\ref{tab:ID} after the
BHs had time to equilibrate ($t/m=150$).}\label{tab:IDr}
\begin{ruledtabular}
\begin{tabular}{llcccccccc}
Run & Config.  & $q^r$ & $m^r_1/m$ & $m^r_2/m$ & $\alpha^r_1$ & $\alpha^r_2$ & $\delta m_r$ & $S_r/m^2_r$ & $\Delta_r/m^2_r$ \\
\hline

1 & Q0.3333\_0.0000\_-0.5000 &	$0.333349$ & $0.250001$ & $0.749967$ & $-0.000002$ & $-0.500063$ & $-0.499982$ & $-0.281279$ & $-0.375042$\\
2 & Q0.3333\_0.0000\_0.5000 &	$0.333349$ & $0.250001$ & $0.749967$ & $-0.000002$ & $0.500054$ & $-0.499982$ & $0.281273$ & $0.375036$\\
3 & Q0.3333\_0.0000\_-0.8500 &	$0.333590$ & $0.250001$ & $0.749425$ & $-0.000002$ & $-0.851337$ & $-0.499711$ & $-0.478692$ & $-0.638379$\\
4 & Q0.3333\_0.0000\_0.8500 &	$0.333588$ & $0.250001$ & $0.749431$ & $-0.000002$ & $0.851293$ & $-0.499713$ & $0.478669$ & $0.638348$\\
5 & Q0.3333\_-0.5000\_-0.2500 &	$0.333346$ & $0.250011$ & $0.750003$ & $-0.500031$ & $-0.250004$ & $-0.499985$ & $-0.171878$ & $-0.062489$\\
6 & Q0.3333\_-0.5000\_0.2500 &	$0.333348$ & $0.250012$ & $0.750003$ & $-0.500027$ & $0.250004$ & $-0.499983$ & $0.109370$ & $0.312512$\\
7 & Q0.3333\_0.5000\_-0.2500 &	$0.333350$ & $0.250013$ & $0.750001$ & $0.499990$ & $-0.250006$ & $-0.499981$ & $-0.109373$ & $-0.312504$\\
8 & Q0.3333\_0.5000\_0.2500 &	$0.333352$ & $0.250014$ & $0.750001$ & $0.499986$ & $0.250006$ & $-0.499979$ & $0.171876$ & $0.062500$\\
9 & Q0.3333\_-0.5000\_-0.6500 &	$0.333404$ & $0.250010$ & $0.749871$ & $-0.500032$ & $-0.650245$ & $-0.499921$ & $-0.396986$ & $-0.362630$\\
10 & Q0.3333\_-0.5000\_0.6500 &	$0.333407$ & $0.250013$ & $0.749874$ & $-0.500027$ & $0.650224$ & $-0.499917$ & $0.334449$ & $0.612669$\\
11 & Q0.3333\_0.5000\_-0.6500 &	$0.333407$ & $0.250012$ & $0.749871$ & $0.499991$ & $-0.650247$ & $-0.499917$ & $-0.334464$ & $-0.612677$\\
12 & Q0.3333\_0.5000\_0.6500 &	$0.333408$ & $0.250013$ & $0.749873$ & $0.499986$ & $0.650231$ & $-0.499916$ & $0.396974$ & $0.362628$\\
13 & Q0.3333\_-0.8000\_0.0000 &	$0.333237$ & $0.249932$ & $0.750011$ & $-0.800604$ & $0.000006$ & $-0.500108$ & $-0.050013$ & $0.200112$\\
14 & Q0.3333\_0.8000\_0.0000 &	$0.333247$ & $0.249936$ & $0.750002$ & $0.800511$ & $0.000002$ & $-0.500097$ & $0.050013$ & $-0.200088$\\
15 & Q0.3333\_-0.8000\_-0.5000 &	$0.333250$ & $0.249930$ & $0.749976$ & $-0.800619$ & $-0.500032$ & $-0.500094$ & $-0.331323$ & $-0.174930$\\
16 & Q0.3333\_-0.8000\_0.5000 &	$0.333261$ & $0.249939$ & $0.749980$ & $-0.800553$ & $0.500026$ & $-0.500082$ & $0.231277$ & $0.575146$\\
17 & Q0.3333\_0.8000\_-0.5000 &	$0.333265$ & $0.249938$ & $0.749969$ & $0.800473$ & $-0.500055$ & $-0.500077$ & $-0.231296$ & $-0.575148$\\
18 & Q0.3333\_0.8000\_0.5000 &	$0.333267$ & $0.249940$ & $0.749969$ & $0.800454$ & $0.500050$ & $-0.500074$ & $0.331319$ & $0.174972$\\
19 & Q0.3333\_-0.8000\_-0.8000 &	$0.333408$ & $0.249924$ & $0.749603$ & $-0.800648$ & $-0.800863$ & $-0.499916$ & $-0.500492$ & $-0.400418$\\
20 & Q0.3333\_0.8000\_0.8000 &	$0.333432$ & $0.249943$ & $0.749607$ & $0.800410$ & $0.800846$ & $-0.499889$ & $0.500457$ & $0.400443$\\
21 & Q0.5000\_0.0000\_-0.8000 &	$0.500261$ & $0.333334$ & $0.666320$ & $-0.000002$ & $-0.800844$ & $-0.333102$ & $-0.355807$ & $-0.533803$\\
22 & Q0.5000\_0.0000\_0.8000 &	$0.500256$ & $0.333333$ & $0.666326$ & $-0.000001$ & $0.800792$ & $-0.333106$ & $0.355786$ & $0.533771$\\
23 & Q0.5000\_-0.5000\_-0.1000 &	$0.500006$ & $0.333338$ & $0.666668$ & $-0.500052$ & $-0.100003$ & $-0.333328$ & $-0.100008$ & $0.100017$\\
24 & Q0.5000\_0.5000\_0.1000 &	$0.500007$ & $0.333338$ & $0.666667$ & $0.500025$ & $0.100004$ & $-0.333327$ & $0.100005$ & $-0.100008$\\
25 & Q0.5000\_-0.5000\_0.5000 &	$0.500031$ & $0.333345$ & $0.666648$ & $-0.500013$ & $0.500027$ & $-0.333306$ & $0.166664$ & $0.500022$\\
26 & Q0.5000\_0.5000\_-0.5000 &	$0.500030$ & $0.333342$ & $0.666643$ & $0.500035$ & $-0.500063$ & $-0.333307$ & $-0.166678$ & $-0.500054$\\
27 & Q0.5000\_-0.5000\_-0.6000 &	$0.500059$ & $0.333338$ & $0.666597$ & $-0.500051$ & $-0.600147$ & $-0.333281$ & $-0.322281$ & $-0.233386$\\
28 & Q0.5000\_0.5000\_0.6000 &	$0.500061$ & $0.333339$ & $0.666596$ & $0.500024$ & $0.600134$ & $-0.333279$ & $0.322272$ & $0.233385$\\
29 & Q0.7500\_0.0000\_0.2500 &	$0.749995$ & $0.428572$ & $0.571433$ & $-0.000002$ & $0.249997$ & $-0.142860$ & $0.081632$ & $0.142857$\\
30 & Q0.7500\_0.0000\_-0.5000 &	$0.750017$ & $0.428572$ & $0.571416$ & $-0.000002$ & $-0.500049$ & $-0.142846$ & $-0.163278$ & $-0.285738$\\
31 & Q0.7500\_0.0000\_0.5000 &	$0.750018$ & $0.428571$ & $0.571415$ & $-0.000001$ & $0.500028$ & $-0.142846$ & $0.163271$ & $0.285728$\\
32 & Q0.7500\_0.0000\_-0.8000 &	$0.750353$ & $0.428575$ & $0.571164$ & $-0.000001$ & $-0.800786$ & $-0.142627$ & $-0.261376$ & $-0.457499$\\
33 & Q0.7500\_0.0000\_0.8000 &	$0.750355$ & $0.428572$ & $0.571159$ & $-0.000001$ & $0.800755$ & $-0.142625$ & $0.261365$ & $0.457482$\\
34 & Q0.7500\_-0.2500\_0.2500 &	$0.750005$ & $0.428577$ & $0.571433$ & $-0.250008$ & $0.249996$ & $-0.142854$ & $0.035711$ & $0.250002$\\
35 & Q0.7500\_0.2500\_-0.2500 &	$0.750006$ & $0.428578$ & $0.571433$ & $0.249993$ & $-0.250006$ & $-0.142853$ & $-0.035716$ & $-0.250000$\\
36 & Q0.7500\_-0.5000\_0.0000 &	$0.750001$ & $0.428573$ & $0.571430$ & $-0.500036$ & $-0.000000$ & $-0.142856$ & $-0.091844$ & $0.214301$\\
37 & Q0.7500\_0.5000\_0.0000 &	$0.750002$ & $0.428572$ & $0.571428$ & $0.500007$ & $-0.000000$ & $-0.142856$ & $0.091838$ & $-0.214289$\\
38 & Q0.7500\_-0.5000\_0.2500 &	$0.750000$ & $0.428576$ & $0.571435$ & $-0.500031$ & $0.249993$ & $-0.142857$ & $-0.010212$ & $0.357152$\\
39 & Q0.7500\_0.5000\_-0.2500 &	$0.750004$ & $0.428576$ & $0.571433$ & $0.499989$ & $-0.250006$ & $-0.142855$ & $0.010201$ & $-0.357142$\\
40 & Q0.7500\_-0.5000\_-0.5000 &	$0.750019$ & $0.428575$ & $0.571419$ & $-0.500031$ & $-0.500038$ & $-0.142845$ & $-0.255119$ & $-0.071431$\\
41 & Q0.7500\_-0.5000\_0.5000 &	$0.750016$ & $0.428572$ & $0.571418$ & $-0.500038$ & $0.500021$ & $-0.142847$ & $0.071423$ & $0.500028$\\
42 & Q0.7500\_0.5000\_-0.5000 &	$0.750021$ & $0.428573$ & $0.571415$ & $0.500002$ & $-0.500050$ & $-0.142844$ & $-0.071438$ & $-0.500029$\\
43 & Q0.7500\_0.5000\_0.5000 &	$0.750018$ & $0.428571$ & $0.571414$ & $0.500010$ & $0.500029$ & $-0.142845$ & $0.255112$ & $0.071435$\\
44 & Q0.7500\_-0.5000\_0.8000 &	$0.750351$ & $0.428572$ & $0.571162$ & $-0.500035$ & $0.800741$ & $-0.142628$ & $0.169469$ & $0.671832$\\
45 & Q0.7500\_-0.8000\_0.0000 &	$0.749701$ & $0.428408$ & $0.571439$ & $-0.800681$ & $0.000002$ & $-0.143053$ & $-0.146996$ & $0.343072$\\
46 & Q0.7500\_0.8000\_0.0000 &	$0.749704$ & $0.428403$ & $0.571430$ & $0.800634$ & $-0.000000$ & $-0.143051$ & $0.146989$ & $-0.343052$\\
47 & Q0.7500\_-0.8000\_0.8000 &	$0.750044$ & $0.428413$ & $0.571185$ & $-0.800638$ & $0.800638$ & $-0.142829$ & $0.114354$ & $0.800638$\\
48 & Q0.7500\_-0.8500\_0.6375 &	$0.749627$ & $0.428318$ & $0.571375$ & $-0.851092$ & $0.637601$ & $-0.143101$ & $0.052051$ & $0.729071$\\
49 & Q0.7500\_0.8500\_-0.6375 &	$0.749648$ & $0.428321$ & $0.571363$ & $0.850998$ & $-0.637676$ & $-0.143087$ & $-0.052082$ & $-0.729075$\\
50 & Q0.8200\_-0.4400\_0.3300 &	$0.819999$ & $0.450552$ & $0.549454$ & $-0.440020$ & $0.329999$ & $-0.098902$ & $0.010304$ & $0.379569$\\
51 & Q1.0000\_0.0000\_-0.5000 &	$1.000013$ & $0.499999$ & $0.499992$ & $-0.000001$ & $-0.499978$ & $0.000007$ & $-0.124993$ & $-0.249987$\\
52 & Q1.0000\_0.0000\_-0.8000 &	$1.000437$ & $0.500005$ & $0.499787$ & $-0.000001$ & $-0.800758$ & $0.000219$ & $-0.200102$ & $-0.400291$\\
53 & Q1.0000\_-0.2500\_0.0000 &	$1.000008$ & $0.500001$ & $0.499997$ & $-0.249958$ & $-0.000001$ & $0.000004$ & $-0.062490$ & $0.124979$\\
54 & Q1.0000\_-0.2500\_-0.2500 &	$1.000000$ & $0.500001$ & $0.500001$ & $-0.249958$ & $-0.249958$ & $0.000000$ & $-0.124979$ & $0.000000$\\
55 & Q1.0000\_-0.2500\_0.2500 &	$1.000000$ & $0.500001$ & $0.500001$ & $-0.249958$ & $0.249950$ & $-0.000000$ & $-0.000002$ & $0.249954$\\
56 & Q1.0000\_0.2500\_0.2500 &	$1.000000$ & $0.500001$ & $0.500001$ & $0.249950$ & $0.249950$ & $-0.000000$ & $0.124975$ & $0.000000$\\
57 & Q1.0000\_-0.5000\_-0.5000 &	$1.000000$ & $0.499994$ & $0.499994$ & $-0.499971$ & $-0.499972$ & $-0.000000$ & $-0.249986$ & $-0.000000$\\
58 & Q1.0000\_0.5000\_0.5000 &	$1.000000$ & $0.499991$ & $0.499991$ & $0.499953$ & $0.499953$ & $-0.000000$ & $0.249976$ & $0.000000$\\
59 & Q1.0000\_-0.8000\_-0.4000 &	$0.999539$ & $0.499773$ & $0.500004$ & $-0.800796$ & $-0.399931$ & $-0.000230$ & $-0.300136$ & $0.200295$\\
60 & Q1.0000\_-0.8000\_-0.8000 &	$1.000000$ & $0.499794$ & $0.499794$ & $-0.800705$ & $-0.800705$ & $-0.000000$ & $-0.400353$ & $-0.000000$\\
61 & Q1.0000\_0.8000\_0.8000 &	$1.000000$ & $0.499788$ & $0.499788$ & $0.800682$ & $0.800682$ & $-0.000000$ & $0.400341$ & $0.000000$\\

\end{tabular}
\end{ruledtabular}
\end{table*}

\begin{table*}
\caption{The mass and spin of the BHBs in Table~\ref{tab:ID} after the
BHs had time to equilibrate ($t/m=150$) for non-spinning systems.}\label{tab:IDr_nonspin}
\begin{ruledtabular}
\begin{tabular}{llcccccccc}
Run & Config.  & $q^r$ & $m^r_1/m$ & $m^r_2/m$ & $\alpha^r_1$ & $\alpha^r_2$ & $\delta m_r$ & $S_r/m^2_r$ & $\Delta_r/m^2_r$ \\
\hline

62 & Q0.1667\_0.0000\_0.0000 &	$0.166674$ & $0.142863$ & $0.857142$ & $0.000008$ & $0.000001$ & $-0.714275$ & $0.000001$ & $-0.000000$\\
63 & Q0.2000\_0.0000\_0.0000 &	$0.200005$ & $0.166671$ & $0.833333$ & $-0.000001$ & $0.000001$ & $-0.666659$ & $0.000001$ & $0.000001$\\
64 & Q0.2500\_0.0000\_0.0000 &	$0.250002$ & $0.200001$ & $0.800000$ & $-0.000002$ & $0.000001$ & $-0.599998$ & $0.000000$ & $0.000001$\\
65 & Q0.3333\_0.0000\_0.0000 &	$0.333335$ & $0.250001$ & $0.750000$ & $-0.000002$ & $0.000000$ & $-0.499998$ & $0.000000$ & $0.000001$\\
66 & Q0.4000\_0.0000\_0.0000 &	$0.400005$ & $0.285718$ & $0.714286$ & $-0.000003$ & $0.000000$ & $-0.428566$ & $-0.000000$ & $0.000001$\\
67 & Q0.5000\_0.0000\_0.0000 &	$0.500001$ & $0.333334$ & $0.666667$ & $-0.000002$ & $-0.000000$ & $-0.333333$ & $-0.000000$ & $0.000001$\\
68 & Q0.6000\_0.0000\_0.0000 &	$0.600000$ & $0.374999$ & $0.624999$ & $-0.000001$ & $-0.000000$ & $-0.250000$ & $-0.000000$ & $0.000000$\\
69 & Q0.6667\_0.0000\_0.0000 &	$0.666687$ & $0.400011$ & $0.599999$ & $-0.000006$ & $-0.000000$ & $-0.199986$ & $-0.000001$ & $0.000002$\\
70 & Q0.7500\_0.0000\_0.0000 &	$0.750011$ & $0.428578$ & $0.571429$ & $-0.000001$ & $-0.000001$ & $-0.142850$ & $-0.000000$ & $0.000000$\\
71 & Q0.8500\_0.0000\_0.0000 &	$0.850013$ & $0.459467$ & $0.540541$ & $0.000001$ & $-0.000001$ & $-0.081073$ & $0.000000$ & $-0.000001$\\

\end{tabular}
\end{ruledtabular}
\end{table*}

In Tables \ref{tab:spinerad} and \ref{tab:spinerad_nonspin}, the energy
radiated and final spins are given.  Both quantities are calculated in
two ways, one locally on the isolated horizon (at the highest resolution
available), and one from the extrapolation
of the gravitational waveforms to infinite observer location.  The errors 
given for the horizon quantities are the drift in the quantity from 
the value calculated just after merger until the end of the simulation,
typically $400-500m$ evolution time.  For the quantities calculated
from the gravitational waves, the error is the difference between
a first and second order polynomial extrapolation to infinite
observer location.
\begin{table*}
\caption{The final energy radiated and spin as measured using the IH formalism
and as measured from the radiation of energy and angular momentum. The
error bars in the radiative quantities are due to radial extrapolation
errors while the error bars in the IH quantities are due to variations
in the measured mass and spin with time.
}\label{tab:spinerad}
\begin{ruledtabular}
\begin{tabular}{llcccc}
Run & Config. & $\delta \mathcal{M}^{IH}$ & $\delta \mathcal{M}^{rad}$ & $\alpha_{\mathrm{rem}}^{IH}$ & $\alpha_{\mathrm{rem}}^{rad}$\\
\hline

1 & Q0.3333\_0.0000\_-0.5000 &	 $0.022851 \pm 0.000002$ & 	$0.022817 \pm 0.000135$ & 	$0.316214 \pm 0.000003$ & 	$0.313624 \pm 0.006158$ \\
2 & Q0.3333\_0.0000\_0.5000 &	 $0.039597 \pm 0.000001$ & 	$0.039168 \pm 0.000145$ & 	$0.755227 \pm 0.000063$ & 	$0.753742 \pm 0.005431$ \\
3 & Q0.3333\_0.0000\_-0.8500 &	 $0.020650 \pm 0.000002$ & 	$0.020650 \pm 0.000166$ & 	$0.154585 \pm 0.000007$ & 	$0.152043 \pm 0.006126$ \\
4 & Q0.3333\_0.0000\_0.8500 &	 $0.056496 \pm 0.000000$ & 	$0.055113 \pm 0.000476$ & 	$0.895880 \pm 0.000003$ & 	$0.895066 \pm 0.007610$ \\
5 & Q0.3333\_-0.5000\_-0.2500 &	 $0.024450 \pm 0.000004$ & 	$0.024341 \pm 0.000149$ & 	$0.412707 \pm 0.000003$ & 	$0.409940 \pm 0.007081$ \\
6 & Q0.3333\_-0.5000\_0.2500 &	 $0.031700 \pm 0.000000$ & 	$0.031438 \pm 0.000128$ & 	$0.634229 \pm 0.000006$ & 	$0.632239 \pm 0.005915$ \\
7 & Q0.3333\_0.5000\_-0.2500 &	 $0.026411 \pm 0.000001$ & 	$0.026301 \pm 0.000105$ & 	$0.445700 \pm 0.000006$ & 	$0.443547 \pm 0.005283$ \\
8 & Q0.3333\_0.5000\_0.2500 &	 $0.034833 \pm 0.000003$ & 	$0.034557 \pm 0.000093$ & 	$0.664090 \pm 0.000002$ & 	$0.662470 \pm 0.004798$ \\
9 & Q0.3333\_-0.5000\_-0.6500 &	 $0.020982 \pm 0.000000$ & 	$0.020936 \pm 0.000159$ & 	$0.230000 \pm 0.000008$ & 	$0.227124 \pm 0.006786$ \\
10 & Q0.3333\_-0.5000\_0.6500 &	 $0.042569 \pm 0.000000$ & 	$0.041919 \pm 0.000216$ & 	$0.803309 \pm 0.000047$ & 	$0.801973 \pm 0.005941$ \\
11 & Q0.3333\_0.5000\_-0.6500 &	 $0.022422 \pm 0.000000$ & 	$0.022373 \pm 0.000138$ & 	$0.264943 \pm 0.000007$ & 	$0.262435 \pm 0.005705$ \\
12 & Q0.3333\_0.5000\_0.6500 &	 $0.048004 \pm 0.000001$ & 	$0.047242 \pm 0.000229$ & 	$0.829632 \pm 0.000030$ & 	$0.828412 \pm 0.005938$ \\
13 & Q0.3333\_-0.8000\_0.0000 &	 $0.026981 \pm 0.000003$ & 	$0.026649 \pm 0.000137$ & 	$0.514639 \pm 0.000005$ & 	$0.513271 \pm 0.006530$ \\
14 & Q0.3333\_0.8000\_0.0000 &	 $0.030886 \pm 0.000002$ & 	$0.030531 \pm 0.000114$ & 	$0.565067 \pm 0.000005$ & 	$0.563548 \pm 0.005781$ \\
15 & Q0.3333\_-0.8000\_-0.5000 &	 $0.021678 \pm 0.000000$ & 	$0.021448 \pm 0.000155$ & 	$0.288018 \pm 0.000003$ & 	$0.286577 \pm 0.006891$ \\
16 & Q0.3333\_-0.8000\_0.5000 &	 $0.036549 \pm 0.000003$ & 	$0.035953 \pm 0.000151$ & 	$0.732134 \pm 0.000024$ & 	$0.731806 \pm 0.005237$ \\
17 & Q0.3333\_0.8000\_-0.5000 &	 $0.024281 \pm 0.000001$ & 	$0.024037 \pm 0.000107$ & 	$0.342874 \pm 0.000001$ & 	$0.341834 \pm 0.004626$ \\
18 & Q0.3333\_0.8000\_0.5000 &	 $0.043435 \pm 0.000005$ & 	$0.042744 \pm 0.000148$ & 	$0.776618 \pm 0.000134$ & 	$0.776230 \pm 0.005133$ \\
19 & Q0.3333\_-0.8000\_-0.8000 &	 $0.019848 \pm 0.000002$ & 	$0.019659 \pm 0.000218$ & 	$0.149516 \pm 0.000003$ & 	$0.147452 \pm 0.008868$ \\
20 & Q0.3333\_0.8000\_0.8000 &	 $0.059611 \pm 0.000125$ & 	$0.057982 \pm 0.000431$ & 	$0.894974 \pm 0.002397$ & 	$0.895480 \pm 0.006129$ \\
21 & Q0.5000\_0.0000\_-0.8000 &	 $0.029011 \pm 0.000005$ & 	$0.028849 \pm 0.000308$ & 	$0.359268 \pm 0.000005$ & 	$0.354958 \pm 0.011629$ \\
22 & Q0.5000\_0.0000\_0.8000 &	 $0.063041 \pm 0.000000$ & 	$0.060434 \pm 0.000791$ & 	$0.865799 \pm 0.000107$ & 	$0.865719 \pm 0.012199$ \\
23 & Q0.5000\_-0.5000\_-0.1000 &	 $0.034843 \pm 0.000003$ & 	$0.034440 \pm 0.000276$ & 	$0.560690 \pm 0.000018$ & 	$0.557023 \pm 0.011230$ \\
24 & Q0.5000\_0.5000\_0.1000 &	 $0.043650 \pm 0.000004$ & 	$0.042936 \pm 0.000332$ & 	$0.683753 \pm 0.000143$ & 	$0.680628 \pm 0.011798$ \\
25 & Q0.5000\_-0.5000\_0.5000 &	 $0.046664 \pm 0.000001$ & 	$0.046314 \pm 0.000143$ & 	$0.750834 \pm 0.000003$ & 	$0.748841 \pm 0.005544$ \\
26 & Q0.5000\_0.5000\_-0.5000 &	 $0.033650 \pm 0.000002$ & 	$0.033372 \pm 0.000165$ & 	$0.491246 \pm 0.000001$ & 	$0.489202 \pm 0.006112$ \\
27 & Q0.5000\_-0.5000\_-0.6000 &	 $0.029090 \pm 0.000000$ & 	$0.028893 \pm 0.000274$ & 	$0.394301 \pm 0.000002$ & 	$0.390023 \pm 0.011285$ \\
28 & Q0.5000\_0.5000\_0.6000 &	 $0.059275 \pm 0.000001$ & 	$0.057310 \pm 0.000639$ & 	$0.833826 \pm 0.000020$ & 	$0.832455 \pm 0.013082$ \\
29 & Q0.7500\_0.0000\_0.2500 &	 $0.051049 \pm 0.000001$ & 	$0.050317 \pm 0.000218$ & 	$0.727731 \pm 0.000003$ & 	$0.726413 \pm 0.006203$ \\
30 & Q0.7500\_0.0000\_-0.5000 &	 $0.039526 \pm 0.000001$ & 	$0.039233 \pm 0.000180$ & 	$0.566042 \pm 0.000006$ & 	$0.563686 \pm 0.006655$ \\
31 & Q0.7500\_0.0000\_0.5000 &	 $0.056774 \pm 0.000001$ & 	$0.055687 \pm 0.000356$ & 	$0.778674 \pm 0.000001$ & 	$0.776580 \pm 0.009746$ \\
32 & Q0.7500\_0.0000\_-0.8000 &	 $0.036685 \pm 0.000000$ & 	$0.036378 \pm 0.000215$ & 	$0.498004 \pm 0.000005$ & 	$0.496150 \pm 0.006218$ \\
33 & Q0.7500\_0.0000\_0.8000 &	 $0.066017 \pm 0.000000$ & 	$0.064170 \pm 0.000608$ & 	$0.836975 \pm 0.000040$ & 	$0.835320 \pm 0.011812$ \\
34 & Q0.7500\_-0.2500\_0.2500 &	 $0.048311 \pm 0.000001$ & 	$0.047681 \pm 0.000204$ & 	$0.701752 \pm 0.000005$ & 	$0.700428 \pm 0.005855$ \\
35 & Q0.7500\_0.2500\_-0.2500 &	 $0.044884 \pm 0.000001$ & 	$0.044437 \pm 0.000173$ & 	$0.647887 \pm 0.000004$ & 	$0.646394 \pm 0.005491$ \\
36 & Q0.7500\_-0.5000\_0.0000 &	 $0.042106 \pm 0.000001$ & 	$0.041679 \pm 0.000197$ & 	$0.621030 \pm 0.000006$ & 	$0.618628 \pm 0.007280$ \\
37 & Q0.7500\_0.5000\_0.0000 &	 $0.051985 \pm 0.000002$ & 	$0.051220 \pm 0.000280$ & 	$0.726504 \pm 0.000002$ & 	$0.723901 \pm 0.009257$ \\
38 & Q0.7500\_-0.5000\_0.2500 &	 $0.045826 \pm 0.000001$ & 	$0.045255 \pm 0.000200$ & 	$0.674934 \pm 0.000007$ & 	$0.673806 \pm 0.005391$ \\
39 & Q0.7500\_0.5000\_-0.2500 &	 $0.047349 \pm 0.000001$ & 	$0.046807 \pm 0.000199$ & 	$0.673966 \pm 0.000005$ & 	$0.672384 \pm 0.005937$ \\
40 & Q0.7500\_-0.5000\_-0.5000 &	 $0.036361 \pm 0.000001$ & 	$0.036099 \pm 0.000171$ & 	$0.509444 \pm 0.000007$ & 	$0.507455 \pm 0.005885$ \\
41 & Q0.7500\_-0.5000\_0.5000 &	 $0.050463 \pm 0.000000$ & 	$0.049660 \pm 0.000300$ & 	$0.727625 \pm 0.000002$ & 	$0.725455 \pm 0.008693$ \\
42 & Q0.7500\_0.5000\_-0.5000 &	 $0.043506 \pm 0.000000$ & 	$0.043094 \pm 0.000222$ & 	$0.619960 \pm 0.000001$ & 	$0.617349 \pm 0.007628$ \\
43 & Q0.7500\_0.5000\_0.5000 &	 $0.065024 \pm 0.000001$ & 	$0.063390 \pm 0.000506$ & 	$0.826599 \pm 0.000017$ & 	$0.824800 \pm 0.011315$ \\
44 & Q0.7500\_-0.5000\_0.8000 &	 $0.057720 \pm 0.000001$ & 	$0.056416 \pm 0.000482$ & 	$0.788343 \pm 0.000002$ & 	$0.786306 \pm 0.010466$ \\
45 & Q0.7500\_-0.8000\_0.0000 &	 $0.039962 \pm 0.000002$ & 	$0.039397 \pm 0.000213$ & 	$0.587671 \pm 0.000007$ & 	$0.586076 \pm 0.006659$ \\
46 & Q0.7500\_0.8000\_0.0000 &	 $0.056197 \pm 0.000002$ & 	$0.055082 \pm 0.000370$ & 	$0.755757 \pm 0.000010$ & 	$0.753416 \pm 0.009901$ \\
47 & Q0.7500\_-0.8000\_0.8000 &	 $0.053885 \pm 0.000005$ & 	$0.053426 \pm 0.000254$ & 	$0.757809 \pm 0.000006$ & 	$0.755944 \pm 0.006548$ \\
48 & Q0.7500\_-0.8500\_0.6375 &	 $0.049693 \pm 0.000001$ & 	$0.048605 \pm 0.000372$ & 	$0.719449 \pm 0.000003$ & 	$0.717813 \pm 0.008919$ \\
49 & Q0.7500\_0.8500\_-0.6375 &	 $0.045036 \pm 0.000001$ & 	$0.044326 \pm 0.000316$ & 	$0.626888 \pm 0.000000$ & 	$0.624688 \pm 0.008358$ \\
50 & Q0.8200\_-0.4400\_0.3300 &	 $0.047979 \pm 0.000002$ & 	$0.047775 \pm 0.000268$ & 	$0.691971 \pm 0.000001$ & 	$0.686578 \pm 0.012997$ \\
51 & Q1.0000\_0.0000\_-0.5000 &	 $0.042523 \pm 0.000001$ & 	$0.042147 \pm 0.000192$ & 	$0.608514 \pm 0.000019$ & 	$0.606031 \pm 0.007032$ \\
52 & Q1.0000\_0.0000\_-0.8000 &	 $0.039828 \pm 0.000002$ & 	$0.039387 \pm 0.000229$ & 	$0.559265 \pm 0.000007$ & 	$0.557558 \pm 0.006678$ \\
53 & Q1.0000\_-0.2500\_0.0000 &	 $0.045246 \pm 0.000003$ & 	$0.044786 \pm 0.000275$ & 	$0.647826 \pm 0.000038$ & 	$0.643983 \pm 0.011363$ \\
54 & Q1.0000\_-0.2500\_-0.2500 &	 $0.042447 \pm 0.000005$ & 	$0.042077 \pm 0.000251$ & 	$0.608476 \pm 0.000019$ & 	$0.604880 \pm 0.010617$ \\
55 & Q1.0000\_-0.2500\_0.2500 &	 $0.048423 \pm 0.000005$ & 	$0.047841 \pm 0.000315$ & 	$0.686208 \pm 0.000087$ & 	$0.682593 \pm 0.012264$ \\
56 & Q1.0000\_0.2500\_0.2500 &	 $0.056224 \pm 0.000010$ & 	$0.055271 \pm 0.000410$ & 	$0.760776 \pm 0.000413$ & 	$0.758590 \pm 0.013925$ \\
57 & Q1.0000\_-0.5000\_-0.5000 &	 $0.037937 \pm 0.000002$ & 	$0.037667 \pm 0.000179$ & 	$0.527590 \pm 0.000016$ & 	$0.525104 \pm 0.006089$ \\
58 & Q1.0000\_0.5000\_0.5000 &	 $0.067306 \pm 0.000009$ & 	$0.065641 \pm 0.000523$ & 	$0.830996 \pm 0.000268$ & 	$0.829616 \pm 0.011876$ \\
59 & Q1.0000\_-0.8000\_-0.4000 &	 $0.036537 \pm 0.000004$ & 	$0.036165 \pm 0.000325$ & 	$0.494027 \pm 0.000013$ & 	$0.489738 \pm 0.012360$ \\
60 & Q1.0000\_-0.8000\_-0.8000 &	 $0.033977 \pm 0.000003$ & 	$0.033485 \pm 0.000243$ & 	$0.426200 \pm 0.000026$ & 	$0.425451 \pm 0.005243$ \\
61 & Q1.0000\_0.8000\_0.8000 &	 $0.089381 \pm 0.000075$ & 	$0.085140 \pm 0.001195$ & 	$0.907691 \pm 0.000042$ & 	$0.908899 \pm 0.013099$ \\

\end{tabular}
\end{ruledtabular}
\end{table*}

\begin{table*}
\caption{The final energy radiated and spin as measured using the IH formalism
and as measured from the radiation of energy and angular momentum. Same as
in Table \ref{tab:spinerad} but for non-spinning systems.
}\label{tab:spinerad_nonspin}
\begin{ruledtabular}
\begin{tabular}{llcccc}
Run & Config. & $\delta \mathcal{M}^{IH}$ & $\delta \mathcal{M}^{rad}$ & $\alpha_{\mathrm{rem}}^{IH}$ & $\alpha_{\mathrm{rem}}^{rad}$\\
\hline

62 & Q0.1667\_0.0000\_0.0000 &	 $0.014711 \pm 0.000008$ & 	$0.014558 \pm 0.000083$ & 	$0.372243 \pm 0.000024$ & 	$0.366107 \pm 0.003115$ \\
63 & Q0.2000\_0.0000\_0.0000 &	 $0.017650 \pm 0.000004$ & 	$0.017461 \pm 0.000094$ & 	$0.416663 \pm 0.000014$ & 	$0.414855 \pm 0.003649$ \\
64 & Q0.2500\_0.0000\_0.0000 &	 $0.022082 \pm 0.000004$ & 	$0.021860 \pm 0.000115$ & 	$0.471618 \pm 0.000002$ & 	$0.470320 \pm 0.004711$ \\
65 & Q0.3333\_0.0000\_0.0000 &	 $0.028719 \pm 0.000001$ & 	$0.028582 \pm 0.000120$ & 	$0.540600 \pm 0.000003$ & 	$0.538311 \pm 0.006163$ \\
66 & Q0.4000\_0.0000\_0.0000 &	 $0.033245 \pm 0.000002$ & 	$0.032988 \pm 0.000157$ & 	$0.580738 \pm 0.000014$ & 	$0.579238 \pm 0.006950$ \\
67 & Q0.5000\_0.0000\_0.0000 &	 $0.038747 \pm 0.000000$ & 	$0.038616 \pm 0.000122$ & 	$0.623456 \pm 0.000000$ & 	$0.620479 \pm 0.006881$ \\
68 & Q0.6000\_0.0000\_0.0000 &	 $0.042781 \pm 0.000002$ & 	$0.042496 \pm 0.000206$ & 	$0.651373 \pm 0.000010$ & 	$0.647754 \pm 0.009099$ \\
69 & Q0.6667\_0.0000\_0.0000 &	 $0.044916 \pm 0.000018$ & 	$0.044685 \pm 0.000254$ & 	$0.663910 \pm 0.000102$ & 	$0.650408 \pm 0.011148$ \\
70 & Q0.7500\_0.0000\_0.0000 &	 $0.046490 \pm 0.000003$ & 	$0.046360 \pm 0.000139$ & 	$0.675108 \pm 0.000040$ & 	$0.671341 \pm 0.007807$ \\
71 & Q0.8500\_0.0000\_0.0000 &	 $0.047783 \pm 0.000004$ & 	$0.047488 \pm 0.000273$ & 	$0.682891 \pm 0.000041$ & 	$0.676581 \pm 0.012176$ \\
\end{tabular}
\end{ruledtabular}
\end{table*}

Finally, in Tables \ref{tab:kicks}, \ref{tab:kicks_nonspin} (for the new runs), 
and \ref{tab:kicks_paper1} (for the previous runs \cite{Healy:2014yta}),
the recoil velocities and peak luminosities are given.  Since the calculation
for the kick uses a new method explained below 
and since the peak luminosities were not given in 
Paper 1 \cite{Healy:2014yta}, we provide \ref{tab:kicks_paper1} for the simulations
of Paper 1.  In this table, we provide both the new run number in the first column, 
along with the Paper 1 run number in column 2.  Error bars for the recoil and
luminosity are described in the next appendix.   
\begin{table*}
\caption{The recoil velocity (in km/s) and peak luminosity as calculated using $\ell_{max}=6$ and
$r_{max}=113.0m$ for spinning systems.
$V_{w}$ and $\mathcal{L}_{w}$ are the weights used in the least-squares fitting.
 The error estimates are detailed in Appendix~\ref{app:errors}.}
\label{tab:kicks}
\begin{ruledtabular}
\begin{tabular}{llrrcc}
Run & Config. &  $V$ [km/s] & $V_{w}$ [km/s] & $\mathcal{L}_{\text{max}}$ & $\mathcal{L}_{w}$\\
\hline
1 & Q0.3333\_0.0000\_-0.5000    &	$269.61 \pm 6.64$ & 6.33 & 4.1989e-04 $\pm$ 5.3772e-06 & -5.2000e-06 \\
2 & Q0.3333\_0.0000\_0.5000    &	$71.54 \pm 9.65$ & 6.33 & 7.0865e-04 $\pm$ 1.0874e-05 & -5.2000e-06 \\
3 & Q0.3333\_0.0000\_-0.8500    &	$325.83 \pm 6.40$ & 6.33 & 3.7435e-04 $\pm$ 5.3254e-06 & -5.2000e-06 \\
4 & Q0.3333\_0.0000\_0.8500    &	$33.54 \pm 6.73$ & 6.33 & 9.4661e-04 $\pm$ 2.3307e-05 & -5.2000e-06 \\
5 & Q0.3333\_-0.5000\_-0.2500    &	$208.71 \pm 6.99$ & 6.33 & 4.5570e-04 $\pm$ 5.9348e-06 & -5.2000e-06 \\
6 & Q0.3333\_-0.5000\_0.2500    &	$117.02 \pm 8.86$ & 6.33 & 5.8663e-04 $\pm$ 7.3102e-06 & -5.2000e-06 \\
7 & Q0.3333\_0.5000\_-0.2500    &	$244.89 \pm 6.80$ & 6.33 & 4.7481e-04 $\pm$ 6.0819e-06 & -5.2000e-06 \\
8 & Q0.3333\_0.5000\_0.2500    &	$133.17 \pm 7.94$ & 6.33 & 6.1982e-04 $\pm$ 6.5282e-06 & -5.2000e-06 \\
9 & Q0.3333\_-0.5000\_-0.6500    &	$274.13 \pm 6.54$ & 6.33 & 3.9091e-04 $\pm$ 5.5952e-06 & -5.2000e-06 \\
10 & Q0.3333\_-0.5000\_0.6500    &	$66.68 \pm 8.57$ & 6.33 & 7.6508e-04 $\pm$ 1.2352e-05 & -5.2000e-06 \\
11 & Q0.3333\_0.5000\_-0.6500    &	$316.55 \pm 6.44$ & 6.33 & 4.0568e-04 $\pm$ 5.4250e-06 & -5.2000e-06 \\
12 & Q0.3333\_0.5000\_0.6500    &	$41.86 \pm 9.39$ & 6.33 & 8.2784e-04 $\pm$ 1.3973e-05 & -5.2000e-06 \\
13 & Q0.3333\_-0.8000\_0.0000    &	$155.41 \pm 7.85$ & 6.33 & 5.0570e-04 $\pm$ 5.8644e-06 & -5.2000e-06 \\
14 & Q0.3333\_0.8000\_0.0000    &	$204.93 \pm 7.00$ & 6.33 & 5.4590e-04 $\pm$ 6.0256e-06 & -5.2000e-06 \\
15 & Q0.3333\_-0.8000\_-0.5000    &	$239.24 \pm 6.66$ & 6.33 & 4.0786e-04 $\pm$ 5.3643e-06 & -5.2000e-06 \\
16 & Q0.3333\_-0.8000\_0.5000    &	$86.30 \pm 8.89$ & 6.33 & 6.6698e-04 $\pm$ 8.4126e-06 & -5.2000e-06 \\
17 & Q0.3333\_0.8000\_-0.5000    &	$302.56 \pm 6.57$ & 6.33 & 4.3127e-04 $\pm$ 5.3695e-06 & -5.2000e-06 \\
18 & Q0.3333\_0.8000\_0.5000    &	$75.11 \pm 8.56$ & 6.33 & 7.4376e-04 $\pm$ 1.1449e-05 & -5.2000e-06 \\
19 & Q0.3333\_-0.8000\_-0.8000    &	$285.61 \pm 6.40$ & 6.33 & 3.6999e-04 $\pm$ 5.2712e-06 & -5.2000e-06 \\
20 & Q0.3333\_0.8000\_0.8000    &	$17.42 \pm 9.22$ & 6.33 & 9.8382e-04 $\pm$ 2.2248e-05 & -5.2000e-06 \\
21 & Q0.5000\_0.0000\_-0.8000    &	$350.67 \pm 9.36$ & 8.99 & 5.9661e-04 $\pm$ 1.5332e-05 & 1.4900e-05 \\
22 & Q0.5000\_0.0000\_0.8000    &	$67.84 \pm 13.08$ & 8.99 & 1.1511e-03 $\pm$ 4.7746e-05 & 1.4900e-05 \\
23 & Q0.5000\_-0.5000\_-0.1000    &	$143.24 \pm 13.32$ & 8.99 & 7.1138e-04 $\pm$ 1.9643e-05 & 1.4900e-05 \\
24 & Q0.5000\_0.5000\_0.1000    &	$161.66 \pm 11.04$ & 8.99 & 8.4802e-04 $\pm$ 2.4327e-05 & 1.4900e-05 \\
25 & Q0.5000\_-0.5000\_0.5000    &	$103.20 \pm 8.11$ & 7.17 & 9.2813e-04 $\pm$ 8.2967e-06 & 4.8400e-06 \\
26 & Q0.5000\_0.5000\_-0.5000    &	$325.60 \pm 9.74$ & 8.99 & 6.7153e-04 $\pm$ 1.7319e-05 & 1.4900e-05 \\
27 & Q0.5000\_-0.5000\_-0.6000    &	$256.73 \pm 9.98$ & 8.99 & 6.0546e-04 $\pm$ 1.6035e-05 & 1.4900e-05 \\
28 & Q0.5000\_0.5000\_0.6000    &	$33.80 \pm 14.28$ & 8.99 & 1.0987e-03 $\pm$ 4.4638e-05 & 1.4900e-05 \\
29 & Q0.7500\_0.0000\_0.2500    &	$39.43 \pm 9.87$ & 6.78 & 1.0528e-03 $\pm$ 2.5374e-05 & 1.9000e-05 \\
30 & Q0.7500\_0.0000\_-0.5000    &	$209.16 \pm 6.89$ & 6.78 & 8.5126e-04 $\pm$ 1.9985e-05 & 1.9000e-05 \\
31 & Q0.7500\_0.0000\_0.5000    &	$70.63 \pm 8.17$ & 6.78 & 1.1504e-03 $\pm$ 3.1034e-05 & 1.9000e-05 \\
32 & Q0.7500\_0.0000\_-0.8000    &	$290.22 \pm 6.86$ & 6.78 & 7.9256e-04 $\pm$ 1.9752e-05 & 1.9000e-05 \\
33 & Q0.7500\_0.0000\_0.8000    &	$111.17 \pm 7.82$ & 6.78 & 1.2986e-03 $\pm$ 3.9935e-05 & 1.9000e-05 \\
34 & Q0.7500\_-0.2500\_0.2500    &	$69.35 \pm 8.64$ & 6.78 & 1.0087e-03 $\pm$ 2.3390e-05 & 1.9000e-05 \\
35 & Q0.7500\_0.2500\_-0.2500    &	$180.96 \pm 7.00$ & 6.78 & 9.3926e-04 $\pm$ 2.1835e-05 & 1.9000e-05 \\
36 & Q0.7500\_-0.5000\_0.0000    &	$66.33 \pm 8.86$ & 6.78 & 9.0246e-04 $\pm$ 2.0991e-05 & 1.9000e-05 \\
37 & Q0.7500\_0.5000\_0.0000    &	$147.79 \pm 6.99$ & 6.78 & 1.0580e-03 $\pm$ 2.3716e-05 & 1.9000e-05 \\
38 & Q0.7500\_-0.5000\_0.2500    &	$115.68 \pm 8.04$ & 6.78 & 9.6913e-04 $\pm$ 2.3444e-05 & 1.9000e-05 \\
39 & Q0.7500\_0.5000\_-0.2500    &	$219.82 \pm 7.00$ & 6.78 & 9.7811e-04 $\pm$ 2.2179e-05 & 1.9000e-05 \\
40 & Q0.7500\_-0.5000\_-0.5000    &	$123.51 \pm 7.07$ & 6.78 & 7.9779e-04 $\pm$ 1.9923e-05 & 1.9000e-05 \\
41 & Q0.7500\_-0.5000\_0.5000    &	$166.02 \pm 7.93$ & 6.78 & 1.0512e-03 $\pm$ 2.4317e-05 & 1.9000e-05 \\
42 & Q0.7500\_0.5000\_-0.5000    &	$292.56 \pm 6.92$ & 6.78 & 9.1151e-04 $\pm$ 2.1213e-05 & 1.9000e-05 \\
43 & Q0.7500\_0.5000\_0.5000    &	$26.22 \pm 8.41$ & 6.78 & 1.2761e-03 $\pm$ 3.4943e-05 & 1.9000e-05 \\
44 & Q0.7500\_-0.5000\_0.8000    &	$216.06 \pm 8.01$ & 6.78 & 1.1755e-03 $\pm$ 2.9614e-05 & 1.9000e-05 \\
45 & Q0.7500\_-0.8000\_0.0000    &	$114.57 \pm 8.24$ & 6.78 & 8.6890e-04 $\pm$ 2.0880e-05 & 1.9000e-05 \\
46 & Q0.7500\_0.8000\_0.0000    &	$183.55 \pm 6.95$ & 6.78 & 1.1198e-03 $\pm$ 2.5957e-05 & 1.9000e-05 \\
47 & Q0.7500\_-0.8000\_0.8000    &	$292.19 \pm 6.82$ & 6.78 & 1.1351e-03 $\pm$ 1.9390e-05 & 1.9000e-05 \\
48 & Q0.7500\_-0.8500\_0.6375    &	$272.78 \pm 8.32$ & 6.78 & 1.0419e-03 $\pm$ 2.3762e-05 & 1.9000e-05 \\
49 & Q0.7500\_0.8500\_-0.6375    &	$384.71 \pm 7.01$ & 6.78 & 9.2953e-04 $\pm$ 2.1158e-05 & 1.9000e-05 \\
50 & Q0.8200\_-0.4400\_0.3300    &	$137.44 \pm 1.12$ & 1.02 & 1.0064e-03 $\pm$ 1.9440e-05 & -1.9300e-05 \\
51 & Q1.0000\_0.0000\_-0.5000    &	$123.50 \pm 2.90$ & 2.83 & 9.1795e-04 $\pm$ 1.1027e-05 & 8.2000e-06 \\
52 & Q1.0000\_0.0000\_-0.8000    &	$201.51 \pm 2.95$ & 2.83 & 8.6553e-04 $\pm$ 1.0178e-05 & 8.2000e-06 \\
53 & Q1.0000\_-0.2500\_0.0000    &	$61.05 \pm 2.84$ & 2.83 & 9.6316e-04 $\pm$ 1.2909e-05 & 8.2000e-06 \\
54 & Q1.0000\_-0.2500\_-0.2500    &	$0.00 \pm 0.00$ & 0.00 & 9.1696e-04 $\pm$ 1.1294e-05 & 8.2000e-06 \\
55 & Q1.0000\_-0.2500\_0.2500    &	$117.02 \pm 2.89$ & 2.83 & 1.0206e-03 $\pm$ 1.3936e-05 & 8.2000e-06 \\
56 & Q1.0000\_0.2500\_0.2500    &	$0.00 \pm 0.00$ & 0.00 & 1.1491e-03 $\pm$ 1.9549e-05 & 8.2000e-06 \\
57 & Q1.0000\_-0.5000\_-0.5000    &	$0.00 \pm 0.00$ & 0.00 & 8.3778e-04 $\pm$ 9.3235e-06 & 8.2000e-06 \\
58 & Q1.0000\_0.5000\_0.5000    &	$0.00 \pm 0.00$ & 0.00 & 1.3261e-03 $\pm$ 3.2405e-05 & 8.2000e-06 \\
59 & Q1.0000\_-0.8000\_-0.4000    &	$102.37 \pm 2.84$ & 2.83 & 8.0836e-04 $\pm$ 9.2692e-06 & 8.2000e-06 \\
60 & Q1.0000\_-0.8000\_-0.8000    &	$0.00 \pm 0.00$ & 0.00 & 7.5828e-04 $\pm$ 8.8100e-06 & 8.2000e-06 \\
61 & Q1.0000\_0.8000\_0.8000    &	$0.00 \pm 0.00$ & 0.00 & 1.6519e-03 $\pm$ 7.2878e-05 & 8.2000e-06 \\
\end{tabular}
\end{ruledtabular}
\end{table*}

\begin{table*}
\caption{Recoil velocities (in km/s) and peak luminosities as in Table \ref{tab:kicks}
 but for non-spinning systems.}
\label{tab:kicks_nonspin}
\begin{ruledtabular}
\begin{tabular}{llrrcc}
Run & Config. &  $V$ [km/s] & $V_{w}$ [km/s] & $\mathcal{L}_{\text{max}}$ & $\mathcal{L}_{w}$\\
\hline
62 & Q0.1667\_0.0000\_0.0000    &	$115.83 \pm 6.91$ & 3.07 & 2.0707e-04 $\pm$ 8.6305e-06 & 6.7400e-06 \\
63 & Q0.2000\_0.0000\_0.0000    &	$136.73 \pm 7.39$ & 3.07 & 2.7679e-04 $\pm$ 7.9392e-06 & 6.7400e-06 \\
64 & Q0.2500\_0.0000\_0.0000    &	$157.63 \pm 8.66$ & 4.72 & 3.6893e-04 $\pm$ 8.7373e-06 & 4.3000e-06 \\
65 & Q0.3333\_0.0000\_0.0000    &	$178.05 \pm 7.38$ & 6.33 & 5.2412e-04 $\pm$ 6.1042e-06 & -5.2000e-06 \\
66 & Q0.4000\_0.0000\_0.0000    &	$169.62 \pm 6.04$ & 2.03 & 6.3407e-04 $\pm$ 6.6580e-06 & 1.0100e-06 \\
67 & Q0.5000\_0.0000\_0.0000    &	$154.81 \pm 5.07$ & 4.50 & 7.7375e-04 $\pm$ 1.6389e-06 & -1.7000e-07 \\
68 & Q0.6000\_0.0000\_0.0000    &	$126.43 \pm 7.56$ & 6.89 & 8.9059e-04 $\pm$ 3.5371e-05 & 3.5100e-05 \\
69 & Q0.6667\_0.0000\_0.0000    &	$103.31 \pm 5.49$ & 4.92 & 9.2086e-04 $\pm$ 6.3996e-06 & -3.0900e-06 \\
70 & Q0.7500\_0.0000\_0.0000    &	$75.66 \pm 1.93$ & 1.74 & 9.7238e-04 $\pm$ 1.7329e-06 & 7.1600e-07 \\
71 & Q0.8500\_0.0000\_0.0000    &	$45.38 \pm 3.82$ & 3.74 & 1.0092e-03 $\pm$ 1.1205e-05 & 1.0300e-05 \\
\end{tabular}
\end{ruledtabular}
\end{table*}

\begin{table*}
\caption{Recoil velocities (in km/s) and peak luminosities as in Table \ref{tab:kicks}
 but for systems in Paper 1 \cite{Healy:2014yta} calculated using $\ell_{max}=6$ and
$r_{max}=113.0m$ and the perturbative extrapolation formulae in Ref.~\cite{Nakano:2015pta}.
}
\label{tab:kicks_paper1}
\begin{ruledtabular}
\begin{tabular}{llrrcc}
Run & Config. &  $V$ [km/s] & $V_{w}$ [km/s] & $\mathcal{L}_{\text{max}}$ & $\mathcal{L}_{w}$\\
\hline
72 & Q1.00\_0.00\_0.00    &	$0.00 \pm 0.00$ & 0.00 & 1.0212e-03 $\pm$ 1.5578e-05 & 8.2000e-06 \\
73 & Q1.00\_0.00\_0.40    &	$86.52 \pm 2.94$ & 2.83 & 1.1222e-03 $\pm$ 1.6170e-05 & 8.2000e-06 \\
74 & Q1.00\_0.00\_0.60    &	$121.67 \pm 3.10$ & 2.83 & 1.1867e-03 $\pm$ 2.2864e-05 & 8.2000e-06 \\
75 & Q1.00\_0.00\_0.80    &	$150.41 \pm 3.36$ & 2.83 & 1.2583e-03 $\pm$ 2.6838e-05 & 8.2000e-06 \\
76 & Q1.00\_0.20\_0.80    &	$100.21 \pm 3.15$ & 2.83 & 1.3354e-03 $\pm$ 2.4063e-05 & 8.2000e-06 \\
77 & Q1.00\_0.40\_-0.40    &	$187.24 \pm 3.13$ & 2.83 & 1.0215e-03 $\pm$ 1.2137e-05 & 8.2000e-06 \\
78 & Q1.00\_0.40\_0.80    &	$56.81 \pm 2.97$ & 2.83 & 1.4303e-03 $\pm$ 4.4059e-05 & 8.2000e-06 \\
79 & Q1.00\_-0.60\_0.60    &	$278.63 \pm 3.61$ & 2.83 & 1.0291e-03 $\pm$ 1.2899e-05 & 8.2000e-06 \\
80 & Q1.00\_-0.80\_0.80    &	$370.96 \pm 4.17$ & 2.83 & 1.0364e-03 $\pm$ 1.3484e-05 & 8.2000e-06 \\
81 & Q1.33\_0.00\_-0.25    &	$141.93 \pm 7.04$ & 6.78 & 9.0970e-04 $\pm$ 2.0439e-05 & 1.9000e-05 \\
82 & Q1.33\_-0.80\_0.45    &	$224.08 \pm 8.04$ & 6.78 & 9.8985e-04 $\pm$ 2.3984e-05 & 1.9000e-05 \\
83 & Q1.33\_0.80\_-0.45    &	$324.51 \pm 7.06$ & 6.78 & 9.7709e-04 $\pm$ 2.1414e-05 & 1.9000e-05 \\
84 & Q1.33\_-0.80\_-0.60    &	$100.95 \pm 7.30$ & 6.78 & 7.5206e-04 $\pm$ 1.9400e-05 & 1.9000e-05 \\
85 & Q1.33\_0.80\_0.60    &	$31.53 \pm 6.81$ & 6.78 & 1.4439e-03 $\pm$ 5.1518e-05 & 1.9000e-05 \\
86 & Q1.33\_0.80\_-0.80    &	$429.46 \pm 6.98$ & 6.78 & 8.9441e-04 $\pm$ 2.0003e-05 & 1.9000e-05 \\
87 & Q2.00\_0.00\_-0.50    &	$280.29 \pm 10.11$ & 8.99 & 6.4881e-04 $\pm$ 1.5696e-05 & 1.4900e-05 \\
88 & Q2.00\_0.00\_0.50    &	$56.14 \pm 18.31$ & 8.99 & 9.8041e-04 $\pm$ 3.6214e-05 & 1.4900e-05 \\
89 & Q2.00\_-0.80\_0.20    &	$106.68 \pm 17.67$ & 8.99 & 7.8122e-04 $\pm$ 2.0382e-05 & 1.4900e-05 \\
90 & Q2.00\_0.80\_-0.20    &	$272.75 \pm 10.36$ & 8.99 & 7.7112e-04 $\pm$ 1.9656e-05 & 1.4900e-05 \\
91 & Q2.00\_-0.80\_-0.40    &	$186.81 \pm 11.35$ & 8.99 & 6.3521e-04 $\pm$ 1.6695e-05 & 1.4900e-05 \\
92 & Q2.00\_0.80\_0.40    &	$90.59 \pm 10.69$ & 8.99 & 1.0226e-03 $\pm$ 2.9113e-05 & 1.4900e-05 \\
93 & Q2.00\_-0.80\_-0.80    &	$275.34 \pm 9.70$ & 8.99 & 5.6439e-04 $\pm$ 1.5505e-05 & 1.4900e-05 \\
94 & Q2.00\_-0.80\_0.80    &	$163.32 \pm 13.49$ & 8.99 & 1.0474e-03 $\pm$ 4.9487e-05 & 1.4900e-05 \\
95 & Q2.00\_0.80\_-0.80    &	$427.96 \pm 9.46$ & 8.99 & 6.2197e-04 $\pm$ 1.5299e-05 & 1.4900e-05 \\
96 & Q2.00\_0.80\_0.80    &	$13.46 \pm 13.54$ & 8.99 & 1.3322e-03 $\pm$ 7.5317e-05 & 1.4900e-05 \\
97 & Q3.00\_0.00\_-0.67    &	$295.28 \pm 8.30$ & 6.33 & 3.9392e-04 $\pm$ 7.5159e-06 & -5.2000e-06 \\
98 & Q3.00\_0.00\_0.67    &	$41.95 \pm 17.23$ & 6.33 & 7.8909e-04 $\pm$ 4.8042e-05 & -5.2000e-06 \\
99 & Q3.00\_-0.80\_0.80    &	$75.01 \pm 7.80$ & 6.33 & 8.4159e-04 $\pm$ 3.6205e-05 & -5.2000e-06 \\
100 & Q3.00\_0.80\_-0.80    &	$353.28 \pm 6.68$ & 6.33 & 3.8485e-04 $\pm$ 5.5913e-06 & -5.2000e-06 \\
101 & Q4.00\_0.00\_-0.75    &	$255.84 \pm 10.72$ & 9.52 & 2.7338e-04 $\pm$ 1.0862e-05 & 1.0400e-05 \\
102 & Q4.00\_0.00\_0.75    &	$24.50 \pm 14.64$ & 9.52 & 6.3280e-04 $\pm$ 5.7880e-05 & 1.0400e-05 \\
103 & Q4.00\_0.80\_-0.80    &	$280.92 \pm 10.50$ & 9.52 & 2.7095e-04 $\pm$ 1.1017e-05 & 1.0400e-05 \\
104 & Q5.00\_0.00\_-0.80    &	$212.13 \pm 8.03$ & 6.77 & 1.9439e-04 $\pm$ 1.0722e-05 & 1.0600e-05 \\
105 & Q5.00\_0.00\_0.80    &	$15.46 \pm 10.17$ & 6.77 & 5.2483e-04 $\pm$ 4.9141e-05 & 1.0600e-05 \\
106 & Q6.00\_0.00\_-0.83    &	$177.69 \pm 7.63$ & 6.77 & 1.4863e-04 $\pm$ 1.0792e-05 & 1.0600e-05 \\
107 & Q6.00\_0.00\_0.83    &	$11.07 \pm 8.55$ & 6.77 & 4.3456e-04 $\pm$ 5.4457e-05 & 1.0600e-05 \\
\end{tabular}
\end{ruledtabular}
\end{table*}

\section{Convergence studies}\label{app:errors}


For the recoil velocity and peak luminosity calculations, 
there are two main sources of error.  The first is the error
from calculating the values at a finite observer location.  
The second, and typically of comparable magnitude, 
is from
finite resolution.  Since the final mass and spin are 
calculated locally on the apparent horizon, where the grid
resolution is highest, the finite resolution error is very low
and there is no associated finite observer error, hence those
quantities can be determined to high accuracy.
In order to get a sense of the error of other radiated quantities,
we can calculate the final mass and spin in an analogous way to how we
calculate the recoil and luminosity.  In this appendix, we will
explore these errors and the convergence of a set of simulations.

For every simulation, we can calculate the finite observer location error.
However, in order to calculate the finite resolution error, we would need
the same binary configuration at three different resolutions.
For 107 runs, this would be very expensive computationally. 
Instead, we choose a simulation from each mass ratio at 
three different resolutions and use this run to calculate 
the finite resolution error.  This error is then used
as a weight in the least-squares fitting for the corresponding family
of the same mass ratio runs.  

The kicks and luminosity values included in the tables 
are calculated using up to $\ell=6$ modes at
an extraction radius of $r_{obs}/m = 113.0$.  To
extrapolate to infinite observer location, we use the
perturbative extrapolation given in Ref. \cite{Nakano:2015pta}.
There is typically still some residual $r$-dependence so to calculate
the extrapolation error we take the difference between
the value calculated at $r_{obs} = 113.0m$ and
extrapolating to $r_{obs}=\infty$ using the average of a first
and second order polynomial in $1/r_{obs}$.  
This is typically on the order of $1-3$ km/s for the kick and 
on the order of 1e-5 to 1e-6  for the luminosity
but for a few runs is higher.

We can estimate the finite resolution error by using three 
different resolutions and a Richardson extrapolation.  We label
our resolutions as $NX$ where $X$ is a global resolution factor
that is related to the resolution in the wavezone.  For example,
a simulation labeled $N140$ would have a wavezone resolution of $m/1.4$.  
Tables \ref{tab:q100conv}-\ref{tab:q033conv} show the values of the kick and luminosity
at each extraction radius and extrapolated to infinite resolution,
along with the order of convergence for each.  We typically observe
a convergence order between 2.5 and 4 with slightly lower convergence
order for the $q=1/3$ luminosity.  The fitting weights are determined
from these tables as the difference between the $(r_{obs},N)=(\infty,\infty)$
and $(113.0m,\mathrm{finite})$, where N is the resolution factor
as described above.  This leads
to finite resolution errors typically on the order of $5-15$ km/s for the
recoil.  The total error reported in 
Tables \ref{tab:kicks}-\ref{tab:kicks_paper1}
is the finite observer and finite resolution errors added in quadrature.

\begin{figure}
\includegraphics[angle=270,width=0.49\columnwidth]{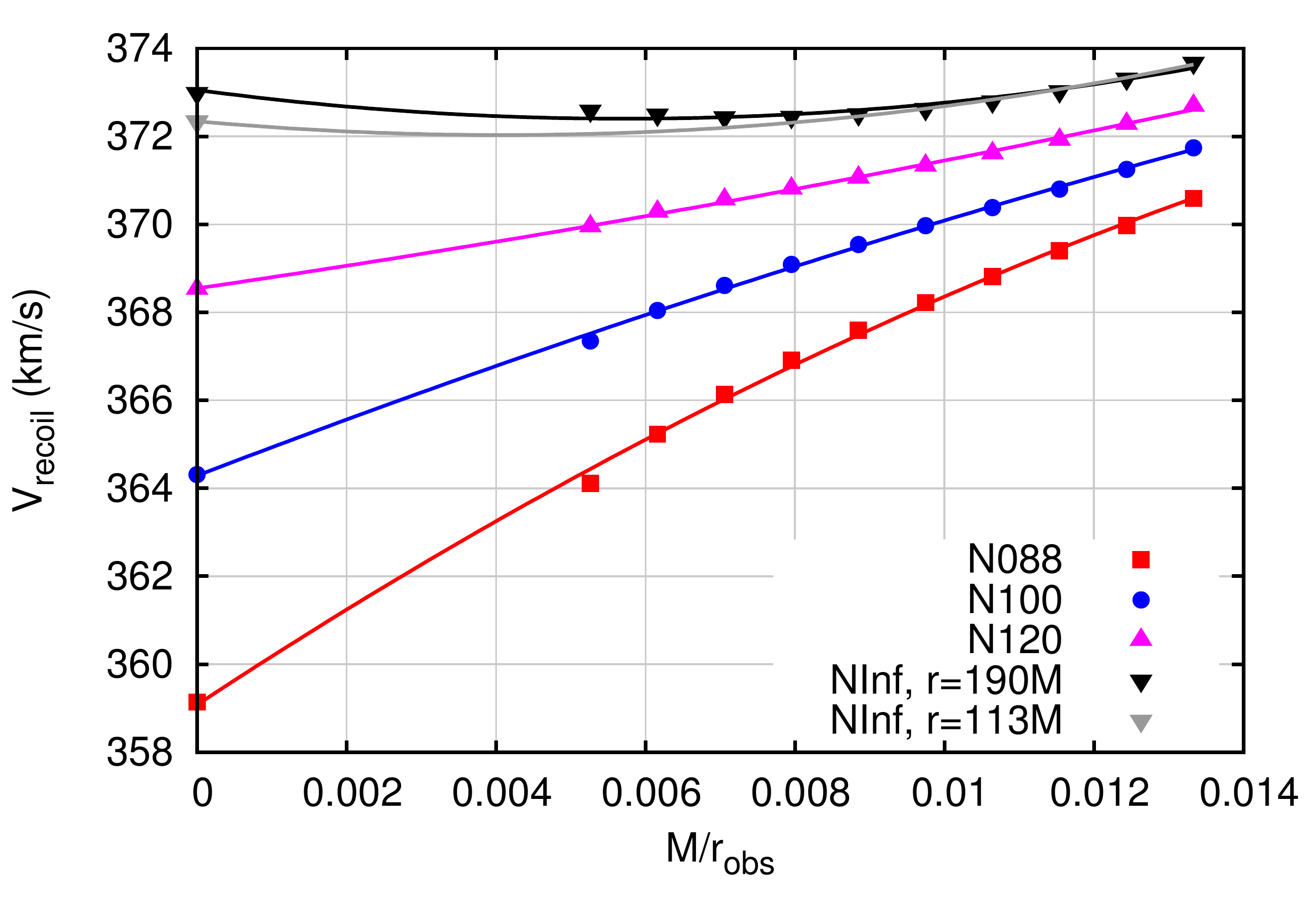}
\includegraphics[angle=270,width=0.49\columnwidth]{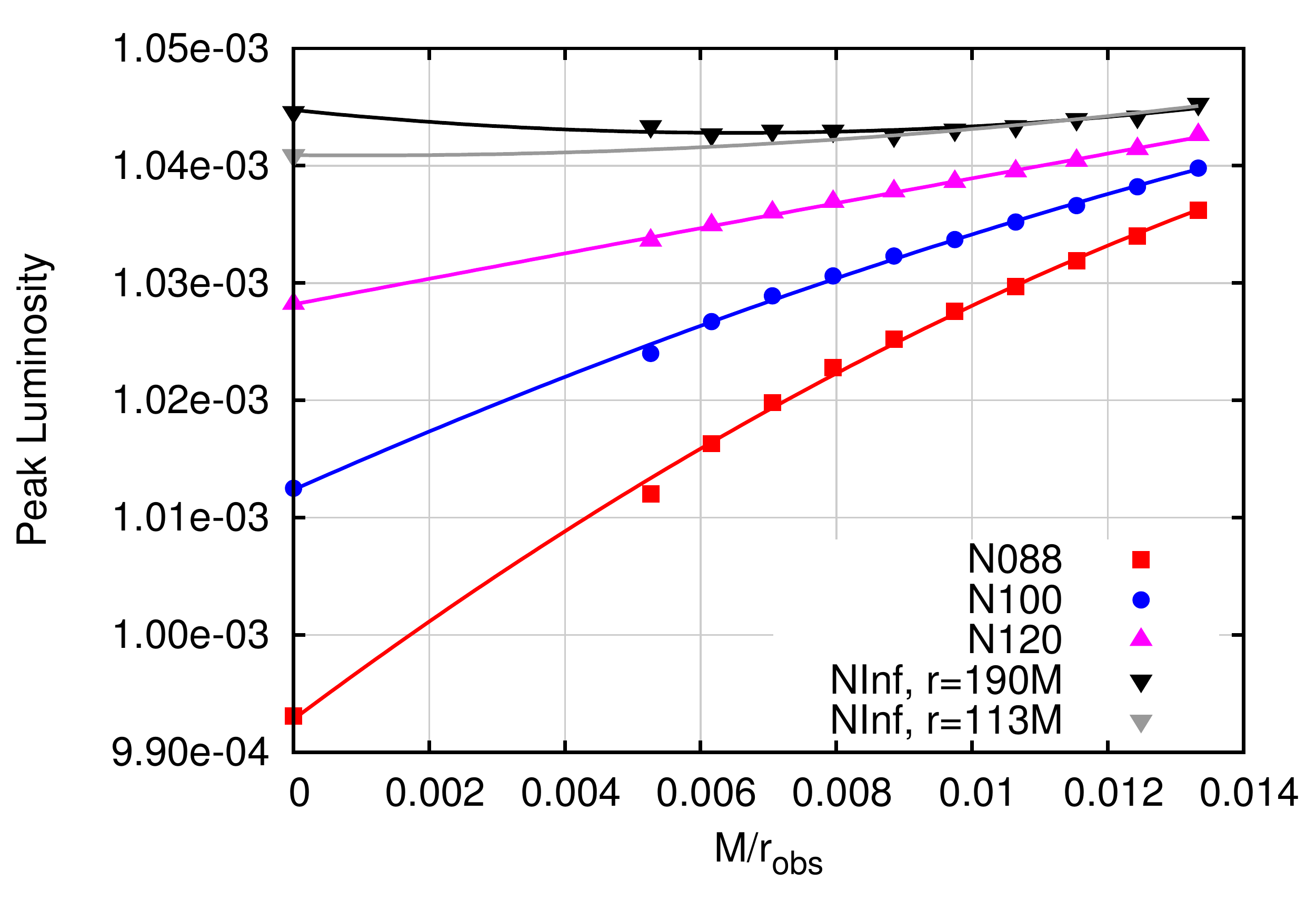}\\
\includegraphics[angle=270,width=0.49\columnwidth]{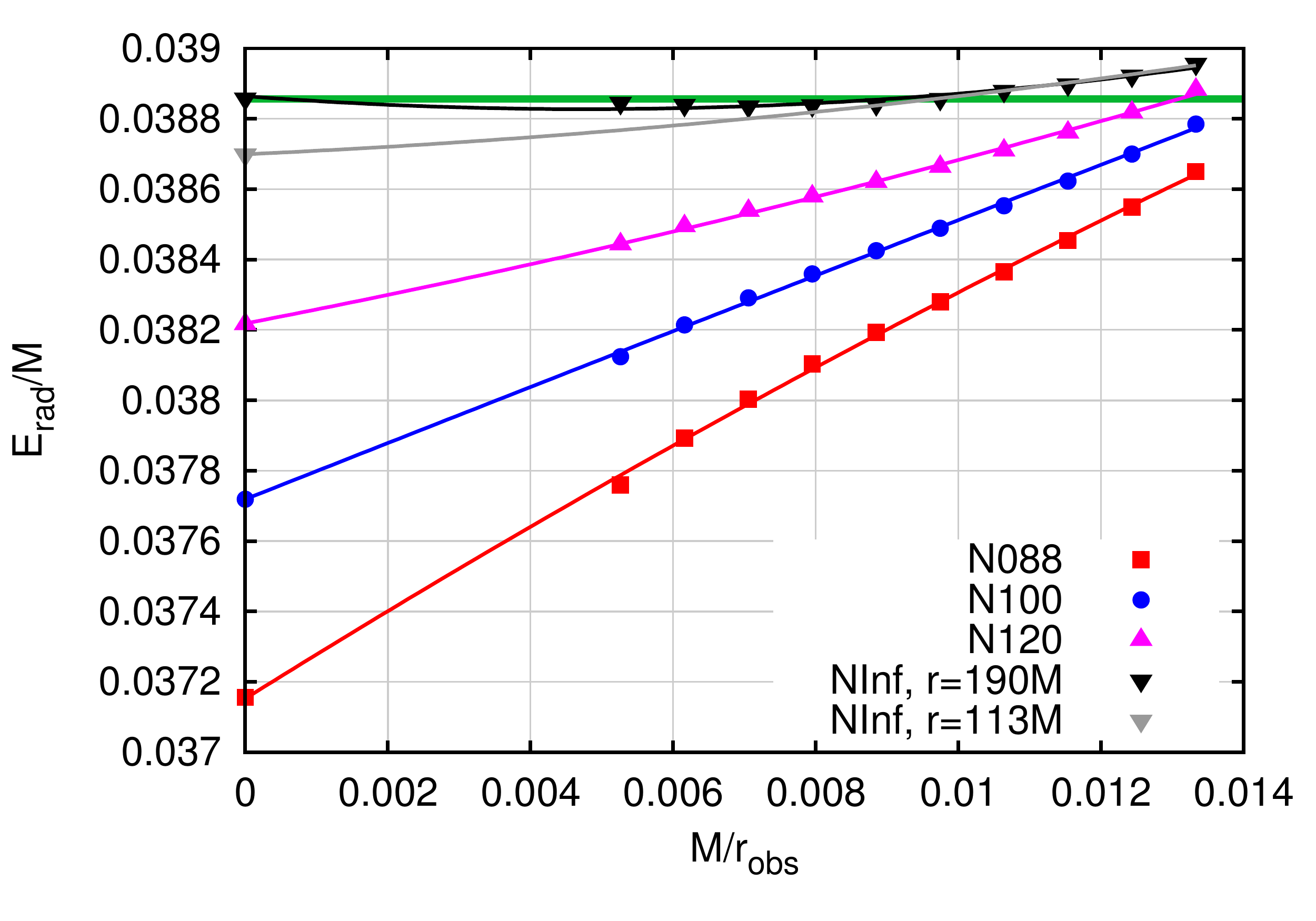}
\includegraphics[angle=270,width=0.49\columnwidth]{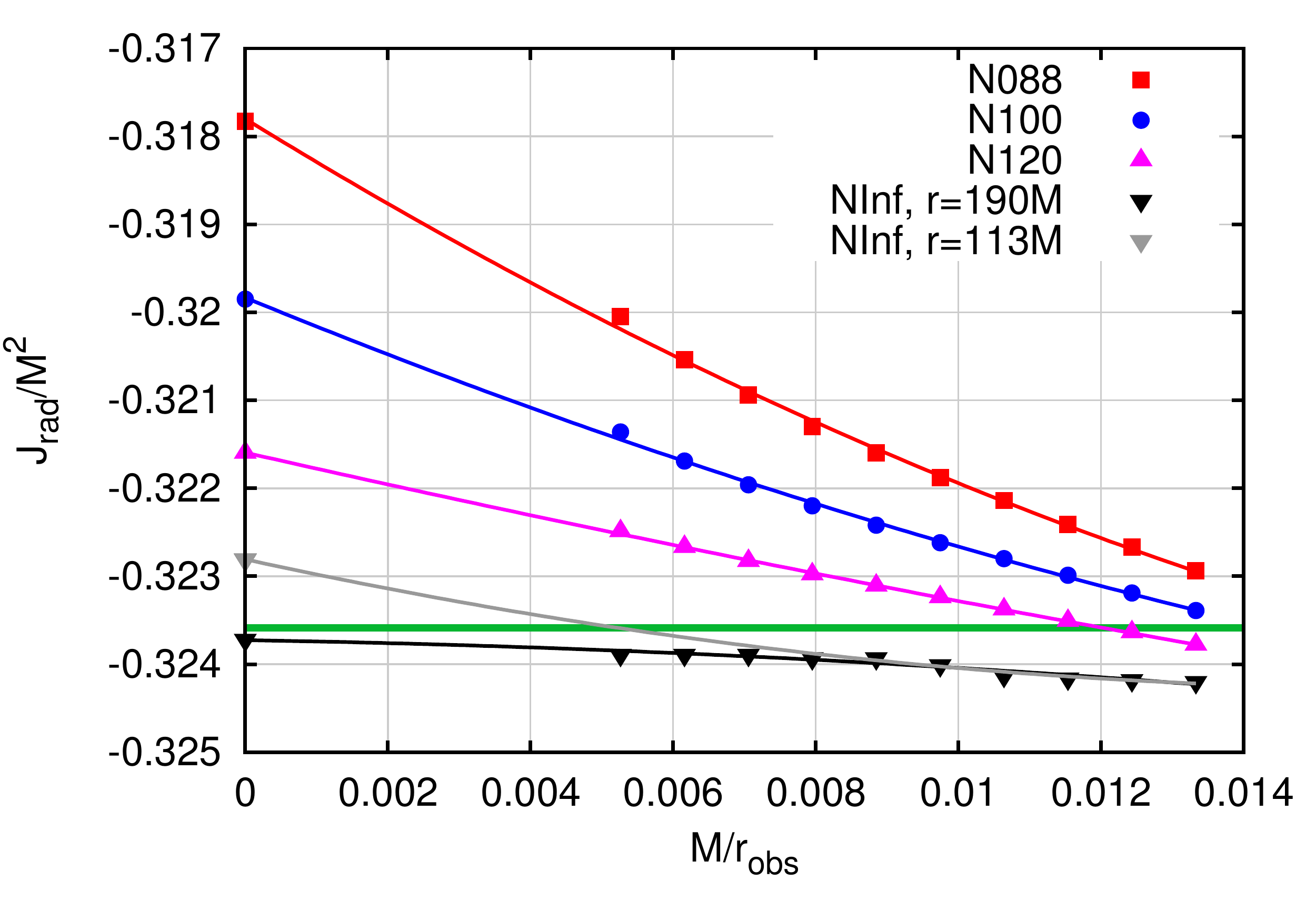}
\caption{ Plots of the convergence of the recoil velocity (top left), 
peak luminosity (top right), energy radiated (bottom left), and angular
momentum radiated (bottom right) as a function of $m/r_{obs}$ 
for case 80 - Q1.0000\_-0.8000\_0.8000. 
Horizontal green solid lines in the bottom row indicate the energy
and angular momentum radiated calculated from the isolated horizon.
The dark gray lines in each plot shows the extrapolation to infinite
observer location using only up to $r=113m$.  }
\label{fig:q100conv}
\end{figure}

\begin{figure}
\includegraphics[angle=270,width=0.49\columnwidth]{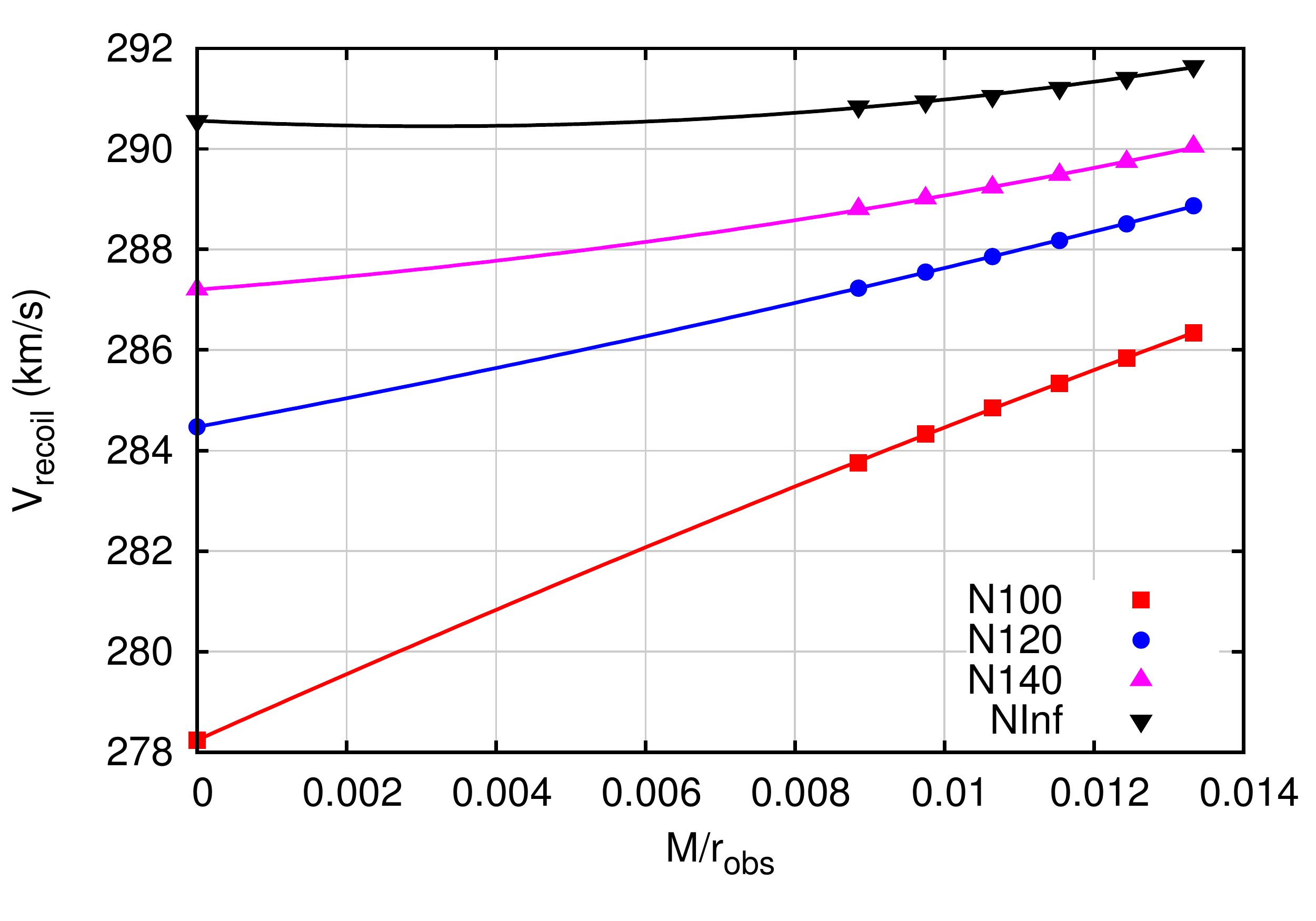}
\includegraphics[angle=270,width=0.49\columnwidth]{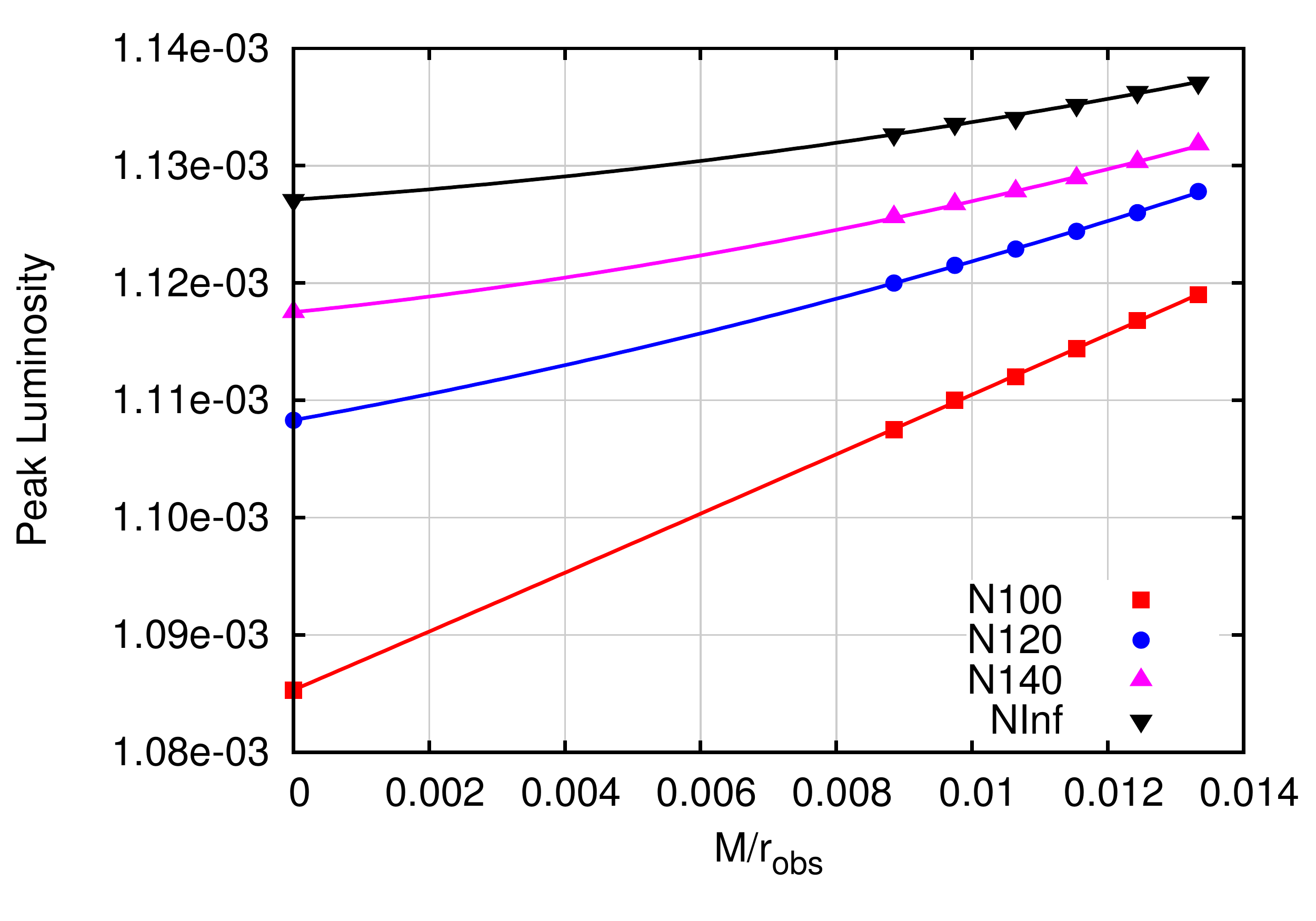}\\
\includegraphics[angle=270,width=0.49\columnwidth]{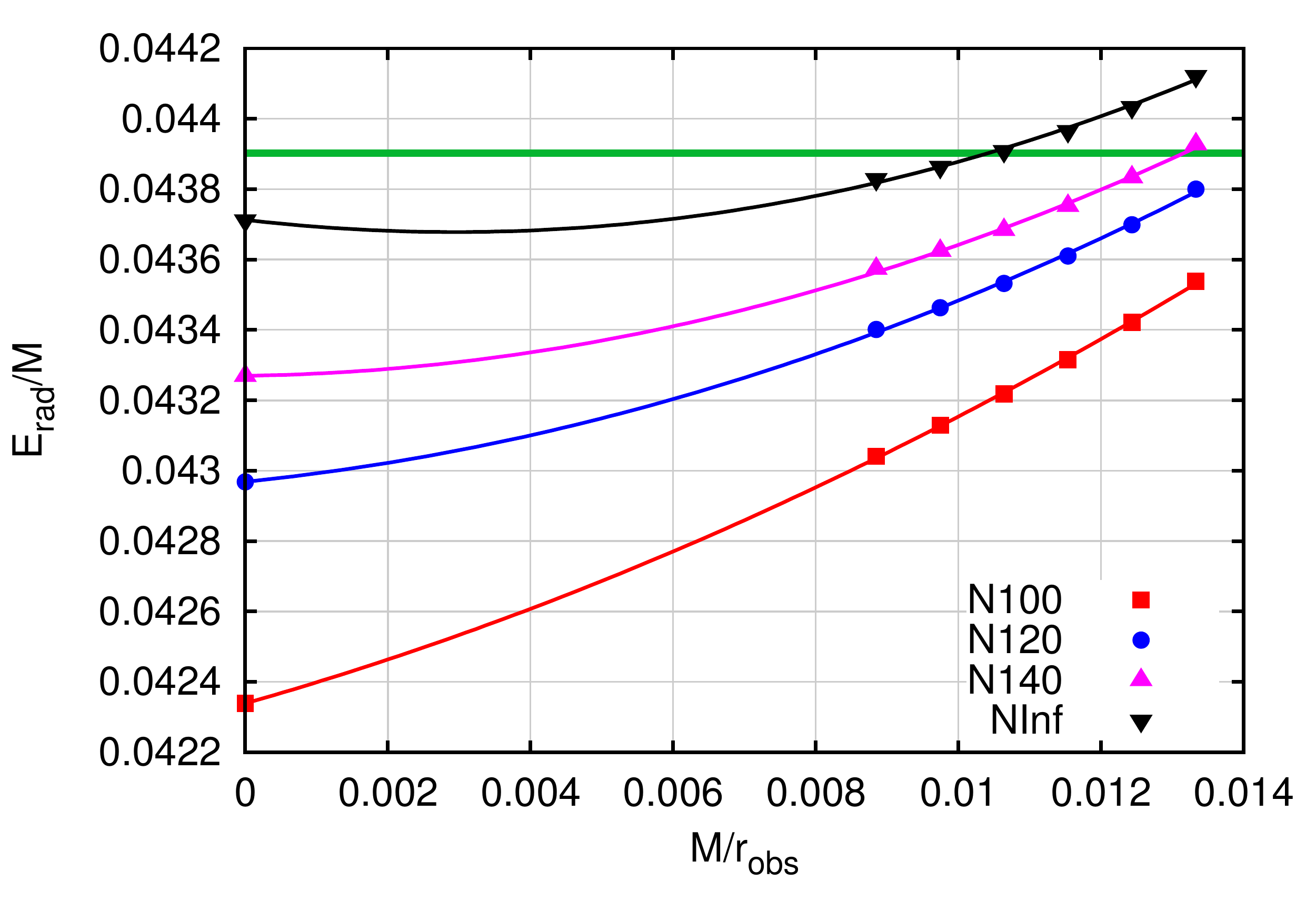}
\includegraphics[angle=270,width=0.49\columnwidth]{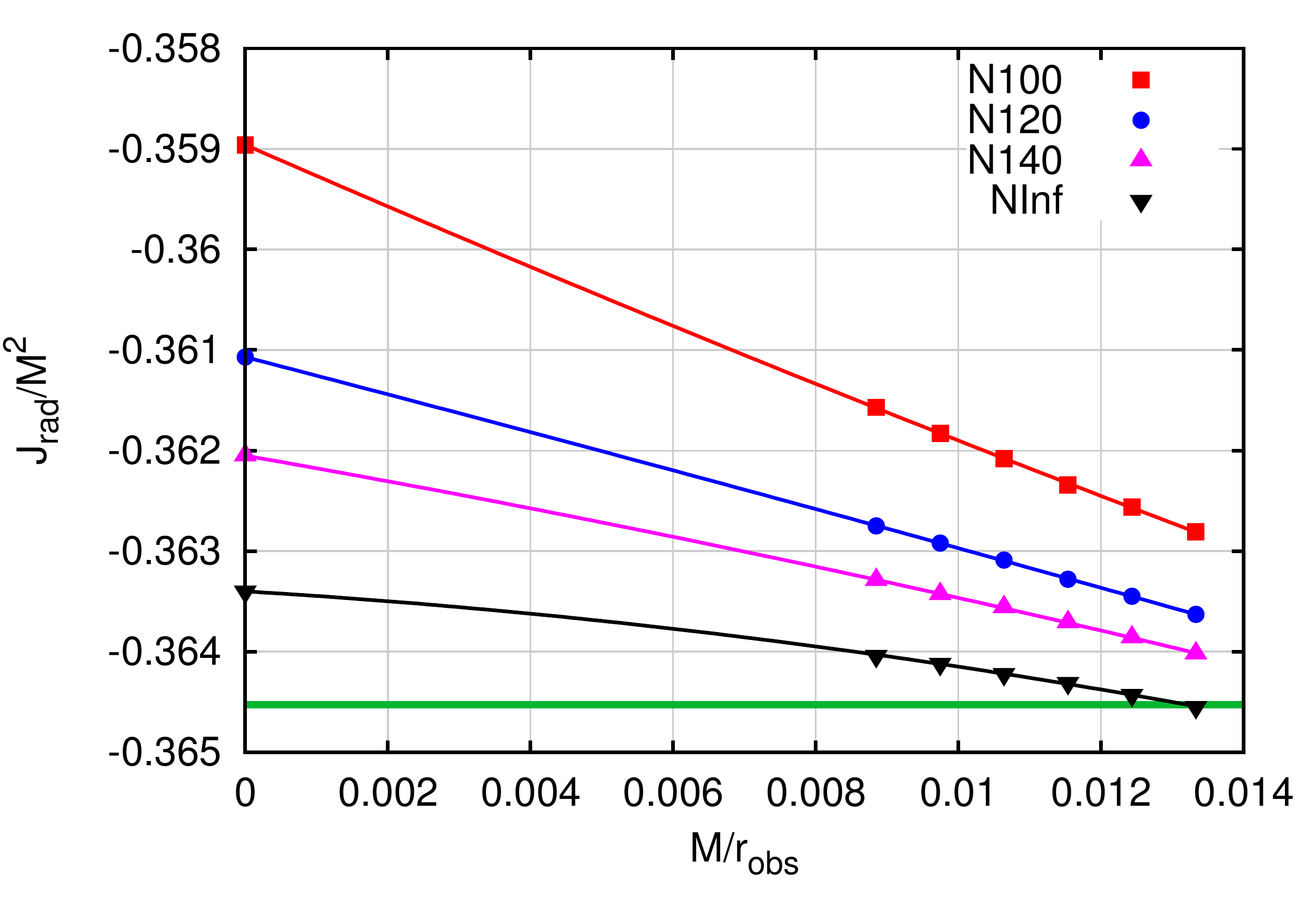}
\caption{Plots of the convergence of the recoil velocity (top left),
peak luminosity (top right), energy radiated (bottom left), and angular
momentum radiated (bottom right) as a function of $m/r_{obs}$ 
for case 47 - Q0.7500\_-0.8000\_0.8000.
Horizontal green solid lines in the bottom row indicate the energy
and angular momentum radiated calculated from the isolated horizon.  }
\label{fig:q075conv}
\end{figure}

\begin{figure}
\includegraphics[angle=270,width=0.49\columnwidth]{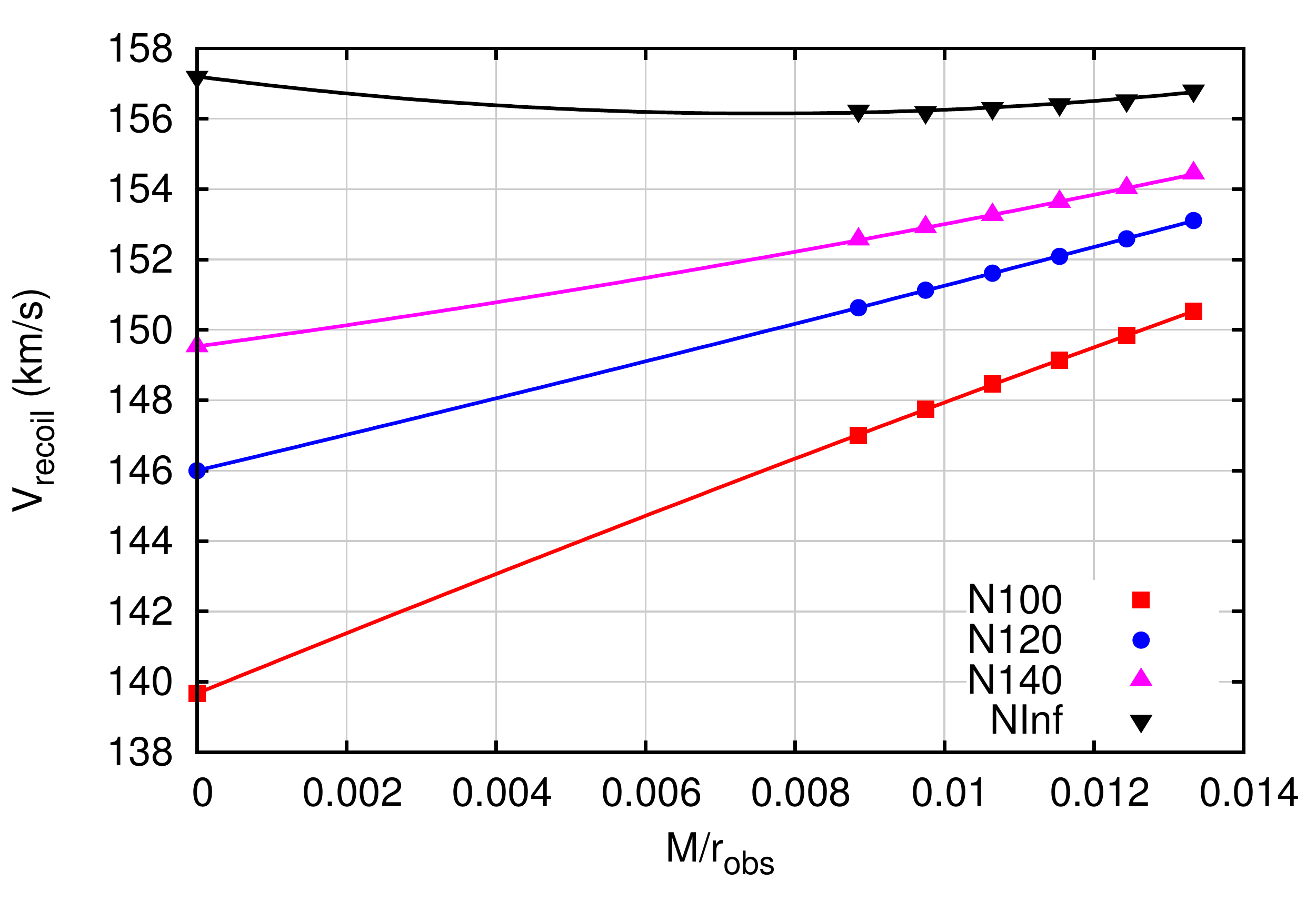}
\includegraphics[angle=270,width=0.49\columnwidth]{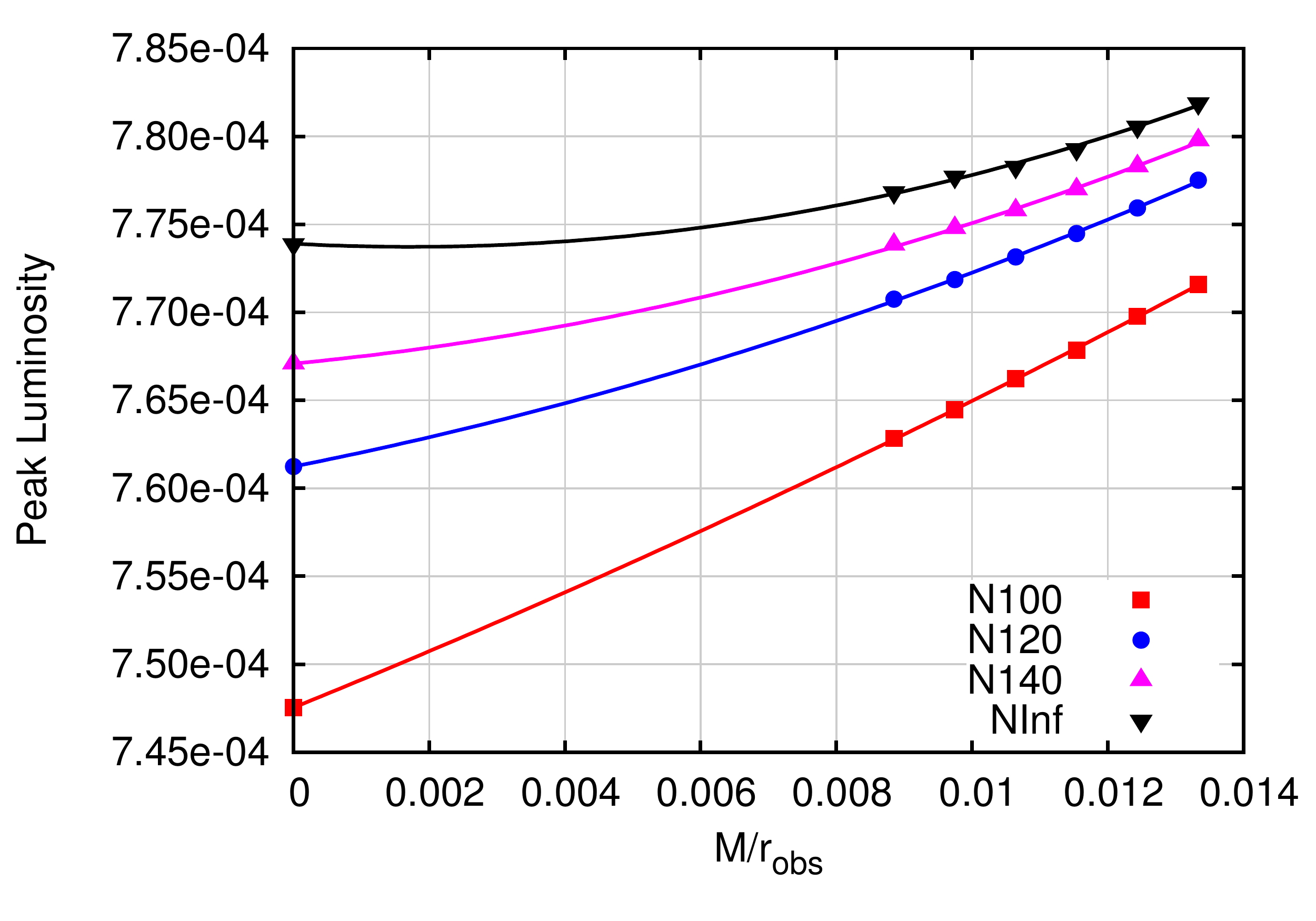}\\
\includegraphics[angle=270,width=0.49\columnwidth]{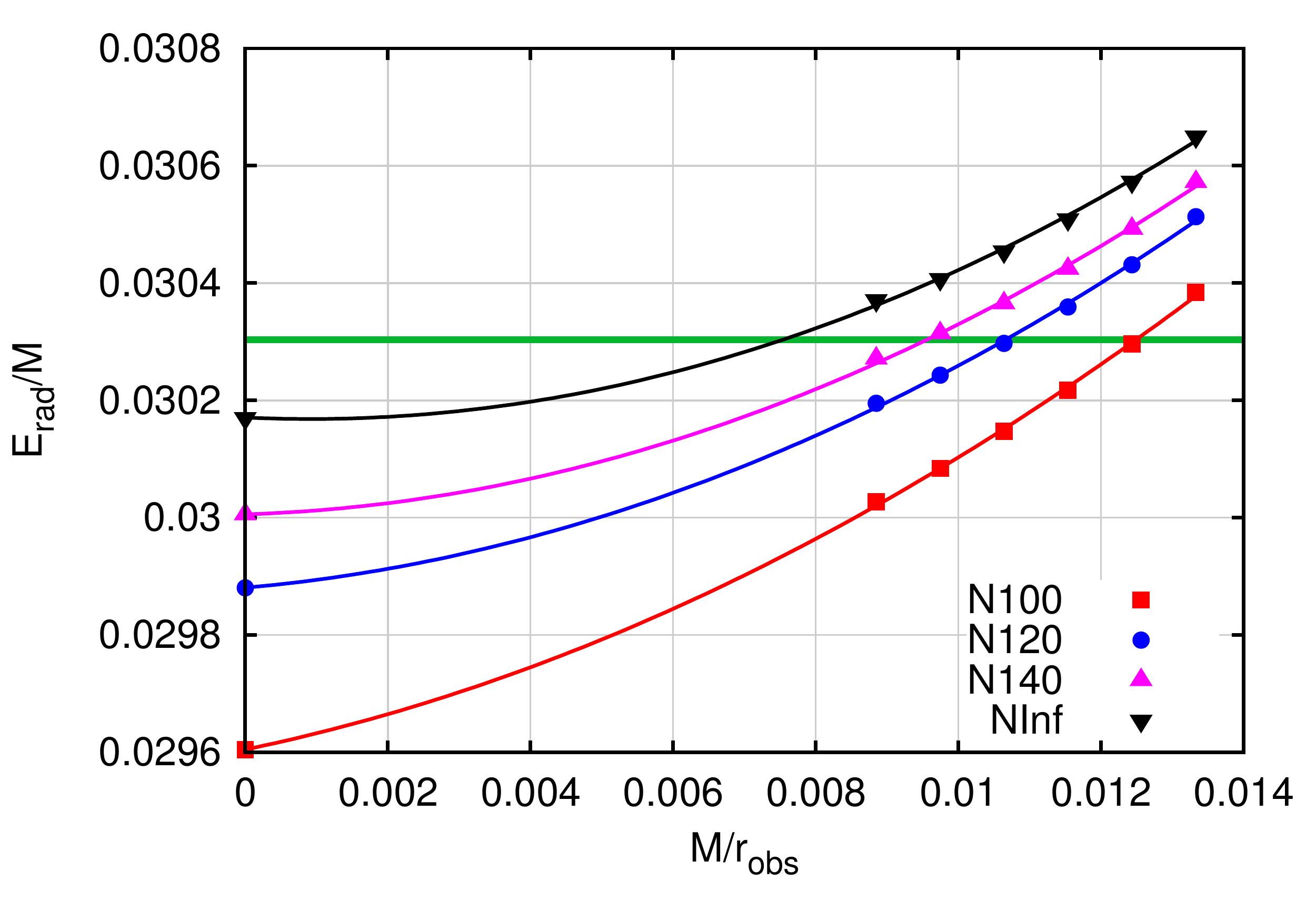}
\includegraphics[angle=270,width=0.49\columnwidth]{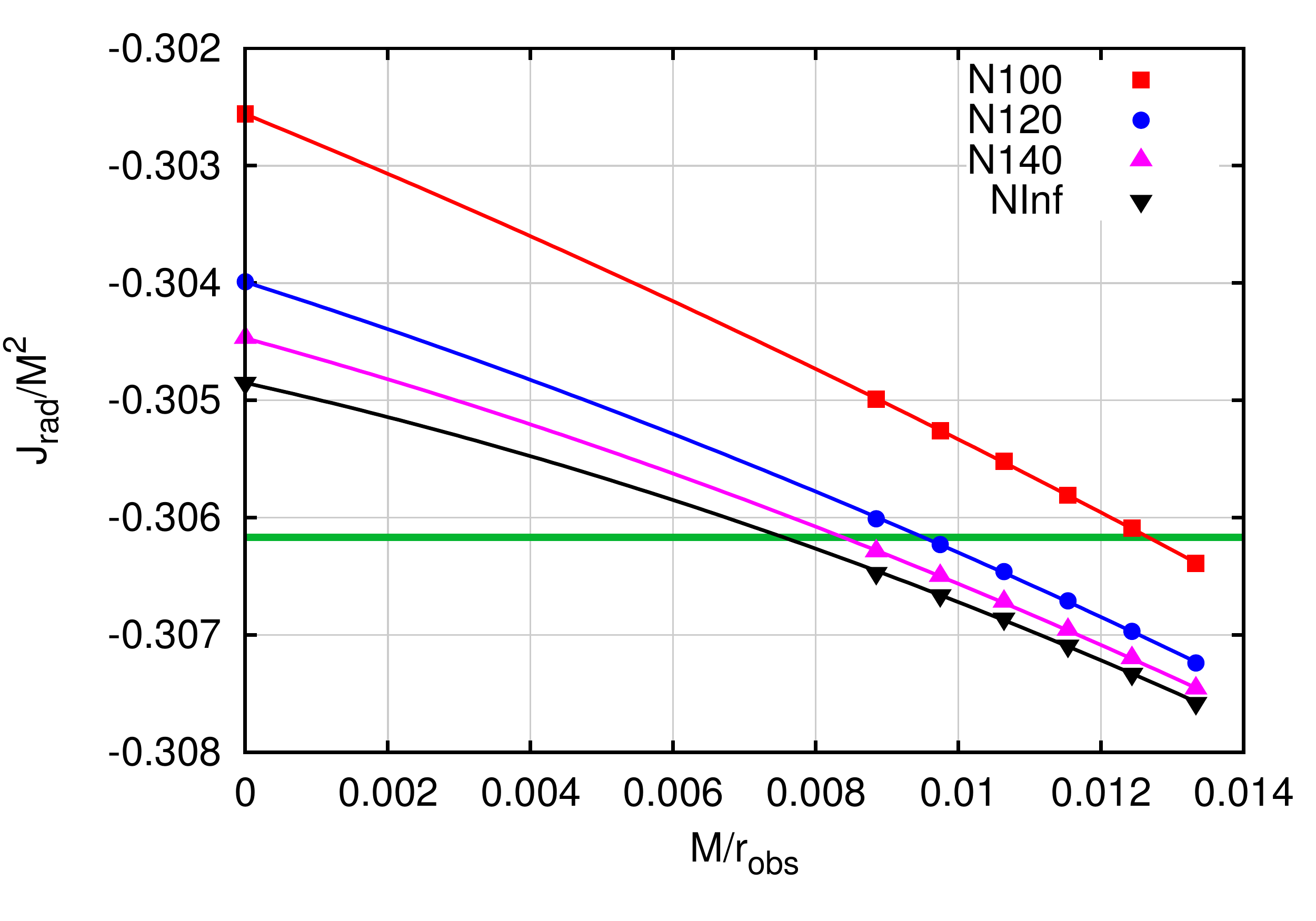}
\caption{ Plots of the convergence of the recoil velocity (top left),
peak luminosity (top right), energy radiated (bottom left), and angular
momentum radiated (bottom right) as a function of $m/r_{obs}$
for case 67 - Q0.5000\_0.0000\_0.0000.
Horizontal green solid lines in the bottom row indicate the energy
and angular momentum radiated calculated from the isolated horizon.  }
\label{fig:q050conv}
\end{figure}

\begin{figure}
\includegraphics[angle=270,width=0.49\columnwidth]{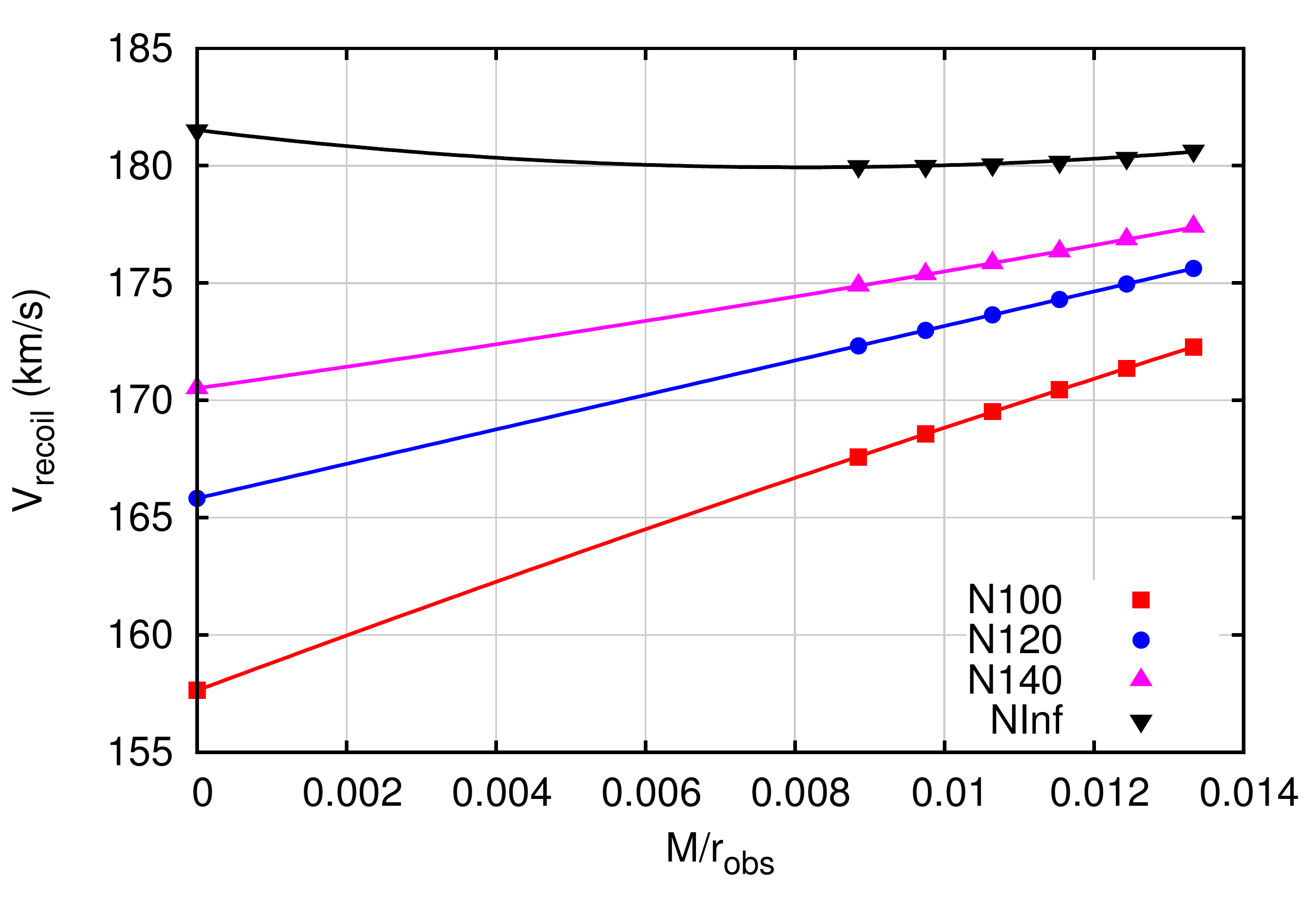}
\includegraphics[angle=270,width=0.49\columnwidth]{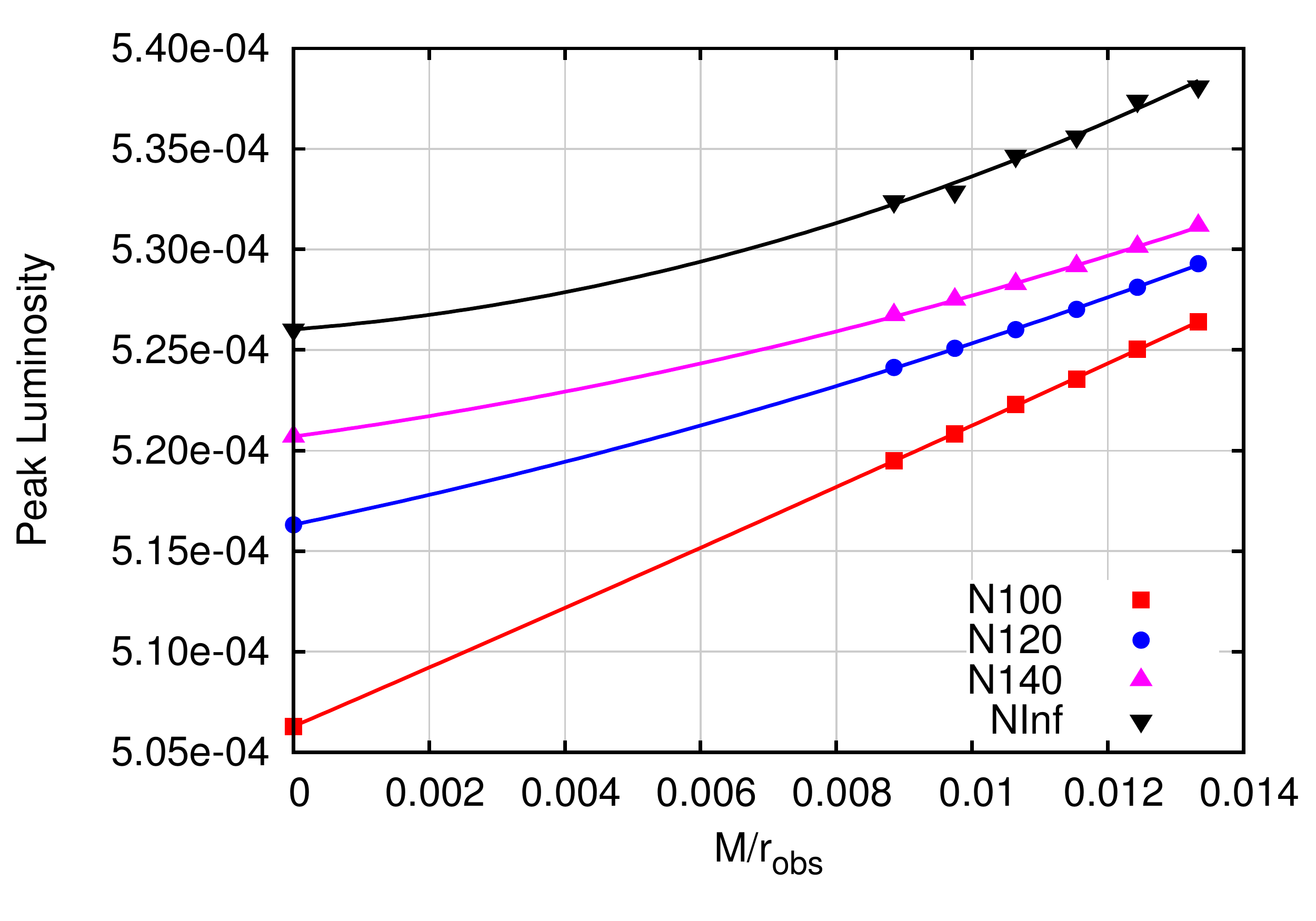}\\
\includegraphics[angle=270,width=0.49\columnwidth]{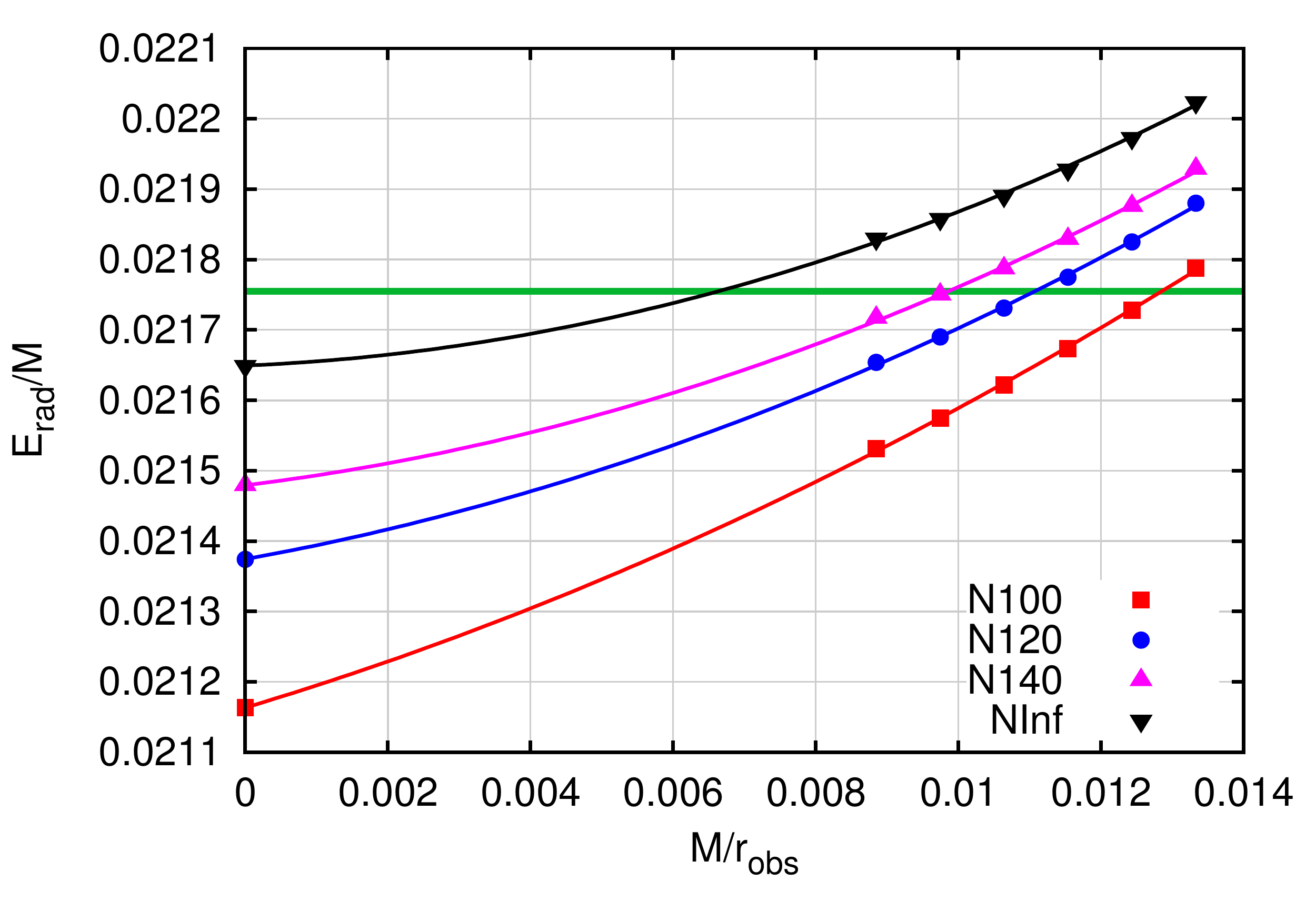}
\includegraphics[angle=270,width=0.49\columnwidth]{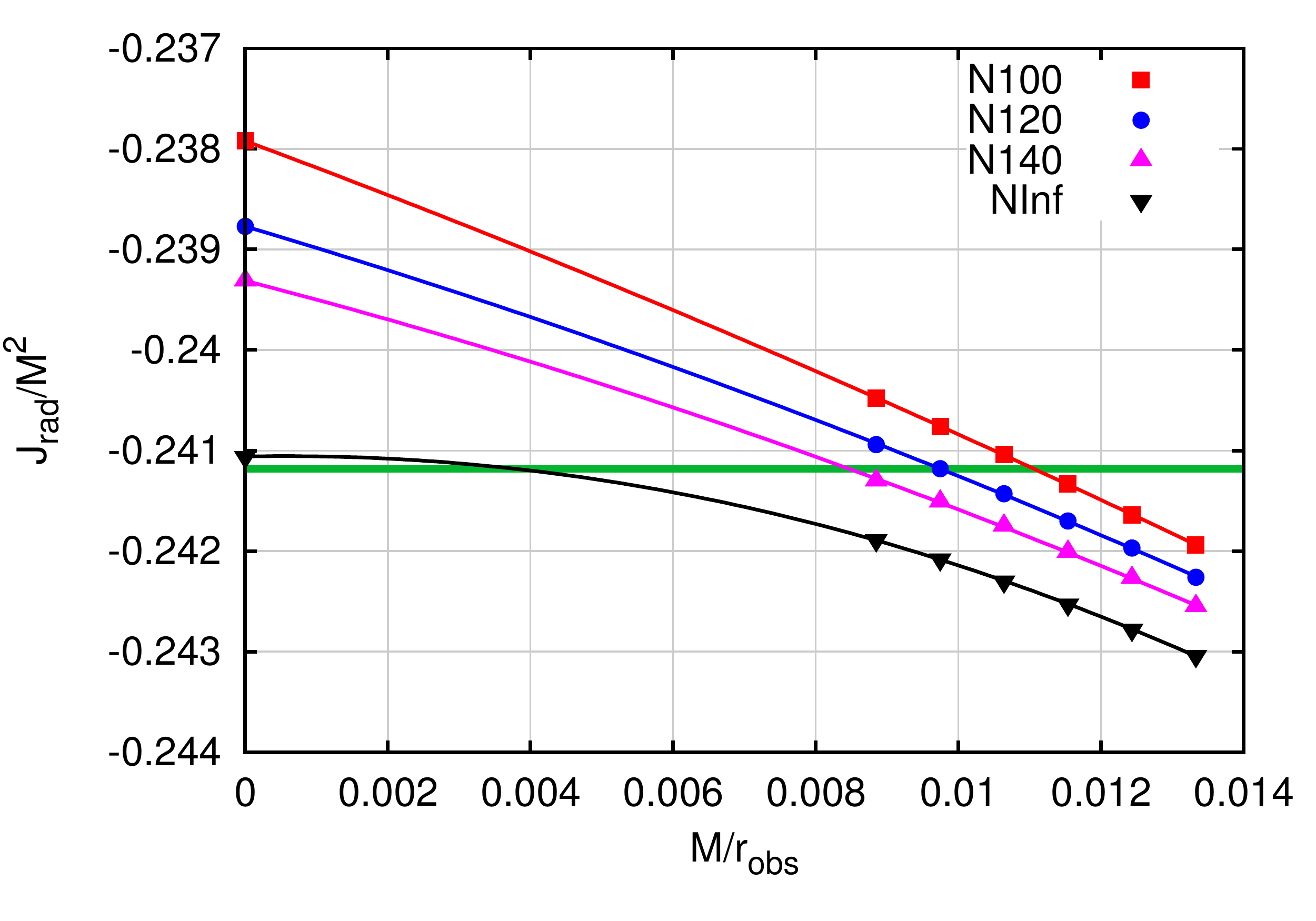}
\caption{ Plots of the convergence of the recoil velocity (top left),
peak luminosity (top right), energy radiated (bottom left), and angular
momentum radiated (bottom right) as a function of $m/r_{obs}$
for case 65 - Q0.3333\_0.0000\_0.0000.
Horizontal green solid lines in the bottom row indicate the energy
and angular momentum radiated calculated from the isolated horizon.  
}
\label{fig:q033conv}
\end{figure}

In Figs. \ref{fig:q100conv}-\ref{fig:q033conv}, we show the recoil velocity,
peak luminosity, radiated energy, and radiated angular momentum for each 
resolution as a function of the inverse of the observer location.  For the 
radiated energy and angular momentum, the value calculated on the isolated 
horizon (at the highest resolution available)
is shown as a green solid line.  In each, the solid lines indicate
a second order polynomial fit in $m/r_{obs}$ to extrapolate to infinite
observer location.  In all cases, we have six extraction radii between
$75m$ and $113m$.  For the equal-mass case in Fig. \ref{fig:q100conv}, we have 
an additional four extraction radii, out to $190m$.  In this figure, we show two
extrapolations to infinite observer, one in black using all available extraction
radii, and one in gray using only up to $r_{obs}=113m$ as in the other figures.
The difference between the two extrapolations is on the order of $1\%$ for
all four quantities.

\begin{table*}
\caption{Convergence of the recoil velocity $V$ (in km/s) and peak luminosity
$L$ for each observer location $r_{obs}/m$
for case 80 - Q1.0000\_-0.8000\_0.8000.
Subscripts on the recoil and luminosity indicate low, medium, high,
and extrapolated to infinite spatial resolution.  For this case, the
three resolutions are N088, N100, and N120.  The order of convergence
is given for both the recoil, $d_V$, and the luminosity $d_L$.
}
\label{tab:q100conv}
\begin{ruledtabular}
\begin{tabular}{ccccccccccc}
$r_{obs}/m$ & $V_{low}$ & $V_{med}$ & $V_{high}$ & $V_{\infty}$ & $d_V$ & $L_{low}$ & $L_{med}$ & $L_{high}$ & $L_{\infty}$ & $d_L$ \\
\hline
75.00 & 370.59 & 371.74 & 372.69 & 373.68 & 3.66 & 1.0362e-03 & 1.0398e-03 & 1.0426e-03 & 1.0453e-03 & 3.97 \\
80.41 & 369.98 & 371.25 & 372.28 & 373.32 & 3.76 & 1.0340e-03 & 1.0382e-03 & 1.0414e-03 & 1.0442e-03 & 4.18 \\
86.66 & 369.40 & 370.80 & 371.92 & 373.03 & 3.83 & 1.0319e-03 & 1.0366e-03 & 1.0404e-03 & 1.0440e-03 & 3.88 \\
93.96 & 368.81 & 370.38 & 371.61 & 372.80 & 3.90 & 1.0297e-03 & 1.0352e-03 & 1.0395e-03 & 1.0434e-03 & 4.00 \\
102.60 & 368.22 & 369.97 & 371.33 & 372.63 & 3.94 & 1.0276e-03 & 1.0337e-03 & 1.0386e-03 & 1.0431e-03 & 3.94 \\
113.00 & 367.59 & 369.54 & 371.06 & 372.51 & 3.95 & 1.0252e-03 & 1.0323e-03 & 1.0378e-03 & 1.0426e-03 & 4.13 \\
125.74 & 366.91 & 369.09 & 370.81 & 372.45 & 3.92 & 1.0228e-03 & 1.0306e-03 & 1.0369e-03 & 1.0430e-03 & 3.84 \\
141.71 & 366.13 & 368.61 & 370.56 & 372.44 & 3.89 & 1.0198e-03 & 1.0289e-03 & 1.0360e-03 & 1.0430e-03 & 3.85 \\
162.34 & 365.23 & 368.04 & 370.28 & 372.50 & 3.82 & 1.0163e-03 & 1.0267e-03 & 1.0349e-03 & 1.0427e-03 & 3.93 \\
190.00 & 364.11 & 367.35 & 369.95 & 372.59 & 3.76 & 1.0120e-03 & 1.0240e-03 & 1.0336e-03 & 1.0434e-03 & 3.75 \\
$\infty$ & 359.14 & 364.31 & 368.53 & 372.99 & 3.65 & 9.9307e-04 & 1.0125e-03 & 1.0282e-03 & 1.0446e-03 & 3.69 \\
\end{tabular}
\end{ruledtabular}
\end{table*}

\begin{table*}
\caption{Convergence of the recoil velocity $V$ (in km/s) and peak luminosity
$L$ for each observer location $r_{obs}/m$
for case 47 - Q0.7500\_-0.8000\_0.8000.  
Subscripts on the recoil and luminosity indicate low, medium, high, 
and extrapolated to infinite spatial resolution.  For this case, the 
three resolutions are N100, N120, and N140.  The order of convergence
is given for both the recoil, $d_V$, and the luminosity $d_L$.  
}
\label{tab:q075conv}
\begin{ruledtabular}
\begin{tabular}{ccccccccccc}
$r_{obs}/m$ & $V_{low}$ & $V_{med}$ & $V_{high}$ & $V_{\infty}$ & $d_V$ & $L_{low}$ & $L_{med}$ & $L_{high}$ & $L_{\infty}$ & $d_L$ \\
\hline
75.00 & 286.34 & 288.87 & 290.04 & 291.65 & 3.55 & 1.1190e-03 & 1.1278e-03 & 1.1318e-03 & 1.1371e-03 & 3.63 \\
80.41 & 285.84 & 288.51 & 289.74 & 291.42 & 3.57 & 1.1168e-03 & 1.1260e-03 & 1.1303e-03 & 1.1363e-03 & 3.50 \\
86.66 & 285.34 & 288.18 & 289.48 & 291.22 & 3.61 & 1.1144e-03 & 1.1244e-03 & 1.1289e-03 & 1.1352e-03 & 3.57 \\
93.96 & 284.85 & 287.86 & 289.23 & 291.06 & 3.64 & 1.1120e-03 & 1.1229e-03 & 1.1278e-03 & 1.1341e-03 & 3.73 \\
102.60 & 284.33 & 287.55 & 289.01 & 290.95 & 3.65 & 1.1100e-03 & 1.1215e-03 & 1.1267e-03 & 1.1336e-03 & 3.65 \\
113.00 & 283.76 & 287.23 & 288.80 & 290.85 & 3.69 & 1.1075e-03 & 1.1200e-03 & 1.1256e-03 & 1.1327e-03 & 3.78 \\
$\infty$ & 278.24 & 284.47 & 287.20 & 290.56 & 3.86 & 1.0853e-03 & 1.1083e-03 & 1.1175e-03 & 1.1271e-03 & 4.39 \\
\end{tabular}
\end{ruledtabular}
\end{table*}

\begin{table*}
\caption{Convergence of the recoil velocity $V$ (in km/s) and peak luminosity
$L$ for each observer location $r_{obs}/m$
for case 67 - Q0.5000\_0.0000\_0.0000.
Subscripts on the recoil and luminosity indicate low, medium, high,
and extrapolated to infinite spatial resolution.  For this case, the
three resolutions are N100, N120, and N140.  The order of convergence
is given for both the recoil, $d_V$, and the luminosity $d_L$.
}
\label{tab:q050conv}
\begin{ruledtabular}
\begin{tabular}{ccccccccccc}
$r_{obs}/m$ & $V_{low}$ & $V_{med}$ & $V_{high}$ & $V_{\infty}$ & $d_V$ & $L_{low}$ & $L_{med}$ & $L_{high}$ & $L_{\infty}$ & $d_L$ \\
\hline
75.00 & 150.53 & 153.11 & 154.44 & 156.81 & 2.90 & 7.7159e-04 & 7.7751e-04 & 7.7976e-04 & 7.8189e-04 & 4.68 \\
80.41 & 149.84 & 152.59 & 154.02 & 156.53 & 2.91 & 7.6977e-04 & 7.7593e-04 & 7.7829e-04 & 7.8056e-04 & 4.63 \\
86.66 & 149.14 & 152.09 & 153.63 & 156.42 & 2.84 & 7.6785e-04 & 7.7448e-04 & 7.7698e-04 & 7.7929e-04 & 4.75 \\
93.96 & 148.46 & 151.61 & 153.26 & 156.31 & 2.81 & 7.6622e-04 & 7.7315e-04 & 7.7579e-04 & 7.7828e-04 & 4.69 \\
102.60 & 147.75 & 151.13 & 152.91 & 156.20 & 2.80 & 7.6446e-04 & 7.7186e-04 & 7.7478e-04 & 7.7772e-04 & 4.48 \\
113.00 & 147.00 & 150.63 & 152.56 & 156.24 & 2.74 & 7.6283e-04 & 7.7075e-04 & 7.7383e-04 & 7.7684e-04 & 4.57 \\
$\infty$ & 139.67 & 146.00 & 149.53 & 157.20 & 2.45 & 7.4753e-04 & 7.6123e-04 & 7.6708e-04 & 7.7389e-04 & 4.02 \\
\end{tabular}
\end{ruledtabular}
\end{table*}

\begin{table*}
\caption{Convergence of the recoil velocity $V$ (in km/s) and peak luminosity
$L$ for each observer location $r_{obs}/m$
for case 65 - Q0.3333\_0.0000\_0.0000.
Subscripts on the recoil and luminosity indicate low, medium, high,
and extrapolated to infinite spatial resolution.  For this case, the
three resolutions are N100, N120, and N140.  The order of convergence
is given for both the recoil, $d_V$, and the luminosity $d_L$.
}
\label{tab:q033conv}
\begin{ruledtabular}
\begin{tabular}{ccccccccccc}
$r_{obs}/m$ & $V_{low}$ & $V_{med}$ & $V_{high}$ & $V_{\infty}$ & $d_V$ & $L_{low}$ & $L_{med}$ & $L_{high}$ & $L_{\infty}$ & $d_L$ \\
\hline
75.00 & 172.27 & 175.62 & 177.39 & 180.65 & 2.80 & 5.2641e-04 & 5.2929e-04 & 5.3117e-04 & 5.3812e-04 & 1.55 \\
80.41 & 171.36 & 174.96 & 176.85 & 180.34 & 2.81 & 5.2503e-04 & 5.2812e-04 & 5.3012e-04 & 5.3741e-04 & 1.58 \\
86.66 & 170.45 & 174.30 & 176.34 & 180.18 & 2.76 & 5.2355e-04 & 5.2702e-04 & 5.2916e-04 & 5.3563e-04 & 1.86 \\
93.96 & 169.52 & 173.64 & 175.84 & 180.06 & 2.72 & 5.2229e-04 & 5.2601e-04 & 5.2827e-04 & 5.3467e-04 & 1.96 \\
102.60 & 168.57 & 172.98 & 175.36 & 180.00 & 2.68 & 5.2082e-04 & 5.2509e-04 & 5.2749e-04 & 5.3289e-04 & 2.39 \\
113.00 & 167.58 & 172.32 & 174.88 & 179.99 & 2.64 & 5.1950e-04 & 5.2413e-04 & 5.2672e-04 & 5.3241e-04 & 2.43 \\
$\infty$ & 157.63 & 165.82 & 170.51 & 181.52 & 2.30 & 5.0628e-04 & 5.1631e-04 & 5.2069e-04 & 5.2603e-04 & 3.89 \\
\end{tabular}
\end{ruledtabular}
\end{table*}


\bibliographystyle{apsrev4-1}
\bibliography{../../../Bibtex/references}

\begin{thebibliography}{64}%
\makeatletter
\providecommand \@ifxundefined [1]{%
 \@ifx{#1\undefined}
}%
\providecommand \@ifnum [1]{%
 \ifnum #1\expandafter \@firstoftwo
 \else \expandafter \@secondoftwo
 \fi
}%
\providecommand \@ifx [1]{%
 \ifx #1\expandafter \@firstoftwo
 \else \expandafter \@secondoftwo
 \fi
}%
\providecommand \natexlab [1]{#1}%
\providecommand \enquote  [1]{``#1''}%
\providecommand \bibnamefont  [1]{#1}%
\providecommand \bibfnamefont [1]{#1}%
\providecommand \citenamefont [1]{#1}%
\providecommand \href@noop [0]{\@secondoftwo}%
\providecommand \href [0]{\begingroup \@sanitize@url \@href}%
\providecommand \@href[1]{\@@startlink{#1}\@@href}%
\providecommand \@@href[1]{\endgroup#1\@@endlink}%
\providecommand \@sanitize@url [0]{\catcode `\\12\catcode `\$12\catcode
  `\&12\catcode `\#12\catcode `\^12\catcode `\_12\catcode `\%12\relax}%
\providecommand \@@startlink[1]{}%
\providecommand \@@endlink[0]{}%
\providecommand \url  [0]{\begingroup\@sanitize@url \@url }%
\providecommand \@url [1]{\endgroup\@href {#1}{\urlprefix }}%
\providecommand \urlprefix  [0]{URL }%
\providecommand \Eprint [0]{\href }%
\providecommand \doibase [0]{http://dx.doi.org/}%
\providecommand \selectlanguage [0]{\@gobble}%
\providecommand \bibinfo  [0]{\@secondoftwo}%
\providecommand \bibfield  [0]{\@secondoftwo}%
\providecommand \translation [1]{[#1]}%
\providecommand \BibitemOpen [0]{}%
\providecommand \bibitemStop [0]{}%
\providecommand \bibitemNoStop [0]{.\EOS\space}%
\providecommand \EOS [0]{\spacefactor3000\relax}%
\providecommand \BibitemShut  [1]{\csname bibitem#1\endcsname}%
\let\auto@bib@innerbib\@empty
\bibitem [{\citenamefont {Pretorius}(2005)}]{Pretorius:2005gq}%
  \BibitemOpen
  \bibfield  {author} {\bibinfo {author} {\bibfnamefont {F.}~\bibnamefont
  {Pretorius}},\ }\href@noop {} {\bibfield  {journal} {\bibinfo  {journal}
  {Phys. Rev. Lett.}\ }\textbf {\bibinfo {volume} {95}},\ \bibinfo {pages}
  {121101} (\bibinfo {year} {2005})},\ \Eprint
  {http://arxiv.org/abs/gr-qc/0507014} {gr-qc/0507014} \BibitemShut {NoStop}%
\bibitem [{\citenamefont {Campanelli}\ \emph
  {et~al.}(2006{\natexlab{a}})\citenamefont {Campanelli}, \citenamefont
  {Lousto}, \citenamefont {Marronetti},\ and\ \citenamefont
  {Zlochower}}]{Campanelli:2005dd}%
  \BibitemOpen
  \bibfield  {author} {\bibinfo {author} {\bibfnamefont {M.}~\bibnamefont
  {Campanelli}}, \bibinfo {author} {\bibfnamefont {C.~O.}\ \bibnamefont
  {Lousto}}, \bibinfo {author} {\bibfnamefont {P.}~\bibnamefont {Marronetti}},
  \ and\ \bibinfo {author} {\bibfnamefont {Y.}~\bibnamefont {Zlochower}},\
  }\href@noop {} {\bibfield  {journal} {\bibinfo  {journal} {Phys. Rev. Lett.}\
  }\textbf {\bibinfo {volume} {96}},\ \bibinfo {pages} {111101} (\bibinfo
  {year} {2006}{\natexlab{a}})},\ \Eprint {http://arxiv.org/abs/gr-qc/0511048}
  {gr-qc/0511048} \BibitemShut {NoStop}%
\bibitem [{\citenamefont {Baker}\ \emph {et~al.}(2006)\citenamefont {Baker},
  \citenamefont {Centrella}, \citenamefont {Choi}, \citenamefont {Koppitz},\
  and\ \citenamefont {van Meter}}]{Baker:2005vv}%
  \BibitemOpen
  \bibfield  {author} {\bibinfo {author} {\bibfnamefont {J.~G.}\ \bibnamefont
  {Baker}}, \bibinfo {author} {\bibfnamefont {J.}~\bibnamefont {Centrella}},
  \bibinfo {author} {\bibfnamefont {D.-I.}\ \bibnamefont {Choi}}, \bibinfo
  {author} {\bibfnamefont {M.}~\bibnamefont {Koppitz}}, \ and\ \bibinfo
  {author} {\bibfnamefont {J.}~\bibnamefont {van Meter}},\ }\href@noop {}
  {\bibfield  {journal} {\bibinfo  {journal} {Phys. Rev. Lett.}\ }\textbf
  {\bibinfo {volume} {96}},\ \bibinfo {pages} {111102} (\bibinfo {year}
  {2006})},\ \Eprint {http://arxiv.org/abs/gr-qc/0511103} {gr-qc/0511103}
  \BibitemShut {NoStop}%
\bibitem [{\citenamefont {Campanelli}\ \emph
  {et~al.}(2009{\natexlab{a}})\citenamefont {Campanelli}, \citenamefont
  {Lousto}, \citenamefont {Nakano},\ and\ \citenamefont
  {Zlochower}}]{Campanelli:2008nk}%
  \BibitemOpen
  \bibfield  {author} {\bibinfo {author} {\bibfnamefont {M.}~\bibnamefont
  {Campanelli}}, \bibinfo {author} {\bibfnamefont {C.~O.}\ \bibnamefont
  {Lousto}}, \bibinfo {author} {\bibfnamefont {H.}~\bibnamefont {Nakano}}, \
  and\ \bibinfo {author} {\bibfnamefont {Y.}~\bibnamefont {Zlochower}},\ }\href
  {\doibase 10.1103/PhysRevD.79.084010} {\bibfield  {journal} {\bibinfo
  {journal} {Phys. Rev.}\ }\textbf {\bibinfo {volume} {D79}},\ \bibinfo {pages}
  {084010} (\bibinfo {year} {2009}{\natexlab{a}})},\ \Eprint
  {http://arxiv.org/abs/0808.0713} {arXiv:0808.0713 [gr-qc]} \BibitemShut
  {NoStop}%
\bibitem [{\citenamefont {Abbott}\ \emph
  {et~al.}(2016{\natexlab{a}})\citenamefont {Abbott} \emph
  {et~al.}}]{Abbott:2016blz}%
  \BibitemOpen
  \bibfield  {author} {\bibinfo {author} {\bibfnamefont {B.}~\bibnamefont
  {Abbott}} \emph {et~al.} (\bibinfo {collaboration} {Virgo, LIGO
  Scientific}),\ }\href {\doibase 10.1103/PhysRevLett.116.061102} {\bibfield
  {journal} {\bibinfo  {journal} {Phys. Rev. Lett.}\ }\textbf {\bibinfo
  {volume} {116}},\ \bibinfo {pages} {061102} (\bibinfo {year}
  {2016}{\natexlab{a}})},\ \Eprint {http://arxiv.org/abs/1602.03837}
  {arXiv:1602.03837 [gr-qc]} \BibitemShut {NoStop}%
\bibitem [{\citenamefont {Abbott}\ \emph
  {et~al.}(2016{\natexlab{b}})\citenamefont {Abbott} \emph
  {et~al.}}]{Abbott:2016nmj}%
  \BibitemOpen
  \bibfield  {author} {\bibinfo {author} {\bibfnamefont {B.~P.}\ \bibnamefont
  {Abbott}} \emph {et~al.} (\bibinfo {collaboration} {Virgo, LIGO
  Scientific}),\ }\href {\doibase 10.1103/PhysRevLett.116.241103} {\bibfield
  {journal} {\bibinfo  {journal} {Phys. Rev. Lett.}\ }\textbf {\bibinfo
  {volume} {116}},\ \bibinfo {pages} {241103} (\bibinfo {year}
  {2016}{\natexlab{b}})},\ \Eprint {http://arxiv.org/abs/1606.04855}
  {arXiv:1606.04855 [gr-qc]} \BibitemShut {NoStop}%
\bibitem [{\citenamefont {Abbott}\ \emph
  {et~al.}(2016{\natexlab{c}})\citenamefont {Abbott} \emph
  {et~al.}}]{TheLIGOScientific:2016pea}%
  \BibitemOpen
  \bibfield  {author} {\bibinfo {author} {\bibfnamefont {B.~P.}\ \bibnamefont
  {Abbott}} \emph {et~al.} (\bibinfo {collaboration} {Virgo, LIGO
  Scientific}),\ }\href {\doibase 10.1103/PhysRevX.6.041015} {\bibfield
  {journal} {\bibinfo  {journal} {Phys. Rev.}\ }\textbf {\bibinfo {volume}
  {X6}},\ \bibinfo {pages} {041015} (\bibinfo {year} {2016}{\natexlab{c}})},\
  \Eprint {http://arxiv.org/abs/1606.04856} {arXiv:1606.04856 [gr-qc]}
  \BibitemShut {NoStop}%
\bibitem [{\citenamefont {Abbott}\ \emph
  {et~al.}(2016{\natexlab{d}})\citenamefont {Abbott} \emph
  {et~al.}}]{Abbott:2016apu}%
  \BibitemOpen
  \bibfield  {author} {\bibinfo {author} {\bibfnamefont {B.~P.}\ \bibnamefont
  {Abbott}} \emph {et~al.} (\bibinfo {collaboration} {Virgo, LIGO
  Scientific}),\ }\href {\doibase 10.1103/PhysRevD.94.064035} {\bibfield
  {journal} {\bibinfo  {journal} {Phys. Rev.}\ }\textbf {\bibinfo {volume}
  {D94}},\ \bibinfo {pages} {064035} (\bibinfo {year} {2016}{\natexlab{d}})},\
  \Eprint {http://arxiv.org/abs/1606.01262} {arXiv:1606.01262 [gr-qc]}
  \BibitemShut {NoStop}%
\bibitem [{\citenamefont {Lovelace}\ \emph {et~al.}(2016)\citenamefont
  {Lovelace} \emph {et~al.}}]{Lovelace:2016uwp}%
  \BibitemOpen
  \bibfield  {author} {\bibinfo {author} {\bibfnamefont {G.}~\bibnamefont
  {Lovelace}} \emph {et~al.},\ }\href {\doibase 10.1088/0264-9381/33/24/244002}
  {\bibfield  {journal} {\bibinfo  {journal} {Class. Quant. Grav.}\ }\textbf
  {\bibinfo {volume} {33}},\ \bibinfo {pages} {244002} (\bibinfo {year}
  {2016})},\ \Eprint {http://arxiv.org/abs/1607.05377} {arXiv:1607.05377
  [gr-qc]} \BibitemShut {NoStop}%
\bibitem [{\citenamefont {Abbott}\ \emph
  {et~al.}(2016{\natexlab{e}})\citenamefont {Abbott} \emph
  {et~al.}}]{TheLIGOScientific:2016src}%
  \BibitemOpen
  \bibfield  {author} {\bibinfo {author} {\bibfnamefont {B.~P.}\ \bibnamefont
  {Abbott}} \emph {et~al.} (\bibinfo {collaboration} {Virgo, LIGO
  Scientific}),\ }\href {\doibase 10.1103/PhysRevLett.116.221101} {\bibfield
  {journal} {\bibinfo  {journal} {Phys. Rev. Lett.}\ }\textbf {\bibinfo
  {volume} {116}},\ \bibinfo {pages} {221101} (\bibinfo {year}
  {2016}{\natexlab{e}})},\ \Eprint {http://arxiv.org/abs/1602.03841}
  {arXiv:1602.03841 [gr-qc]} \BibitemShut {NoStop}%
\bibitem [{\citenamefont {Campanelli}\ \emph
  {et~al.}(2006{\natexlab{b}})\citenamefont {Campanelli}, \citenamefont
  {Lousto},\ and\ \citenamefont {Zlochower}}]{Campanelli:2006uy}%
  \BibitemOpen
  \bibfield  {author} {\bibinfo {author} {\bibfnamefont {M.}~\bibnamefont
  {Campanelli}}, \bibinfo {author} {\bibfnamefont {C.~O.}\ \bibnamefont
  {Lousto}}, \ and\ \bibinfo {author} {\bibfnamefont {Y.}~\bibnamefont
  {Zlochower}},\ }\href@noop {} {\bibfield  {journal} {\bibinfo  {journal}
  {Phys. Rev.}\ }\textbf {\bibinfo {volume} {D74}},\ \bibinfo {pages}
  {041501(R)} (\bibinfo {year} {2006}{\natexlab{b}})},\ \Eprint
  {http://arxiv.org/abs/gr-qc/0604012} {gr-qc/0604012} \BibitemShut {NoStop}%
\bibitem [{\citenamefont {Campanelli}\ \emph
  {et~al.}(2007{\natexlab{a}})\citenamefont {Campanelli}, \citenamefont
  {Lousto}, \citenamefont {Zlochower},\ and\ \citenamefont
  {Merritt}}]{Campanelli:2007ew}%
  \BibitemOpen
  \bibfield  {author} {\bibinfo {author} {\bibfnamefont {M.}~\bibnamefont
  {Campanelli}}, \bibinfo {author} {\bibfnamefont {C.~O.}\ \bibnamefont
  {Lousto}}, \bibinfo {author} {\bibfnamefont {Y.}~\bibnamefont {Zlochower}}, \
  and\ \bibinfo {author} {\bibfnamefont {D.}~\bibnamefont {Merritt}},\
  }\href@noop {} {\bibfield  {journal} {\bibinfo  {journal} {Astrophys. J.}\
  }\textbf {\bibinfo {volume} {659}},\ \bibinfo {pages} {L5} (\bibinfo {year}
  {2007}{\natexlab{a}})},\ \Eprint {http://arxiv.org/abs/gr-qc/0701164}
  {gr-qc/0701164} \BibitemShut {NoStop}%
\bibitem [{\citenamefont {Campanelli}\ \emph
  {et~al.}(2007{\natexlab{b}})\citenamefont {Campanelli}, \citenamefont
  {Lousto}, \citenamefont {Zlochower},\ and\ \citenamefont
  {Merritt}}]{Campanelli:2007cga}%
  \BibitemOpen
  \bibfield  {author} {\bibinfo {author} {\bibfnamefont {M.}~\bibnamefont
  {Campanelli}}, \bibinfo {author} {\bibfnamefont {C.~O.}\ \bibnamefont
  {Lousto}}, \bibinfo {author} {\bibfnamefont {Y.}~\bibnamefont {Zlochower}}, \
  and\ \bibinfo {author} {\bibfnamefont {D.}~\bibnamefont {Merritt}},\
  }\href@noop {} {\bibfield  {journal} {\bibinfo  {journal} {Phys. Rev. Lett.}\
  }\textbf {\bibinfo {volume} {98}},\ \bibinfo {pages} {231102} (\bibinfo
  {year} {2007}{\natexlab{b}})},\ \Eprint {http://arxiv.org/abs/gr-qc/0702133}
  {gr-qc/0702133} \BibitemShut {NoStop}%
\bibitem [{\citenamefont {Lousto}\ and\ \citenamefont
  {Zlochower}(2011)}]{Lousto:2011kp}%
  \BibitemOpen
  \bibfield  {author} {\bibinfo {author} {\bibfnamefont {C.~O.}\ \bibnamefont
  {Lousto}}\ and\ \bibinfo {author} {\bibfnamefont {Y.}~\bibnamefont
  {Zlochower}},\ }\href {\doibase 10.1103/PhysRevLett.107.231102} {\bibfield
  {journal} {\bibinfo  {journal} {Phys. Rev. Lett.}\ }\textbf {\bibinfo
  {volume} {107}},\ \bibinfo {pages} {231102} (\bibinfo {year} {2011})},\
  \Eprint {http://arxiv.org/abs/1108.2009} {arXiv:1108.2009 [gr-qc]}
  \BibitemShut {NoStop}%
\bibitem [{\citenamefont {Lousto}\ and\ \citenamefont
  {Healy}(2015)}]{Lousto:2014ida}%
  \BibitemOpen
  \bibfield  {author} {\bibinfo {author} {\bibfnamefont {C.~O.}\ \bibnamefont
  {Lousto}}\ and\ \bibinfo {author} {\bibfnamefont {J.}~\bibnamefont {Healy}},\
  }\href {\doibase 10.1103/PhysRevLett.114.141101} {\bibfield  {journal}
  {\bibinfo  {journal} {Phys. Rev. Lett.}\ }\textbf {\bibinfo {volume} {114}},\
  \bibinfo {pages} {141101} (\bibinfo {year} {2015})},\ \Eprint
  {http://arxiv.org/abs/1410.3830} {arXiv:1410.3830 [gr-qc]} \BibitemShut
  {NoStop}%
\bibitem [{\citenamefont {Lousto}\ \emph {et~al.}(2016)\citenamefont {Lousto},
  \citenamefont {Healy},\ and\ \citenamefont {Nakano}}]{Lousto:2015uwa}%
  \BibitemOpen
  \bibfield  {author} {\bibinfo {author} {\bibfnamefont {C.~O.}\ \bibnamefont
  {Lousto}}, \bibinfo {author} {\bibfnamefont {J.}~\bibnamefont {Healy}}, \
  and\ \bibinfo {author} {\bibfnamefont {H.}~\bibnamefont {Nakano}},\ }\href
  {\doibase 10.1103/PhysRevD.93.044031} {\bibfield  {journal} {\bibinfo
  {journal} {Phys. Rev.}\ }\textbf {\bibinfo {volume} {D93}},\ \bibinfo {pages}
  {044031} (\bibinfo {year} {2016})},\ \Eprint
  {http://arxiv.org/abs/1506.04768} {arXiv:1506.04768 [gr-qc]} \BibitemShut
  {NoStop}%
\bibitem [{\citenamefont {Lousto}\ and\ \citenamefont
  {Healy}(2016)}]{Lousto:2016nlp}%
  \BibitemOpen
  \bibfield  {author} {\bibinfo {author} {\bibfnamefont {C.~O.}\ \bibnamefont
  {Lousto}}\ and\ \bibinfo {author} {\bibfnamefont {J.}~\bibnamefont {Healy}},\
  }\href {\doibase 10.1103/PhysRevD.93.124074} {\bibfield  {journal} {\bibinfo
  {journal} {Phys. Rev.}\ }\textbf {\bibinfo {volume} {D93}},\ \bibinfo {pages}
  {124074} (\bibinfo {year} {2016})},\ \Eprint
  {http://arxiv.org/abs/1601.05086} {arXiv:1601.05086 [gr-qc]} \BibitemShut
  {NoStop}%
\bibitem [{\citenamefont {Lovelace}\ \emph {et~al.}(2015)\citenamefont
  {Lovelace}, \citenamefont {Scheel}, \citenamefont {Owen}, \citenamefont
  {Giesler}, \citenamefont {Katebi}, \citenamefont {Szil{\'a}gyi},
  \citenamefont {Chu}, \citenamefont {Demos}, \citenamefont {Hemberger},
  \citenamefont {Kidder}, \citenamefont {Pfeiffer},\ and\ \citenamefont
  {Afshari}}]{Lovelace:2014twa}%
  \BibitemOpen
  \bibfield  {author} {\bibinfo {author} {\bibfnamefont {G.}~\bibnamefont
  {Lovelace}}, \bibinfo {author} {\bibfnamefont {M.~A.}\ \bibnamefont
  {Scheel}}, \bibinfo {author} {\bibfnamefont {R.}~\bibnamefont {Owen}},
  \bibinfo {author} {\bibfnamefont {M.}~\bibnamefont {Giesler}}, \bibinfo
  {author} {\bibfnamefont {R.}~\bibnamefont {Katebi}}, \bibinfo {author}
  {\bibfnamefont {B.}~\bibnamefont {Szil{\'a}gyi}}, \bibinfo {author}
  {\bibfnamefont {T.}~\bibnamefont {Chu}}, \bibinfo {author} {\bibfnamefont
  {N.}~\bibnamefont {Demos}}, \bibinfo {author} {\bibfnamefont {D.~A.}\
  \bibnamefont {Hemberger}}, \bibinfo {author} {\bibfnamefont {L.~E.}\
  \bibnamefont {Kidder}}, \bibinfo {author} {\bibfnamefont {H.~P.}\
  \bibnamefont {Pfeiffer}}, \ and\ \bibinfo {author} {\bibfnamefont
  {N.}~\bibnamefont {Afshari}},\ }\href {\doibase
  10.1088/0264-9381/32/6/065007} {\bibfield  {journal} {\bibinfo  {journal}
  {Class. Quant. Grav.}\ }\textbf {\bibinfo {volume} {32}},\ \bibinfo {pages}
  {065007} (\bibinfo {year} {2015})},\ \Eprint {http://arxiv.org/abs/1411.7297}
  {arXiv:1411.7297 [gr-qc]} \BibitemShut {NoStop}%
\bibitem [{\citenamefont {Lousto}\ and\ \citenamefont
  {Zlochower}(2014{\natexlab{a}})}]{Lousto:2013vpa}%
  \BibitemOpen
  \bibfield  {author} {\bibinfo {author} {\bibfnamefont {C.~O.}\ \bibnamefont
  {Lousto}}\ and\ \bibinfo {author} {\bibfnamefont {Y.}~\bibnamefont
  {Zlochower}},\ }\href {\doibase 10.1103/PhysRevD.89.021501} {\bibfield
  {journal} {\bibinfo  {journal} {Phys. Rev.}\ }\textbf {\bibinfo {volume}
  {D89}},\ \bibinfo {pages} {021501} (\bibinfo {year} {2014}{\natexlab{a}})},\
  \Eprint {http://arxiv.org/abs/1307.6237} {arXiv:1307.6237 [gr-qc]}
  \BibitemShut {NoStop}%
\bibitem [{\citenamefont {Healy}\ \emph {et~al.}(2014)\citenamefont {Healy},
  \citenamefont {Lousto},\ and\ \citenamefont {Zlochower}}]{Healy:2014yta}%
  \BibitemOpen
  \bibfield  {author} {\bibinfo {author} {\bibfnamefont {J.}~\bibnamefont
  {Healy}}, \bibinfo {author} {\bibfnamefont {C.~O.}\ \bibnamefont {Lousto}}, \
  and\ \bibinfo {author} {\bibfnamefont {Y.}~\bibnamefont {Zlochower}},\ }\href
  {\doibase 10.1103/PhysRevD.90.104004} {\bibfield  {journal} {\bibinfo
  {journal} {Phys. Rev.}\ }\textbf {\bibinfo {volume} {D90}},\ \bibinfo {pages}
  {104004} (\bibinfo {year} {2014})},\ \Eprint {http://arxiv.org/abs/1406.7295}
  {arXiv:1406.7295 [gr-qc]} \BibitemShut {NoStop}%
\bibitem [{\citenamefont {Mroue}\ \emph {et~al.}(2013)\citenamefont {Mroue},
  \citenamefont {Scheel}, \citenamefont {Szilagyi}, \citenamefont {Pfeiffer},
  \citenamefont {Boyle} \emph {et~al.}}]{Mroue:2013xna}%
  \BibitemOpen
  \bibfield  {author} {\bibinfo {author} {\bibfnamefont {A.~H.}\ \bibnamefont
  {Mroue}}, \bibinfo {author} {\bibfnamefont {M.~A.}\ \bibnamefont {Scheel}},
  \bibinfo {author} {\bibfnamefont {B.}~\bibnamefont {Szilagyi}}, \bibinfo
  {author} {\bibfnamefont {H.~P.}\ \bibnamefont {Pfeiffer}}, \bibinfo {author}
  {\bibfnamefont {M.}~\bibnamefont {Boyle}},  \emph {et~al.},\ }\href {\doibase
  10.1103/PhysRevLett.111.241104} {\bibfield  {journal} {\bibinfo  {journal}
  {Phys. Rev. Lett.}\ }\textbf {\bibinfo {volume} {111}},\ \bibinfo {pages}
  {241104} (\bibinfo {year} {2013})},\ \Eprint {http://arxiv.org/abs/1304.6077}
  {arXiv:1304.6077 [gr-qc]} \BibitemShut {NoStop}%
\bibitem [{\citenamefont {Jani}\ \emph {et~al.}(2016)\citenamefont {Jani},
  \citenamefont {Healy}, \citenamefont {Clark}, \citenamefont {London},
  \citenamefont {Laguna},\ and\ \citenamefont {Shoemaker}}]{Jani:2016wkt}%
  \BibitemOpen
  \bibfield  {author} {\bibinfo {author} {\bibfnamefont {K.}~\bibnamefont
  {Jani}}, \bibinfo {author} {\bibfnamefont {J.}~\bibnamefont {Healy}},
  \bibinfo {author} {\bibfnamefont {J.~A.}\ \bibnamefont {Clark}}, \bibinfo
  {author} {\bibfnamefont {L.}~\bibnamefont {London}}, \bibinfo {author}
  {\bibfnamefont {P.}~\bibnamefont {Laguna}}, \ and\ \bibinfo {author}
  {\bibfnamefont {D.}~\bibnamefont {Shoemaker}},\ }\href {\doibase
  10.1088/0264-9381/33/20/204001} {\bibfield  {journal} {\bibinfo  {journal}
  {Class. Quant. Grav.}\ }\textbf {\bibinfo {volume} {33}},\ \bibinfo {pages}
  {204001} (\bibinfo {year} {2016})},\ \Eprint
  {http://arxiv.org/abs/1605.03204} {arXiv:1605.03204 [gr-qc]} \BibitemShut
  {NoStop}%
\bibitem [{\citenamefont {Brandt}\ and\ \citenamefont
  {Br{\"u}gmann}(1997)}]{Brandt97b}%
  \BibitemOpen
  \bibfield  {author} {\bibinfo {author} {\bibfnamefont {S.}~\bibnamefont
  {Brandt}}\ and\ \bibinfo {author} {\bibfnamefont {B.}~\bibnamefont
  {Br{\"u}gmann}},\ }\href@noop {} {\bibfield  {journal} {\bibinfo  {journal}
  {Phys. Rev. Lett.}\ }\textbf {\bibinfo {volume} {78}},\ \bibinfo {pages}
  {3606} (\bibinfo {year} {1997})},\ \Eprint
  {http://arxiv.org/abs/gr-qc/9703066} {gr-qc/9703066} \BibitemShut {NoStop}%
\bibitem [{\citenamefont {Ansorg}\ \emph {et~al.}(2004)\citenamefont {Ansorg},
  \citenamefont {Br\"ugmann},\ and\ \citenamefont {Tichy}}]{Ansorg:2004ds}%
  \BibitemOpen
  \bibfield  {author} {\bibinfo {author} {\bibfnamefont {M.}~\bibnamefont
  {Ansorg}}, \bibinfo {author} {\bibfnamefont {B.}~\bibnamefont {Br\"ugmann}},
  \ and\ \bibinfo {author} {\bibfnamefont {W.}~\bibnamefont {Tichy}},\
  }\href@noop {} {\bibfield  {journal} {\bibinfo  {journal} {Phys. Rev.}\
  }\textbf {\bibinfo {volume} {D70}},\ \bibinfo {pages} {064011} (\bibinfo
  {year} {2004})},\ \Eprint {http://arxiv.org/abs/gr-qc/0404056}
  {gr-qc/0404056} \BibitemShut {NoStop}%
\bibitem [{\citenamefont {Zlochower}\ \emph {et~al.}(2005)\citenamefont
  {Zlochower}, \citenamefont {Baker}, \citenamefont {Campanelli},\ and\
  \citenamefont {Lousto}}]{Zlochower:2005bj}%
  \BibitemOpen
  \bibfield  {author} {\bibinfo {author} {\bibfnamefont {Y.}~\bibnamefont
  {Zlochower}}, \bibinfo {author} {\bibfnamefont {J.~G.}\ \bibnamefont
  {Baker}}, \bibinfo {author} {\bibfnamefont {M.}~\bibnamefont {Campanelli}}, \
  and\ \bibinfo {author} {\bibfnamefont {C.~O.}\ \bibnamefont {Lousto}},\
  }\href {\doibase 10.1103/PhysRevD.72.024021} {\bibfield  {journal} {\bibinfo
  {journal} {Phys. Rev.}\ }\textbf {\bibinfo {volume} {D72}},\ \bibinfo {pages}
  {024021} (\bibinfo {year} {2005})},\ \Eprint
  {http://arxiv.org/abs/gr-qc/0505055} {arXiv:gr-qc/0505055} \BibitemShut
  {NoStop}%
\bibitem [{\citenamefont {Marronetti}\ \emph {et~al.}(2008)\citenamefont
  {Marronetti}, \citenamefont {Tichy}, \citenamefont {Br{\"u}gmann},
  \citenamefont {Gonzalez},\ and\ \citenamefont
  {Sperhake}}]{Marronetti:2007wz}%
  \BibitemOpen
  \bibfield  {author} {\bibinfo {author} {\bibfnamefont {P.}~\bibnamefont
  {Marronetti}}, \bibinfo {author} {\bibfnamefont {W.}~\bibnamefont {Tichy}},
  \bibinfo {author} {\bibfnamefont {B.}~\bibnamefont {Br{\"u}gmann}}, \bibinfo
  {author} {\bibfnamefont {J.}~\bibnamefont {Gonzalez}}, \ and\ \bibinfo
  {author} {\bibfnamefont {U.}~\bibnamefont {Sperhake}},\ }\href {\doibase
  10.1103/PhysRevD.77.064010} {\bibfield  {journal} {\bibinfo  {journal} {Phys.
  Rev.}\ }\textbf {\bibinfo {volume} {D77}},\ \bibinfo {pages} {064010}
  (\bibinfo {year} {2008})},\ \Eprint {http://arxiv.org/abs/0709.2160}
  {arXiv:0709.2160 [gr-qc]} \BibitemShut {NoStop}%
\bibitem [{\citenamefont {Lousto}\ and\ \citenamefont
  {Zlochower}(2008)}]{Lousto:2007rj}%
  \BibitemOpen
  \bibfield  {author} {\bibinfo {author} {\bibfnamefont {C.~O.}\ \bibnamefont
  {Lousto}}\ and\ \bibinfo {author} {\bibfnamefont {Y.}~\bibnamefont
  {Zlochower}},\ }\href {\doibase 10.1103/PhysRevD.77.024034} {\bibfield
  {journal} {\bibinfo  {journal} {Phys. Rev.}\ }\textbf {\bibinfo {volume}
  {D77}},\ \bibinfo {pages} {024034} (\bibinfo {year} {2008})},\ \Eprint
  {http://arxiv.org/abs/0711.1165} {arXiv:0711.1165 [gr-qc]} \BibitemShut
  {NoStop}%
\bibitem [{\citenamefont {Zlochower}\ \emph {et~al.}(2012)\citenamefont
  {Zlochower}, \citenamefont {Ponce},\ and\ \citenamefont
  {Lousto}}]{Zlochower:2012fk}%
  \BibitemOpen
  \bibfield  {author} {\bibinfo {author} {\bibfnamefont {Y.}~\bibnamefont
  {Zlochower}}, \bibinfo {author} {\bibfnamefont {M.}~\bibnamefont {Ponce}}, \
  and\ \bibinfo {author} {\bibfnamefont {C.~O.}\ \bibnamefont {Lousto}},\
  }\href {\doibase 10.1103/PhysRevD.86.104056} {\bibfield  {journal} {\bibinfo
  {journal} {Phys. Rev.}\ }\textbf {\bibinfo {volume} {D86}},\ \bibinfo {pages}
  {104056} (\bibinfo {year} {2012})},\ \Eprint {http://arxiv.org/abs/1208.5494}
  {arXiv:1208.5494 [gr-qc]} \BibitemShut {NoStop}%
\bibitem [{Note1()}]{Note1}%
  \BibitemOpen
  \bibinfo {note} {Https://portal.xsede.org/sdsc-comet}\BibitemShut {NoStop}%
\bibitem [{\citenamefont {Lousto}\ and\ \citenamefont
  {Zlochower}(2013{\natexlab{a}})}]{Lousto:2013oza}%
  \BibitemOpen
  \bibfield  {author} {\bibinfo {author} {\bibfnamefont {C.~O.}\ \bibnamefont
  {Lousto}}\ and\ \bibinfo {author} {\bibfnamefont {Y.}~\bibnamefont
  {Zlochower}},\ }\href {\doibase 10.1103/PhysRevD.88.024001} {\bibfield
  {journal} {\bibinfo  {journal} {Phys. Rev.}\ }\textbf {\bibinfo {volume}
  {D88}},\ \bibinfo {pages} {024001} (\bibinfo {year} {2013}{\natexlab{a}})},\
  \Eprint {http://arxiv.org/abs/1304.3937} {arXiv:1304.3937 [gr-qc]}
  \BibitemShut {NoStop}%
\bibitem [{\citenamefont {L{\"o}ffler}\ \emph {et~al.}(2012)\citenamefont
  {L{\"o}ffler}, \citenamefont {Faber}, \citenamefont {Bentivegna},
  \citenamefont {Bode}, \citenamefont {Diener}, \citenamefont {Haas},
  \citenamefont {Hinder}, \citenamefont {Mundim}, \citenamefont {Ott},
  \citenamefont {Schnetter}, \citenamefont {Allen}, \citenamefont
  {Campanelli},\ and\ \citenamefont {Laguna}}]{Loffler:2011ay}%
  \BibitemOpen
  \bibfield  {author} {\bibinfo {author} {\bibfnamefont {F.}~\bibnamefont
  {L{\"o}ffler}}, \bibinfo {author} {\bibfnamefont {J.}~\bibnamefont {Faber}},
  \bibinfo {author} {\bibfnamefont {E.}~\bibnamefont {Bentivegna}}, \bibinfo
  {author} {\bibfnamefont {T.}~\bibnamefont {Bode}}, \bibinfo {author}
  {\bibfnamefont {P.}~\bibnamefont {Diener}}, \bibinfo {author} {\bibfnamefont
  {R.}~\bibnamefont {Haas}}, \bibinfo {author} {\bibfnamefont {I.}~\bibnamefont
  {Hinder}}, \bibinfo {author} {\bibfnamefont {B.~C.}\ \bibnamefont {Mundim}},
  \bibinfo {author} {\bibfnamefont {C.~D.}\ \bibnamefont {Ott}}, \bibinfo
  {author} {\bibfnamefont {E.}~\bibnamefont {Schnetter}}, \bibinfo {author}
  {\bibfnamefont {G.}~\bibnamefont {Allen}}, \bibinfo {author} {\bibfnamefont
  {M.}~\bibnamefont {Campanelli}}, \ and\ \bibinfo {author} {\bibfnamefont
  {P.}~\bibnamefont {Laguna}},\ }\href@noop {} {\bibfield  {journal} {\bibinfo
  {journal} {Class. Quant. Grav.}\ }\textbf {\bibinfo {volume} {29}},\ \bibinfo
  {pages} {115001} (\bibinfo {year} {2012})},\ \Eprint
  {http://arxiv.org/abs/1111.3344} {arXiv:1111.3344 [gr-qc]} \BibitemShut
  {NoStop}%
\bibitem [{ein()}]{einsteintoolkit}%
  \BibitemOpen
  \href@noop {} {}\bibinfo {note} {Einstein Toolkit home page: {\tt
  http://einsteintoolkit.org}}\BibitemShut {NoStop}%
\bibitem [{cac()}]{cactus_web}%
  \BibitemOpen
  \href@noop {} {}\bibinfo {note} {Cactus Computational Toolkit home page: {\tt
  http://cactuscode.org}}\BibitemShut {NoStop}%
\bibitem [{\citenamefont {Schnetter}\ \emph {et~al.}(2004)\citenamefont
  {Schnetter}, \citenamefont {Hawley},\ and\ \citenamefont
  {Hawke}}]{Schnetter-etal-03b}%
  \BibitemOpen
  \bibfield  {author} {\bibinfo {author} {\bibfnamefont {E.}~\bibnamefont
  {Schnetter}}, \bibinfo {author} {\bibfnamefont {S.~H.}\ \bibnamefont
  {Hawley}}, \ and\ \bibinfo {author} {\bibfnamefont {I.}~\bibnamefont
  {Hawke}},\ }\href@noop {} {\bibfield  {journal} {\bibinfo  {journal} {Class.
  Quant. Grav.}\ }\textbf {\bibinfo {volume} {21}},\ \bibinfo {pages} {1465}
  (\bibinfo {year} {2004})},\ \Eprint {http://arxiv.org/abs/gr-qc/0310042}
  {gr-qc/0310042} \BibitemShut {NoStop}%
\bibitem [{\citenamefont {Thornburg}(2004)}]{Thornburg2003:AH-finding}%
  \BibitemOpen
  \bibfield  {author} {\bibinfo {author} {\bibfnamefont {J.}~\bibnamefont
  {Thornburg}},\ }\href {\doibase 10.1088/0264-9381/21/2/026} {\bibfield
  {journal} {\bibinfo  {journal} {Class. Quant. Grav.}\ }\textbf {\bibinfo
  {volume} {21}},\ \bibinfo {pages} {743} (\bibinfo {year} {2004})},\ \Eprint
  {http://arxiv.org/abs/gr-qc/0306056} {gr-qc/0306056} \BibitemShut {NoStop}%
\bibitem [{\citenamefont {Dreyer}\ \emph {et~al.}(2003)\citenamefont {Dreyer},
  \citenamefont {Krishnan}, \citenamefont {Shoemaker},\ and\ \citenamefont
  {Schnetter}}]{Dreyer02a}%
  \BibitemOpen
  \bibfield  {author} {\bibinfo {author} {\bibfnamefont {O.}~\bibnamefont
  {Dreyer}}, \bibinfo {author} {\bibfnamefont {B.}~\bibnamefont {Krishnan}},
  \bibinfo {author} {\bibfnamefont {D.}~\bibnamefont {Shoemaker}}, \ and\
  \bibinfo {author} {\bibfnamefont {E.}~\bibnamefont {Schnetter}},\ }\href@noop
  {} {\bibfield  {journal} {\bibinfo  {journal} {Phys. Rev.}\ }\textbf
  {\bibinfo {volume} {D67}},\ \bibinfo {pages} {024018} (\bibinfo {year}
  {2003})},\ \Eprint {http://arxiv.org/abs/gr-qc/0206008} {gr-qc/0206008}
  \BibitemShut {NoStop}%
\bibitem [{\citenamefont {Campanelli}\ \emph
  {et~al.}(2007{\natexlab{c}})\citenamefont {Campanelli}, \citenamefont
  {Lousto}, \citenamefont {Zlochower}, \citenamefont {Krishnan},\ and\
  \citenamefont {Merritt}}]{Campanelli:2006fy}%
  \BibitemOpen
  \bibfield  {author} {\bibinfo {author} {\bibfnamefont {M.}~\bibnamefont
  {Campanelli}}, \bibinfo {author} {\bibfnamefont {C.~O.}\ \bibnamefont
  {Lousto}}, \bibinfo {author} {\bibfnamefont {Y.}~\bibnamefont {Zlochower}},
  \bibinfo {author} {\bibfnamefont {B.}~\bibnamefont {Krishnan}}, \ and\
  \bibinfo {author} {\bibfnamefont {D.}~\bibnamefont {Merritt}},\ }\href@noop
  {} {\bibfield  {journal} {\bibinfo  {journal} {Phys. Rev.}\ }\textbf
  {\bibinfo {volume} {D75}},\ \bibinfo {pages} {064030} (\bibinfo {year}
  {2007}{\natexlab{c}})},\ \Eprint {http://arxiv.org/abs/gr-qc/0612076}
  {gr-qc/0612076} \BibitemShut {NoStop}%
\bibitem [{\citenamefont {Campanelli}\ and\ \citenamefont
  {Lousto}(1999)}]{Campanelli:1998jv}%
  \BibitemOpen
  \bibfield  {author} {\bibinfo {author} {\bibfnamefont {M.}~\bibnamefont
  {Campanelli}}\ and\ \bibinfo {author} {\bibfnamefont {C.~O.}\ \bibnamefont
  {Lousto}},\ }\href {\doibase 10.1103/PhysRevD.59.124022} {\bibfield
  {journal} {\bibinfo  {journal} {Phys. Rev.}\ }\textbf {\bibinfo {volume}
  {D59}},\ \bibinfo {pages} {124022} (\bibinfo {year} {1999})},\ \Eprint
  {http://arxiv.org/abs/gr-qc/9811019} {arXiv:gr-qc/9811019 [gr-qc]}
  \BibitemShut {NoStop}%
\bibitem [{\citenamefont {Lousto}\ and\ \citenamefont
  {Zlochower}(2007)}]{Lousto:2007mh}%
  \BibitemOpen
  \bibfield  {author} {\bibinfo {author} {\bibfnamefont {C.~O.}\ \bibnamefont
  {Lousto}}\ and\ \bibinfo {author} {\bibfnamefont {Y.}~\bibnamefont
  {Zlochower}},\ }\href@noop {} {\bibfield  {journal} {\bibinfo  {journal}
  {Phys. Rev.}\ }\textbf {\bibinfo {volume} {D76}},\ \bibinfo {pages}
  {041502(R)} (\bibinfo {year} {2007})},\ \Eprint
  {http://arxiv.org/abs/gr-qc/0703061} {gr-qc/0703061} \BibitemShut {NoStop}%
\bibitem [{\citenamefont {Nakano}\ \emph {et~al.}(2015)\citenamefont {Nakano},
  \citenamefont {Healy}, \citenamefont {Lousto},\ and\ \citenamefont
  {Zlochower}}]{Nakano:2015pta}%
  \BibitemOpen
  \bibfield  {author} {\bibinfo {author} {\bibfnamefont {H.}~\bibnamefont
  {Nakano}}, \bibinfo {author} {\bibfnamefont {J.}~\bibnamefont {Healy}},
  \bibinfo {author} {\bibfnamefont {C.~O.}\ \bibnamefont {Lousto}}, \ and\
  \bibinfo {author} {\bibfnamefont {Y.}~\bibnamefont {Zlochower}},\ }\href
  {\doibase 10.1103/PhysRevD.91.104022} {\bibfield  {journal} {\bibinfo
  {journal} {Phys. Rev.}\ }\textbf {\bibinfo {volume} {D91}},\ \bibinfo {pages}
  {104022} (\bibinfo {year} {2015})},\ \Eprint
  {http://arxiv.org/abs/1503.00718} {arXiv:1503.00718 [gr-qc]} \BibitemShut
  {NoStop}%
\bibitem [{\citenamefont {Baker}\ \emph {et~al.}(2004)\citenamefont {Baker},
  \citenamefont {Campanelli}, \citenamefont {Lousto},\ and\ \citenamefont
  {Takahashi}}]{Baker:2003ds}%
  \BibitemOpen
  \bibfield  {author} {\bibinfo {author} {\bibfnamefont {J.~G.}\ \bibnamefont
  {Baker}}, \bibinfo {author} {\bibfnamefont {M.}~\bibnamefont {Campanelli}},
  \bibinfo {author} {\bibfnamefont {C.~O.}\ \bibnamefont {Lousto}}, \ and\
  \bibinfo {author} {\bibfnamefont {R.}~\bibnamefont {Takahashi}},\ }\href
  {\doibase 10.1103/PhysRevD.69.027505} {\bibfield  {journal} {\bibinfo
  {journal} {Phys. Rev.}\ }\textbf {\bibinfo {volume} {D69}},\ \bibinfo {pages}
  {027505} (\bibinfo {year} {2004})},\ \Eprint
  {http://arxiv.org/abs/astro-ph/0305287} {arXiv:astro-ph/0305287} \BibitemShut
  {NoStop}%
\bibitem [{\citenamefont {Barausse}\ \emph {et~al.}(2012)\citenamefont
  {Barausse}, \citenamefont {Morozova},\ and\ \citenamefont
  {Rezzolla}}]{Barausse:2012qz}%
  \BibitemOpen
  \bibfield  {author} {\bibinfo {author} {\bibfnamefont {E.}~\bibnamefont
  {Barausse}}, \bibinfo {author} {\bibfnamefont {V.}~\bibnamefont {Morozova}},
  \ and\ \bibinfo {author} {\bibfnamefont {L.}~\bibnamefont {Rezzolla}},\
  }\href {\doibase 10.1088/0004-637X/758/1/63} {\bibfield  {journal} {\bibinfo
  {journal} {Astrophys. J.}\ }\textbf {\bibinfo {volume} {758}},\ \bibinfo
  {pages} {63} (\bibinfo {year} {2012})},\ \Eprint
  {http://arxiv.org/abs/1206.3803} {arXiv:1206.3803 [gr-qc]} \BibitemShut
  {NoStop}%
\bibitem [{\citenamefont {Rezzolla}\ \emph {et~al.}(2008)\citenamefont
  {Rezzolla}, \citenamefont {Barausse}, \citenamefont {Dorband}, \citenamefont
  {Pollney}, \citenamefont {Reisswig}, \citenamefont {Seiler},\ and\
  \citenamefont {Husa}}]{Rezzolla:2007rz}%
  \BibitemOpen
  \bibfield  {author} {\bibinfo {author} {\bibfnamefont {L.}~\bibnamefont
  {Rezzolla}}, \bibinfo {author} {\bibfnamefont {E.}~\bibnamefont {Barausse}},
  \bibinfo {author} {\bibfnamefont {E.~N.}\ \bibnamefont {Dorband}}, \bibinfo
  {author} {\bibfnamefont {D.}~\bibnamefont {Pollney}}, \bibinfo {author}
  {\bibfnamefont {C.}~\bibnamefont {Reisswig}}, \bibinfo {author}
  {\bibfnamefont {J.}~\bibnamefont {Seiler}}, \ and\ \bibinfo {author}
  {\bibfnamefont {S.}~\bibnamefont {Husa}},\ }\href {\doibase
  10.1103/PhysRevD.78.044002} {\bibfield  {journal} {\bibinfo  {journal} {Phys.
  Rev.}\ }\textbf {\bibinfo {volume} {D78}},\ \bibinfo {pages} {044002}
  (\bibinfo {year} {2008})},\ \Eprint {http://arxiv.org/abs/0712.3541}
  {arXiv:0712.3541 [gr-qc]} \BibitemShut {NoStop}%
\bibitem [{\citenamefont {Hofmann}\ \emph {et~al.}(2016)\citenamefont
  {Hofmann}, \citenamefont {Barausse},\ and\ \citenamefont
  {Rezzolla}}]{Hofmann:2016yih}%
  \BibitemOpen
  \bibfield  {author} {\bibinfo {author} {\bibfnamefont {F.}~\bibnamefont
  {Hofmann}}, \bibinfo {author} {\bibfnamefont {E.}~\bibnamefont {Barausse}}, \
  and\ \bibinfo {author} {\bibfnamefont {L.}~\bibnamefont {Rezzolla}},\ }\href
  {\doibase 10.3847/2041-8205/825/2/L19} {\bibfield  {journal} {\bibinfo
  {journal} {Astrophys. J.}\ }\textbf {\bibinfo {volume} {825}},\ \bibinfo
  {pages} {L19} (\bibinfo {year} {2016})},\ \Eprint
  {http://arxiv.org/abs/1605.01938} {arXiv:1605.01938 [gr-qc]} \BibitemShut
  {NoStop}%
\bibitem [{\citenamefont {Jim{\'e}nez-Forteza}\ \emph
  {et~al.}(2016)\citenamefont {Jim{\'e}nez-Forteza}, \citenamefont {Keitel},
  \citenamefont {Husa}, \citenamefont {Hannam}, \citenamefont {Khan},\ and\
  \citenamefont {P{\"u}rrer}}]{Jimenez-Forteza:2016oae}%
  \BibitemOpen
  \bibfield  {author} {\bibinfo {author} {\bibfnamefont {X.}~\bibnamefont
  {Jim{\'e}nez-Forteza}}, \bibinfo {author} {\bibfnamefont {D.}~\bibnamefont
  {Keitel}}, \bibinfo {author} {\bibfnamefont {S.}~\bibnamefont {Husa}},
  \bibinfo {author} {\bibfnamefont {M.}~\bibnamefont {Hannam}}, \bibinfo
  {author} {\bibfnamefont {S.}~\bibnamefont {Khan}}, \ and\ \bibinfo {author}
  {\bibfnamefont {M.}~\bibnamefont {P{\"u}rrer}},\ }\href@noop {} {\  (\bibinfo
  {year} {2016})},\ \Eprint {http://arxiv.org/abs/1611.00332} {arXiv:1611.00332
  [gr-qc]} \BibitemShut {NoStop}%
\bibitem [{\citenamefont {Lousto}\ \emph {et~al.}(2010)\citenamefont {Lousto},
  \citenamefont {Campanelli}, \citenamefont {Zlochower},\ and\ \citenamefont
  {Nakano}}]{Lousto:2009mf}%
  \BibitemOpen
  \bibfield  {author} {\bibinfo {author} {\bibfnamefont {C.~O.}\ \bibnamefont
  {Lousto}}, \bibinfo {author} {\bibfnamefont {M.}~\bibnamefont {Campanelli}},
  \bibinfo {author} {\bibfnamefont {Y.}~\bibnamefont {Zlochower}}, \ and\
  \bibinfo {author} {\bibfnamefont {H.}~\bibnamefont {Nakano}},\ }\href
  {\doibase 10.1088/0264-9381/27/11/114006} {\bibfield  {journal} {\bibinfo
  {journal} {Class. Quant. Grav.}\ }\textbf {\bibinfo {volume} {27}},\ \bibinfo
  {pages} {114006} (\bibinfo {year} {2010})},\ \Eprint
  {http://arxiv.org/abs/0904.3541} {arXiv:0904.3541 [gr-qc]} \BibitemShut
  {NoStop}%
\bibitem [{\citenamefont {Lousto}\ and\ \citenamefont
  {Zlochower}(2014{\natexlab{b}})}]{Lousto:2013wta}%
  \BibitemOpen
  \bibfield  {author} {\bibinfo {author} {\bibfnamefont {C.~O.}\ \bibnamefont
  {Lousto}}\ and\ \bibinfo {author} {\bibfnamefont {Y.}~\bibnamefont
  {Zlochower}},\ }\href {\doibase 10.1103/PhysRevD.89.104052} {\bibfield
  {journal} {\bibinfo  {journal} {Phys. Rev.}\ }\textbf {\bibinfo {volume}
  {D89}},\ \bibinfo {pages} {104052} (\bibinfo {year} {2014}{\natexlab{b}})},\
  \Eprint {http://arxiv.org/abs/1312.5775} {arXiv:1312.5775 [gr-qc]}
  \BibitemShut {NoStop}%
\bibitem [{\citenamefont {Zlochower}\ and\ \citenamefont
  {Lousto}(2015)}]{Zlochower:2015wga}%
  \BibitemOpen
  \bibfield  {author} {\bibinfo {author} {\bibfnamefont {Y.}~\bibnamefont
  {Zlochower}}\ and\ \bibinfo {author} {\bibfnamefont {C.~O.}\ \bibnamefont
  {Lousto}},\ }\href {\doibase 10.1103/PhysRevD.92.024022} {\bibfield
  {journal} {\bibinfo  {journal} {Phys. Rev.}\ }\textbf {\bibinfo {volume}
  {D92}},\ \bibinfo {pages} {024022} (\bibinfo {year} {2015})},\ \Eprint
  {http://arxiv.org/abs/1503.07536} {arXiv:1503.07536 [gr-qc]} \BibitemShut
  {NoStop}%
\bibitem [{\citenamefont {Campanelli}\ \emph
  {et~al.}(2009{\natexlab{b}})\citenamefont {Campanelli}, \citenamefont
  {Lousto},\ and\ \citenamefont {Zlochower}}]{Campanelli:2008dv}%
  \BibitemOpen
  \bibfield  {author} {\bibinfo {author} {\bibfnamefont {M.}~\bibnamefont
  {Campanelli}}, \bibinfo {author} {\bibfnamefont {C.~O.}\ \bibnamefont
  {Lousto}}, \ and\ \bibinfo {author} {\bibfnamefont {Y.}~\bibnamefont
  {Zlochower}},\ }\href {\doibase 10.1103/PhysRevD.79.084012} {\bibfield
  {journal} {\bibinfo  {journal} {Phys. Rev.}\ }\textbf {\bibinfo {volume}
  {D79}},\ \bibinfo {pages} {084012} (\bibinfo {year} {2009}{\natexlab{b}})},\
  \Eprint {http://arxiv.org/abs/0811.3006} {arXiv:0811.3006 [gr-qc]}
  \BibitemShut {NoStop}%
\bibitem [{\citenamefont {Owen}(2010)}]{Owen:2010vw}%
  \BibitemOpen
  \bibfield  {author} {\bibinfo {author} {\bibfnamefont {R.}~\bibnamefont
  {Owen}},\ }\href {\doibase 10.1103/PhysRevD.81.124042} {\bibfield  {journal}
  {\bibinfo  {journal} {Phys. Rev.}\ }\textbf {\bibinfo {volume} {D81}},\
  \bibinfo {pages} {124042} (\bibinfo {year} {2010})},\ \Eprint
  {http://arxiv.org/abs/1004.3768} {arXiv:1004.3768 [gr-qc]} \BibitemShut
  {NoStop}%
\bibitem [{\citenamefont {Abbott}\ \emph
  {et~al.}(2016{\natexlab{f}})\citenamefont {Abbott} \emph
  {et~al.}}]{TheLIGOScientific:2016wfe}%
  \BibitemOpen
  \bibfield  {author} {\bibinfo {author} {\bibfnamefont {B.~P.}\ \bibnamefont
  {Abbott}} \emph {et~al.} (\bibinfo {collaboration} {Virgo, LIGO
  Scientific}),\ }\href {\doibase 10.1103/PhysRevLett.116.241102} {\bibfield
  {journal} {\bibinfo  {journal} {Phys. Rev. Lett.}\ }\textbf {\bibinfo
  {volume} {116}},\ \bibinfo {pages} {241102} (\bibinfo {year}
  {2016}{\natexlab{f}})},\ \Eprint {http://arxiv.org/abs/1602.03840}
  {arXiv:1602.03840 [gr-qc]} \BibitemShut {NoStop}%
\bibitem [{\citenamefont {Keitel}\ \emph {et~al.}(2016)\citenamefont {Keitel}
  \emph {et~al.}}]{Keitel:2016krm}%
  \BibitemOpen
  \bibfield  {author} {\bibinfo {author} {\bibfnamefont {D.}~\bibnamefont
  {Keitel}} \emph {et~al.},\ }\href@noop {} {\  (\bibinfo {year} {2016})},\
  \Eprint {http://arxiv.org/abs/1612.09566} {arXiv:1612.09566 [gr-qc]}
  \BibitemShut {NoStop}%
\bibitem [{\citenamefont {Lousto}\ and\ \citenamefont
  {Zlochower}(2013{\natexlab{b}})}]{Lousto:2012gt}%
  \BibitemOpen
  \bibfield  {author} {\bibinfo {author} {\bibfnamefont {C.~O.}\ \bibnamefont
  {Lousto}}\ and\ \bibinfo {author} {\bibfnamefont {Y.}~\bibnamefont
  {Zlochower}},\ }\href {\doibase 10.1103/PhysRevD.87.084027} {\bibfield
  {journal} {\bibinfo  {journal} {Phys. Rev.}\ }\textbf {\bibinfo {volume}
  {D87}},\ \bibinfo {pages} {084027} (\bibinfo {year} {2013}{\natexlab{b}})},\
  \Eprint {http://arxiv.org/abs/1211.7099} {arXiv:1211.7099 [gr-qc]}
  \BibitemShut {NoStop}%
\bibitem [{\citenamefont {Baker}\ \emph {et~al.}(2002)\citenamefont {Baker},
  \citenamefont {Campanelli},\ and\ \citenamefont {Lousto}}]{Baker:2001sf}%
  \BibitemOpen
  \bibfield  {author} {\bibinfo {author} {\bibfnamefont {J.~G.}\ \bibnamefont
  {Baker}}, \bibinfo {author} {\bibfnamefont {M.}~\bibnamefont {Campanelli}}, \
  and\ \bibinfo {author} {\bibfnamefont {C.~O.}\ \bibnamefont {Lousto}},\
  }\href {\doibase 10.1103/PhysRevD.65.044001} {\bibfield  {journal} {\bibinfo
  {journal} {Phys. Rev.}\ }\textbf {\bibinfo {volume} {D65}},\ \bibinfo {pages}
  {044001} (\bibinfo {year} {2002})},\ \Eprint
  {http://arxiv.org/abs/gr-qc/0104063} {arXiv:gr-qc/0104063 [gr-qc]}
  \BibitemShut {NoStop}%
\bibitem [{\citenamefont {Husa}\ \emph {et~al.}(2008)\citenamefont {Husa},
  \citenamefont {Hannam}, \citenamefont {Gonzalez}, \citenamefont {Sperhake},\
  and\ \citenamefont {Brugmann}}]{Husa:2007rh}%
  \BibitemOpen
  \bibfield  {author} {\bibinfo {author} {\bibfnamefont {S.}~\bibnamefont
  {Husa}}, \bibinfo {author} {\bibfnamefont {M.}~\bibnamefont {Hannam}},
  \bibinfo {author} {\bibfnamefont {J.~A.}\ \bibnamefont {Gonzalez}}, \bibinfo
  {author} {\bibfnamefont {U.}~\bibnamefont {Sperhake}}, \ and\ \bibinfo
  {author} {\bibfnamefont {B.}~\bibnamefont {Brugmann}},\ }\href {\doibase
  10.1103/PhysRevD.77.044037} {\bibfield  {journal} {\bibinfo  {journal} {Phys.
  Rev.}\ }\textbf {\bibinfo {volume} {D77}},\ \bibinfo {pages} {044037}
  (\bibinfo {year} {2008})},\ \Eprint {http://arxiv.org/abs/0706.0904}
  {arXiv:0706.0904 [gr-qc]} \BibitemShut {NoStop}%
\bibitem [{\citenamefont {Pfeiffer}\ \emph {et~al.}(2007)\citenamefont
  {Pfeiffer}, \citenamefont {Brown}, \citenamefont {Kidder}, \citenamefont
  {Lindblom}, \citenamefont {Lovelace},\ and\ \citenamefont
  {Scheel}}]{Pfeiffer:2007yz}%
  \BibitemOpen
  \bibfield  {author} {\bibinfo {author} {\bibfnamefont {H.~P.}\ \bibnamefont
  {Pfeiffer}}, \bibinfo {author} {\bibfnamefont {D.~A.}\ \bibnamefont {Brown}},
  \bibinfo {author} {\bibfnamefont {L.~E.}\ \bibnamefont {Kidder}}, \bibinfo
  {author} {\bibfnamefont {L.}~\bibnamefont {Lindblom}}, \bibinfo {author}
  {\bibfnamefont {G.}~\bibnamefont {Lovelace}}, \ and\ \bibinfo {author}
  {\bibfnamefont {M.~A.}\ \bibnamefont {Scheel}},\ }\href {\doibase
  10.1088/0264-9381/24/12/S06} {\bibfield  {journal} {\bibinfo  {journal}
  {Class. Quant. Grav.}\ }\textbf {\bibinfo {volume} {24}},\ \bibinfo {pages}
  {S59} (\bibinfo {year} {2007})},\ \Eprint
  {http://arxiv.org/abs/gr-qc/0702106} {arXiv:gr-qc/0702106 [gr-qc]}
  \BibitemShut {NoStop}%
\bibitem [{\citenamefont {Ori}\ and\ \citenamefont
  {Thorne}(2000)}]{Ori:2000zn}%
  \BibitemOpen
  \bibfield  {author} {\bibinfo {author} {\bibfnamefont {A.}~\bibnamefont
  {Ori}}\ and\ \bibinfo {author} {\bibfnamefont {K.~S.}\ \bibnamefont
  {Thorne}},\ }\href {\doibase 10.1103/PhysRevD.62.124022} {\bibfield
  {journal} {\bibinfo  {journal} {Phys. Rev.}\ }\textbf {\bibinfo {volume}
  {D62}},\ \bibinfo {pages} {124022} (\bibinfo {year} {2000})},\ \Eprint
  {http://arxiv.org/abs/gr-qc/0003032} {arXiv:gr-qc/0003032} \BibitemShut
  {NoStop}%
\bibitem [{SXS()}]{SXS:catalog}%
  \BibitemOpen
  \href@noop {} {}\bibinfo {howpublished}
  {\url{http://www.black-holes.org/waveforms}}\BibitemShut {NoStop}%
\bibitem [{\citenamefont {Fujita}(2015)}]{Fujita:2014eta}%
  \BibitemOpen
  \bibfield  {author} {\bibinfo {author} {\bibfnamefont {R.}~\bibnamefont
  {Fujita}},\ }\href {\doibase 10.1093/ptep/ptv012} {\bibfield  {journal}
  {\bibinfo  {journal} {PTEP}\ }\textbf {\bibinfo {volume} {2015}},\ \bibinfo
  {pages} {033E01} (\bibinfo {year} {2015})},\ \Eprint
  {http://arxiv.org/abs/1412.5689} {arXiv:1412.5689 [gr-qc]} \BibitemShut
  {NoStop}%
\bibitem [{Note2()}]{Note2}%
  \BibitemOpen
  \bibinfo {note} {F. Jim\'enez Forteza et al., LIGO Document T1600018,
  https://dcc.ligo.org/LIGO-T1600018/public}\BibitemShut {NoStop}%
\bibitem [{\citenamefont {Gonz\'alez}\ \emph {et~al.}(2007)\citenamefont
  {Gonz\'alez}, \citenamefont {Hannam}, \citenamefont {Sperhake}, \citenamefont
  {Brugmann},\ and\ \citenamefont {Husa}}]{Gonzalez:2007hi}%
  \BibitemOpen
  \bibfield  {author} {\bibinfo {author} {\bibfnamefont {J.~A.}\ \bibnamefont
  {Gonz\'alez}}, \bibinfo {author} {\bibfnamefont {M.~D.}\ \bibnamefont
  {Hannam}}, \bibinfo {author} {\bibfnamefont {U.}~\bibnamefont {Sperhake}},
  \bibinfo {author} {\bibfnamefont {B.}~\bibnamefont {Brugmann}}, \ and\
  \bibinfo {author} {\bibfnamefont {S.}~\bibnamefont {Husa}},\ }\href@noop {}
  {\bibfield  {journal} {\bibinfo  {journal} {Phys. Rev. Lett.}\ }\textbf
  {\bibinfo {volume} {98}},\ \bibinfo {pages} {231101} (\bibinfo {year}
  {2007})},\ \Eprint {http://arxiv.org/abs/gr-qc/0702052} {gr-qc/0702052}
  \BibitemShut {NoStop}%
\bibitem [{\citenamefont {Krishnan}\ \emph {et~al.}(2007)\citenamefont
  {Krishnan}, \citenamefont {Lousto},\ and\ \citenamefont
  {Zlochower}}]{Krishnan:2007pu}%
  \BibitemOpen
  \bibfield  {author} {\bibinfo {author} {\bibfnamefont {B.}~\bibnamefont
  {Krishnan}}, \bibinfo {author} {\bibfnamefont {C.~O.}\ \bibnamefont
  {Lousto}}, \ and\ \bibinfo {author} {\bibfnamefont {Y.}~\bibnamefont
  {Zlochower}},\ }\href {\doibase 10.1103/PhysRevD.76.081501} {\bibfield
  {journal} {\bibinfo  {journal} {Phys. Rev.}\ }\textbf {\bibinfo {volume}
  {D76}},\ \bibinfo {pages} {081501} (\bibinfo {year} {2007})},\ \Eprint
  {http://arxiv.org/abs/0707.0876} {arXiv:0707.0876 [gr-qc]} \BibitemShut
  {NoStop}%
\bibitem [{\citenamefont {Abbott}\ \emph
  {et~al.}(2016{\natexlab{g}})\citenamefont {Abbott} \emph
  {et~al.}}]{Abbott:2016izl}%
  \BibitemOpen
  \bibfield  {author} {\bibinfo {author} {\bibfnamefont {B.~P.}\ \bibnamefont
  {Abbott}} \emph {et~al.} (\bibinfo {collaboration} {Virgo, LIGO
  Scientific}),\ }\href {\doibase 10.1103/PhysRevX.6.041014} {\bibfield
  {journal} {\bibinfo  {journal} {Phys. Rev.}\ }\textbf {\bibinfo {volume}
  {X6}},\ \bibinfo {pages} {041014} (\bibinfo {year} {2016}{\natexlab{g}})},\
  \Eprint {http://arxiv.org/abs/1606.01210} {arXiv:1606.01210 [gr-qc]}
  \BibitemShut {NoStop}%
\bibitem [{Note3()}]{Note3}%
  \BibitemOpen
  \bibinfo {note} {N.K. Johnson-McDaniel et al., LIGO Document T1600168,
  https://dcc.ligo.org/LIGO-T1600168/public}\BibitemShut {NoStop}%
\end{thebibliography}%

\end{document}